\documentclass[a4paper,11pt]{article}
 \pdfoutput=1
 \usepackage{jheppub} 
 \usepackage{graphicx}
 \usepackage{amssymb}
\usepackage{dcolumn}
\usepackage{bm}
\usepackage{color}
\usepackage{multirow}
 
\newcommand{\met}{{/\!\!\! E_T}} 
\newcommand{\mpt}{{\;/\!\!\!\! \vec{P}_T}} 

 \newcommand{\lsim}{{\;\raise0.3ex\hbox{$<$\kern-0.75em\raise-1.1ex\hbox{$\sim$}}\;}}
\newcommand{\gsim}{{\;\raise0.3ex\hbox{$>$\kern-0.75em\raise-1.1ex\hbox{$\sim$}}\;}}
\newcommand{\beq}{\begin{equation}}
\newcommand{\eeq}{\end{equation}}
\newcommand{\bea}{\begin{eqnarray}}
\newcommand{\eea}{\end{eqnarray}}

\def\baa{\begin{array}}
\def\eaa{\end{array}}

\mathchardef\minus="002D

\def\met{E_T\hspace{-0.45cm}/\hspace{0.25cm}}

\title{\boldmath On-shell constrained $M_2$ variables with applications to mass measurements and topology disambiguation}
 
\author[a]{Won Sang Cho,}  
\author[a]{James S.~Gainer,}
\author[a]{Doojin Kim,}
\author[a]{Konstantin T.~Matchev,} 
\author[b]{Filip Moortgat,}
\author[b]{Luc Pape,}
\author[c]{Myeonghun Park\footnote{Corresponding author: myeonghun.park@ipmu.jp}} 

\affiliation[a]{Physics Department, University of Florida, Gainesville, FL 32611, USA}
\affiliation[b]{CERN, Geneva CH-1211, Switzerland}
\affiliation[c]{Kavli Institute for the Physics and Mathematics of the Universe (WPI), Todai Institutes for Advanced Study, The University of Tokyo, Japan}

\abstract{We consider a class of on-shell constrained mass variables that are 
3+1 dimensional generalizations of the Cambridge $M_{T2}$ variable and that
automatically incorporate various assumptions about the underlying event topology.
The presence of additional on-shell constraints causes their kinematic distributions 
to exhibit sharper endpoints than the usual $M_{T2}$ distribution. 
We study the mathematical properties of these new variables, 
e.g.,~the uniqueness of the solution selected by the minimization 
over the invisible particle 4-momenta. 
We then use this solution to reconstruct the masses of various particles along the decay chain.
We propose several tests for validating the assumed event topology 
in missing energy events from new physics. The tests
are able to determine: 1) whether the decays in the event are two-body
or three-body,
2) if the decay is two-body, whether the intermediate resonances in
the two decay chains are the same,
and 3) the exact sequence in which the visible particles are emitted
from each decay chain. }
 
\preprint{IPMU14-0006} 
\date{January 7, 2014}

\begin{document} 
\maketitle
\flushbottom

\section{Introduction}
\label{sec:introduction}

The measurement of particle properties in events with missing energy at hadron colliders 
is a challenging problem which has been receiving increased attention as of late
(see~\cite{Barr:2010zj} and~\cite{Wang:2008sw} for reviews on mass and spin measurement methods, respectively).
The difficulty arises because in most new physics models with dark matter candidates, 
some conserved, often $Z_2$, parity is needed to make the dark matter stable.  
Particles which are charged with respect to this parity are pair produced; 
each such event contains at least two invisible (dark matter) particles whose energy and momenta are not measured.
It is precisely this lack of information which makes the straightforward application of standard 
mass reconstruction techniques impossible.

In order to deal with the lack of knowledge about the invisible particle momenta, the following
three approaches have been suggested:
\begin{itemize}
\item {\em Use variables built from measured momenta only.}

The best known example is the invariant mass of (sets of) visible particles observed
in the detector. The measurement of kinematic endpoints in various invariant mass distributions 
is the classic method for mass determination in supersymmetry 
~\cite{Hinchliffe:1996iu,Bachacou:1999zb,Allanach:2000kt,Gjelsten:2004ki,Gjelsten:2005aw,Matchev:2009iw}.
Other recently proposed variables include the contransverse mass variable $M_{CT}$ 
~\cite{Tovey:2008ui,Polesello:2009rn} and its variants $M_{CT_\perp}$  and $M_{CT_\parallel}$~\cite{Matchev:2009ad},
the ratio of visible transverse energies~\cite{Nojiri:2000wq,Cheng:2011ya},
and the energy itself~\cite{Agashe:2012bn,Agashe:2012fs,Agashe:2013eba}.

Of course, while the {\em individual} invisible momenta are unknown, the sum of their transverse components
is measured as the missing transverse momentum $\mpt$ of the event. Thus one could also consider 
variables which are functions of the visible momenta and $\mpt$, e.g.,~the 
transverse mass~\cite{Smith:1983aa,Barger:1983wf}, the effective mass $M_{eff}$
\cite{Hinchliffe:1996iu,Tovey:2000wk}, the minimum partonic center-of-mass energy 
$\sqrt{\hat{s}}_{min}$~\cite{Konar:2008ei,Konar:2010ma,Robens:2011zm},
the razor variables~\cite{Rogan:2010kb,Buckley:2013kua}, etc. Such variables provide a good global 
characterization of the event, and are useful for discriminating signal from background, measuring an 
overall scale, or determining a signal rate. However, since they are not very sensitive to the particular details of the event,
they are far from ideal for the purposes of precision studies of the signal. 
\item {\em Calculate exactly the unknown individual momenta of the invisible particles.}

This is generally done by assuming a specific event topology and imposing a sufficient 
number of on-shell constraints~\cite{Kawagoe:2004rz,Nojiri:2007pq,Cheng:2008mg,Cheng:2009fw}.
If applicable, this method is very powerful, since the event kinematics is fully determined
and one can easily move on to precision studies~\cite{Cheng:2010yy}. The main disadvantage of
exact reconstruction techniques is that they require sufficiently long decay chains in order to provide
the required number of mass-shell constraints. Otherwise, the system is underconstrained,
and mass measurements are only possible on a statistical basis, by testing for consistency over
the whole ensemble of signal events~\cite{Cheng:2007xv,Cheng:2008hk}.
\item {\em Use a compromise approach.}

The third approach is a compromise between the previous two --- one still constructs kinematic variables 
which depend on the invisible momenta, but one gives up on trying to determine those momenta exactly 
on an event-per-event basis. Instead, some kind of ansatz is used to assign values (consistent with the measured $\mpt$)
to the individual momenta of the invisible particles in each event. The most celebrated variable of this 
class is the Cambridge $M_{T2}$ variable~\cite{Lester:1999tx,Barr:2003rg}, which is calculated by fixing the 
{\em transverse} momenta of the invisible particles to minimize the resulting transverse mass of the 
(larger of the two) parent particles. The idea of fixing the unknown invisible momenta by minimizing
a suitable mass function is very powerful, and many of the kinematic variables proposed in the literature can be
reinterpreted that way~\cite{Barr:2011xt}. The $M_{T2}$ approach is very well developed by now ---
analytical formulas exist for the calculation of $M_{T2}$ in a given event and for the interpretation of its endpoint 
\cite{Cho:2007qv,Gripaios:2007is,Barr:2007hy,Cho:2007dh,Burns:2008va,Lester:2011nj,Mahbubani:2012kx}. 
Since the original $M_{T2}$ proposal~\cite{Lester:1999tx,Barr:2003rg}, 
several other related variables have been suggested as well, 
e.g.~$M_{T2\perp}$ and $M_{T2\parallel}$ \cite{Konar:2009wn},
the asymmetric $M_{T2}$ \cite{Barr:2009jv,Konar:2009qr},
$M_{CT2}$ \cite{Cho:2009ve,Cho:2010vz}, and
$M_{T2}^{approx}$ \cite{Lally:2012uj}.

Note that the $M_{T2}$ prescription determines only the transverse
components of the invisible momenta. In order to fix the longitudinal components, one could rely on additional
measurements or assumptions. For example, in the $M_{T2}$-assisted on-shell (MAOS) reconstruction 
method, one uses the measured $M_{T2}$ kinematic endpoint and enforces the on-shell condition for the 
mother particle, which allows one to solve for the longitudinal momenta~\cite{Cho:2008tj,Park:2011uz}. (The idea behind the
$M_{2C}$ variable~\cite{Ross:2007rm,Barr:2008ba} is very similar.) A variation of this method
arises if the invisible particles are neutrinos from $W$ (or $\tau$) decays --- then one can use the known $W$-boson 
(or $\tau$-lepton) 
mass as a constraint and again solve for the longitudinal momenta~\cite{Choi:2009hn,Choi:2010dw,Barr:2011he,Barr:2011ux,Barr:2011si,Guadagnoli:2013xia}.
Since the on-shell constraints are nonlinear functions, the MAOS approach typically yields 
multiple solutions for the longitudinal momentum components, so one must also specify a prescription for handling
this multiplicity. 

An alternative approach to MAOS, which may avoid this ambiguity, was outlined in Ref.~\cite{Barr:2011xt},
which pointed out that the $M_{T2}$ variable and its friends allow a 3+1 dimensional formulation,
in which one always deals with the {\em actual} instead of the transverse masses.  The corresponding
3+1 dimensional analogue of $M_{T2}$ was denoted simply as $M_2$, omitting the transverse 
index\footnote{Supersymmetry aficionados should not confuse $M_2$ with the wino mass parameter.}.
The actual mass, being 3+1 dimensional, already carries dependence on both transverse and 
longitudinal momentum components, thus the minimization procedure required to obtain $M_2$ is expected
to automatically assign unique values for {\em all} momentum components of {\em each} individual invisible particle.
Since much of our discussion below will make crucial use of this property, 
we will discuss carefully the minimization procedure for the different $M_2$-type variables 
and the uniqueness of the resulting solutions for the invisible momenta in Sec.~\ref{sec:relation}.
\end{itemize}

An important benefit from extending the transverse $M_{T2}$ formalism to the
3+1-dimen\-sional $M_2$ language was recently emphasized in~\cite{Mahbubani:2012kx}. 
In many practical applications of $M_{T2}$ and similar kinematic variables, one 
has in mind a very specific signal topology, which in turn implies additional kinematic 
constraints on the (unknown) individual invisible momenta. For example, SUSY decay chains 
often proceed through intermediate on-shell resonances, the classic example being 
the decay of a heavy gluino through a lighter on-shell intermediate squark.
While the mass of the intermediate resonance is {\em a priori} unknown, in symmetric event
topologies the two decay chains are identical, so one may still impose the condition that the
mass of the intermediate resonance (whatever its value) ends up being equal in the two 
decay chains~\cite{Mahbubani:2012kx} (for specific applications to 
$H\to \tau^+\tau^-$ and $H\to WW$ decay, see \cite{Barr:2011he} and \cite{Barr:2011si}, respectively). 
Adding such on-shell constraints further restricts the allowed 
domain of values for the components of the individual invisible momenta and in general leads to 
a different outcome from the minimization procedure, resulting in 
a new set of kinematic 
variables\footnote{Note that it is not possible to add such constraints in the case of
transverse variables like $M_{T}$, $M_{T2}$, $M_{CT}$, etc.}. 

In this paper we shall extend the 
discussion from~\cite{Mahbubani:2012kx}, which focused only on
intermediate resonances, i.e.,~particles
appearing in the decay chain in between the decaying parent and the corresponding daughter.
In particular, we shall allow ourselves to also consider resonances which appear ``outside" the parent-daughter system,
e.g.,~progenitor particles upstream from the parents, or descendant particles downstream from the 
daughters. The benefits from this generalization will become clear in the physics examples studied below.

In the paper,
we study the mathematical properties of these on-shell constrained $M_2$ kinematic 
variables and propose several novel techniques for 
mass measurements and for disambiguating alternative event topologies. Our main results are:
\begin{itemize}
\item {\em We find that differential distributions of the constrained $M_2$ variables exhibit sharper 
kinematic endpoints, making them easier to measure in the presence of backgrounds.}  

This is because, as expected, the addition of on-shell kinematic constraints generally increases 
the value of the corresponding $M_2$ variable, thus providing a more stringent lower bound on the mass
of the parent. The sharper endpoints would ultimately lead to an improvement 
in the precision with which the parent masses can be determined experimentally.
\item {\em We propose a new method for measuring the mass of a heavy resonance in 
a SUSY decay chain, by using the invisible momenta found during the $M_2$ minimization.} 

The standard procedure so far has been to treat that resonance as a parent particle
in a suitably defined subsystem of the event~\cite{Burns:2008va}, then measure the upper 
kinematic endpoint of the corresponding $M_{T2}$ distribution. Instead, here we treat 
the resonance as an on-shell constraint to be applied during the minimization process while calculating 
the $M_2$ variable for a suitably defined subsystem (which may or may not extend over the resonance itself).
Since the $M_2$ minimization procedure selects a unique configuration for the individual invisible 
momenta, one has all the information required to reconstruct the mass of this hypothetical resonance directly. 
The key observation, supported in our examples in Sec.~\ref{sec:peak} below, is that the peak of that mass distribution
is very well correlated with the true mass of the resonance. The spirit of our method is similar 
to MAOS reconstruction~\cite{Cho:2008tj,Choi:2009hn,Choi:2010dw,Park:2011uz,Choi:2011ys,Guadagnoli:2013xia} 
and the $M_{2C}$ approach~\cite{Ross:2007rm}.
The difference is that we do not rely on preliminary measurements of kinematic endpoints;
the measurement is instead done from first principles.
\item {\em We find that this new method, in combination with other standard techniques, can
be used to determine the mass of the invisible (dark matter) particles.}

An interesting feature of the method just described is that the result exhibits a different functional dependence on the test 
daughter mass than the results from analogous methods based on $M_{T2}$ or invariant mass kinematic endpoints.
This means that one is able to obtain the true daughter mass by simply putting together the functional parent-daughter
mass relationship obtained from our method and from the other canonical methods in the literature --- the true
answer is given by the crossing point of the different curves. This technique is complementary to the $M_{T2}$
``kink" method where one looks for a kink instead of a crossing point~\cite{Cho:2007qv,Gripaios:2007is,Barr:2007hy,Cho:2007dh,Barr:2009jv}
(other techniques for measuring the absolute daughter mass are described in 
\cite{Matchev:2009ad,Matchev:2009fh,Alwall:2009sv,Konar:2009wn,Cohen:2010wv}). 
\item {\em We propose methods for identifying the event topology and resolving 
combinatorial ambiguities.}

The large variety of on-shell constrained $M_2$ variables allows us to address a long standing problem
in SUSY phenomenology, namely, the question of identifying the correct event topology.
There are two aspects of the problem --- first, resolving the combinatorial ambiguities in assigning
the observed final state particles to the hypothesized event topology~\cite{Rajaraman:2010hy,Baringer:2011nh,Choi:2011ys},
and second, validation of the hypothesized event topology itself, 
e.g.,~the partitioning into two decay chains~\cite{Bai:2010hd,Cho:2012er}, 
the number of invisible particles~\cite{Agashe:2010gt,Agashe:2010tu,Giudice:2011ib,Cho:2012er},
the number of intermediate on-shell resonances~\cite{Bai:2010hd,Cho:2012er}, etc.
We can use the fact that the different versions of our on-shell constrained $M_2$ variables 
have different assumptions about the underlying event topology built in. Thus, by comparing results 
obtained with different $M_2$ variables, we can test those assumptions, for example:
\begin{enumerate}
\item In Sec.~\ref{sec:endpoint} we design a method which tests for the presence of 
intermediate on-shell resonances in the SUSY decay chain, i.e.,~distinguishes between 
a sequence of two 2-body decays and a single 3-body decay.
\item In Sec.~\ref{sec:Dalitz} we address the question of the proper sequence in which the visible
particles get emitted along a SUSY decay chain. We use the invisible particle momenta selected by the 
$M_2$ minimization procedure to construct Dalitz-type plots involving invariant masses of suitable particle pairs.
The correct ordering of the visible particle is then determined by comparing the characteristic shapes of 
those plots. 
\item A similar idea, illustrated in Sec.~\ref{sec:scatter}, can be used to test whether the events 
are symmetric, i.e.,~whether the two decay chains are the same~\cite{Konar:2009qr}. 
\end{enumerate}
\end{itemize}

\begin{figure}[t]
\centering
\includegraphics[scale=1]{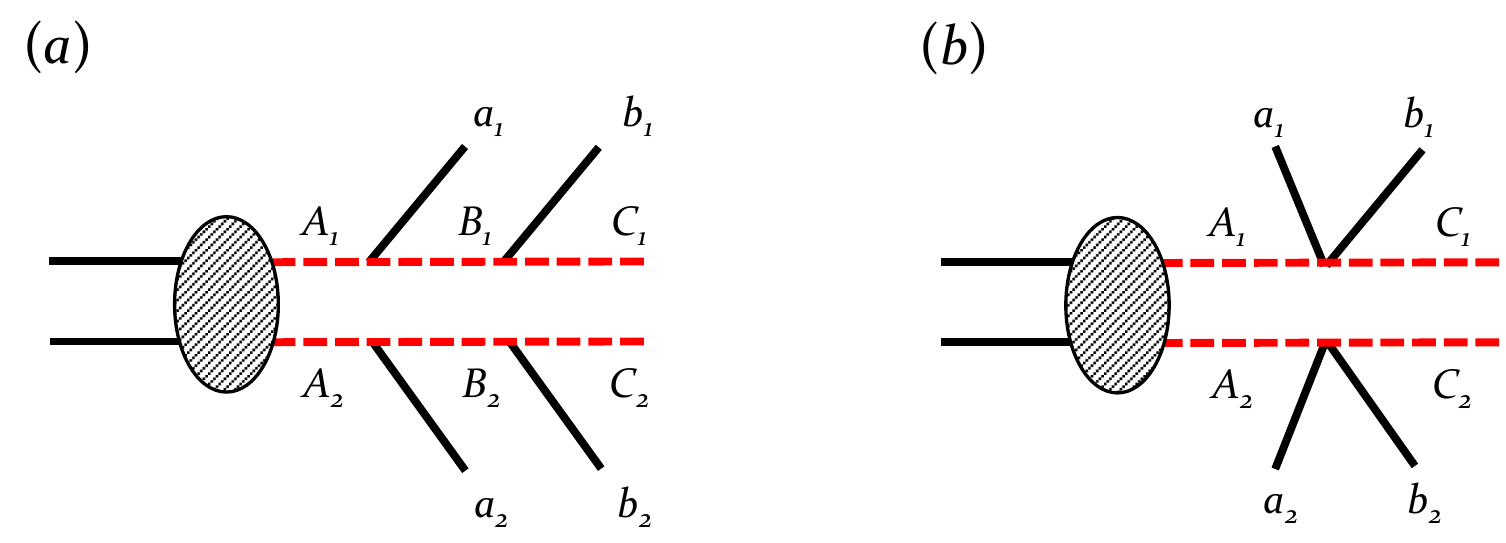}
\caption{\label{fig:process} The decay topologies under consideration in this paper. 
In diagram (a), each parent particle, $A_i$, ($i=1,2$) decays to two visible particles, $a_i$ and $b_i$,
and an invisible daughter particle, $C_i$, through an intermediate on-shell resonance, $B_i$.
In diagram (b), the intermediate state, $B_i$, is absent (or very heavy) and the 
$A_i\to a_i b_i C_i$ decay is a three body process.}
\end{figure}

The paper is organized as follows. In Sec.~\ref{sec:notation}, we specify the process studied 
(depicted in Fig.~\ref{fig:process}) and introduce our conventions and terminology. 
We then proceed to define all possible on-shell constrained $M_2$ variables
for that process (a total of 12 variables altogether, listed in Table~\ref{tab:M2variables}).
However, not all of those variables are independent --- Sec.~\ref{sec:relation} discusses
the existing relationships among them, including the connection to the Cambridge $M_{T2}$ variable\footnote{Readers 
who are mostly interested in the practical applications of the $M_2$ variables 
and wish to skip over the math are invited to jump straight to Sec.~\ref{sec:summary}, where they 
will find a summary of the main results from Sec.~\ref{sec:relation}.}.
The subsequent sections demonstrate the utility of those variables for practical applications:
mass measurements from kinematic endpoints (Sec. \ref{sec:mass}),
mass measurements from $M_2$-assisted peak reconstruction (Sec. \ref{sec:peak}),
and topology disambiguation (Sec. \ref{sec:topology}).
Sec.~\ref{sec:conclusions} is reserved for our conclusions.

\section{Notations and setup}
\label{sec:notation}

\subsection{The physics process}

In this paper we shall consider the generic processes depicted in Fig.~\ref{fig:process}. We assume the pair production
of two heavy particles, $A_1$ and $A_2$, which decay in a similar fashion: 
\beq
A_i\to a_ib_iC_i, \quad (i=1,2).
\label{eq:process}
\eeq
The process (\ref{eq:process}) may occur either through on-shell intermediate resonances, $B_i$, as in Fig.~\ref{fig:process}(a),
or as a genuine three-body decay, as in Fig.~\ref{fig:process}(b).
The particles, $C_i$, are invisible in the detector --- in realistic models, their role is typically played by some dark matter 
candidate, e.g.,~the lightest supersymmetric particle (LSP) in supersymmetry. The particles, $a_i$ and $b_i$, are SM particles 
which are visible in the detector, thus their 4-momenta $p_{a_1}^\mu$, $p_{b_1}^\mu$, $p_{a_2}^\mu$, and $p_{b_2}^\mu$
are measured known quantities. In contrast, the 4-momenta of the $C_i$, which we shall denote by $q_i^\mu$, are 
{\em a priori} unknown\footnote{Note that in our notation, the letter ``p" is used for measured momenta, while the 
letter ``q" refers to the unknown momenta of invisible particles. Since for the process of Fig.~\ref{fig:process}
there are only two invisible particles in the final state, we simplify the notation by using $\vec{q}_i$
instead of the clumsier $\vec{q}_{C_i}$.}, and are only constrained by the $\mpt$ measurement:
\beq
\vec{q}_{1T}+\vec{q}_{2T}=\mpt. 
\label{eq:mpt}
\eeq
The masses of the particles along the red dashed lines in Fig.~\ref{fig:process}
are denoted by $m_{A_1}$, $m_{B_1}$, $\cdots$, $m_{C_2}$.
The process (\ref{eq:process}) depicted in Fig.~\ref{fig:process} covers a large class of physically interesting 
and motivated scenarios, including dilepton events from top pair production and decay, 
stop decays in supersymmetry ($\tilde t\to b\ell\tilde\nu_\ell$),
and many more.

In what follows, we shall assume that all four visible particles $a_i$ and $b_i$
in Fig.~\ref{fig:process} are distinguishable. As already mentioned in the introduction, 
depending on the nature of the visible particles $a_i$ and $b_i$, various combinatorial issues may arise, e.g.:
\begin{enumerate}
\item Should the four visible particles be partitioned as $2+2$, $1+3$, or $0+4$? This question
can be answered relatively easily by studying suitable invariant mass distributions of the 
visible particles \cite{Bai:2010hd}.
\item Another question is, {\em which} visible particles belong to the first decay chain ($a_1$, $b_1$) and which 
belong to the second ($a_2$, $b_2$). Two possible approaches have been pursued: first, by applying suitable cuts,
one could try to increase the chances of picking the correct pairwise assignment 
\cite{Rajaraman:2010hy,Baringer:2011nh,Choi:2011ys}.
Alternatively, one could consider all possible assignments and then try to subtract out the 
contributions from wrong assignments (e.g., by the mixed event subtraction technique
\cite{Hinchliffe:1996iu}). 
\item Finally, when $a_i$ is distinguishable from $b_i$, one could also ask
which of these two particles was emitted first and which came second.
The answer to this question will be the subject of Sec.~\ref{sec:Dalitz}.
\end{enumerate}

\subsection{$M_2$ subsystems and the particle family tree}
 
\begin{figure}[t]
\centering
\includegraphics[scale=1]{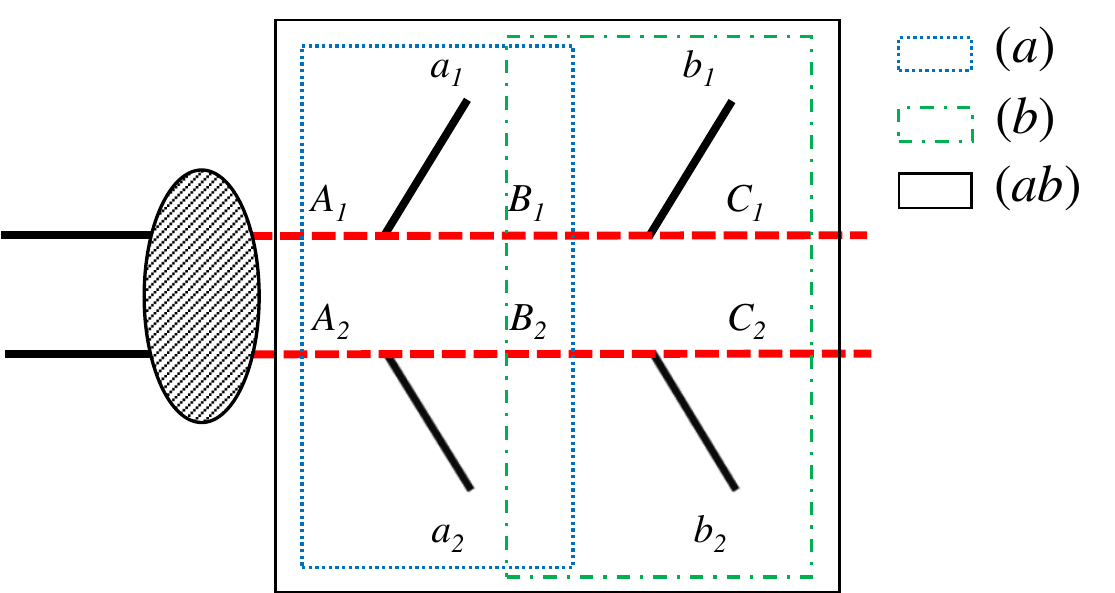}
\caption{\label{fig:DecaySubsystem} The decay process from Fig.~\ref{fig:process}(a)
with the corresponding subsystems explicitly delineated. The blue dotted, green dot-dashed, and black solid
lines indicate the subsystems $(a)$, $(b)$, and $(ab)$, respectively.}
\end{figure}

As first discussed in the context of the $M_{T2}$ variable~\cite{Burns:2008va}, one can proliferate the number of useful
measurements by considering different subsystems within the original event. The subsystems are defined by the sets of visible particles 
which are used to construct an $M_{T2}$ variable (see Fig.~\ref{fig:DecaySubsystem}):
\begin{itemize}
\item The $(ab)$ subsystem, indicated by the solid black box in Fig.~\ref{fig:DecaySubsystem}.
Here one uses both types of visible particles, $a_i$ and $b_i$, treating $A_i$ as parent particles
and $C_i$ as daughter particles.
\item The $(a)$ subsystem, shown by the blue dotted box in Fig.~\ref{fig:DecaySubsystem}.
Now one uses only the visible particles, $a_i$, but not $b_i$. The $A_i$ particles are again 
treated as parents, but the daughters are now the $B_i$ particles.
\item The $(b)$ subsystem, depicted by the green dot-dashed box in Fig.~\ref{fig:DecaySubsystem}.
Now the visible particles, $b_i$, are used, but not $a_i$. The parents are the $B_i$ particles and the 
daughters are the $C_i$ particles.
\end{itemize}
In this paper, the $M_{T2}$ variables corresponding to these three subsystems will be denoted 
as\footnote{Contrast this to the superscript notation previously used in 
\cite{Burns:2008va,Chatrchyan:2013boa}: $M_{T2}^{220}$, $M_{T2}^{221}$, and $M_{T2}^{210}$.}
$M_{T2}(ab)$, $M_{T2}(a)$,  and $M_{T2}(b)$; the same convention will be used for the $M_2$ variables defined below.
 
We see that, depending on our choice of subsystem, each particle from Fig.~\ref{fig:process}(a)
can be classified into one of the following three categories (summarized also in Table~\ref{tab:tree}):
\begin{table}[t]
\centering
\begin{tabular}{|c|c|c|c|}
\hline
Subsystem & Parents $P_i$     & Daughters $D_i$ & Relatives $R_i$\\ 
\hline \hline
$(ab)$  &   $A_i$   &    $C_i$     &     $B_i$    \\
$(a)$   &   $A_i$   &    $B_i$     &     $C_i$    \\
$(b)$   &   $B_i$   &    $C_i$     &     $A_i$    \\
\hline
\end{tabular}
\caption{\label{tab:tree} The roles played by different particles depending on the subsystem under consideration. }
\end{table}
\begin{itemize}
\item {\bf Parents}. These are the two particles at the top of the decay chains in a given subsystem.
In the following, we shall denote the parents by $P_i, (i=1,2)$ and their masses by $M_{P_i}$. 
The $M_2$ kinematic variables in Sec.~\ref{sec:M2def} below will be defined by a suitable minimization of the 
parent masses, $M_{P_i}$, over the unknown components of the invisible momenta~\cite{Barr:2011xt}.
\item {\bf Daughters}. These are the two particles at the end of the decay chains in a given subsystem.
They may or may not be LSPs; see Table~\ref{tab:tree}. The daughters will be denoted by $D_i$
and their masses by $M_{D_i}$. Each parent mass, $M_{P_i}$, is a function of the corresponding
daughter mass, $M_{D_i}$, which is {\em a priori} unknown. Thus when calculating parent masses, one
must always specify a test daughter mass parameter, which will be denoted by $\tilde{m}$ throughout this paper.
For the most part, we shall be considering ``symmetric" events, i.e.,~events in which the two decay chains are 
the same, and thus there is a single test mass $\tilde m$. The generalization to the asymmetric case is 
straightforward~\cite{Konar:2009qr} --- one simply needs to introduce separate test masses, $\tilde m_i$, 
for the upper and the lower decay chains in Fig.~\ref{fig:process}.
\item {\bf Relatives}. These are particles which are neither parents nor daughters; see Table~\ref{tab:tree}.
The relatives will be denoted by $R_i$ and their masses by $M_{R_i}$. Since the decay chains 
in Fig.~\ref{fig:process}(a) involve only 3 new particles, there is always only one possible relative, 
which may appear upstream (as in the case of subsystem $(b)$), downstream (as in the case of subsystem $(a)$),
or midstream (as in the case of subsystem $(ab)$). In other words, for the simple example
of Fig.~\ref{fig:process}(a), the identity of the relative is uniquely fixed once we specify the subsystem under 
consideration, so we do not need to introduce any additional notation regarding the relatives. However, in
more complicated examples with longer decay chains, there will be several relatives, and one would have
to invent some notation to distinguish among them.
\end{itemize}

\subsection{Definition of the on-shell constrained $M_2$ variables}
\label{sec:M2def}

We start by reviewing the standard definition of the canonical $M_{T2}$ variable~\cite{Lester:1999tx}. 
Consider the transverse masses $M_{TP_i}(\vec{q}_{iT},\tilde m)$ of the two parent particles and then minimize the 
larger of them with respect to the transverse\footnote{The longitudinal components 
$q_{1z}$ and $q_{2z}$ are irrelevant since they do not enter the definition 
of the transverse masses $M_{TP_i}$.} components of the invisible momenta, subject to the 
$\mpt$ constraint, (\ref{eq:mpt}):
\bea
M_{T2} (\tilde m) &\equiv& \min_{\vec{q}_{1T},\vec{q}_{2T}}\left\{\max\left[M_{TP_1}(\vec{q}_{1T},\tilde m),\;M_{TP_2} (\vec{q}_{2T},\tilde m)\right] \right\}.  
\label{eq:mt2def}\\
\vec{q}_{1T}+\vec{q}_{2T} &=& \mpt   \nonumber
\eea 

Following~\cite{Barr:2011xt}, one could instead start with the actual parent masses, $M_{P_i}$, 
and define the 3+1-dimensional analogue of (\ref{eq:mt2def}) as
\bea
M_{2} (\tilde m) &\equiv& \min_{\vec{q}_{1},\vec{q}_{2}}\left\{\max\left[M_{P_1}(\vec{q}_{1},\tilde m),\;M_{P_2} (\vec{q}_{2},\tilde m)\right] \right\},  
\label{eq:m2def}\\
\vec{q}_{1T}+\vec{q}_{2T} &=& \mpt   \nonumber
\eea 
where the minimization is performed over the 3-component momentum vectors $\vec{q}_{1}$ and $\vec{q}_{2}$.
As stated in~\cite{Ross:2007rm,Barr:2011xt}, the two definitions (\ref{eq:mt2def}) and (\ref{eq:m2def})
are equivalent, in the sense that the resulting two variables, $M_{T2}$ and $M_2$, will have the same numerical 
value (a proof of this claim can be found in Section~\ref{sec:equivalence} below). 
Nevertheless, for our purposes here, the definition (\ref{eq:m2def}) is much more convenient, for the following reasons:
\begin{itemize}
\item The minimization in (\ref{eq:m2def}) is done over the full 3-momentum vectors $\vec{q}_{1}$ and $\vec{q}_{2}$,
and thus it also selects their longitudinal components $q_{1z}$ and $q_{2z}$. This completely fixes the kinematics of the event.
\item The 3+1-dimensional language of Eq.~(\ref{eq:m2def}) makes it very easy to impose the additional on-shell 
constraints that arise in specific event topologies~\cite{Mahbubani:2012kx}. 
\end{itemize}

Given that here we are interested in the specific event topology of Fig.~\ref{fig:process}(a), it makes sense to consider 
additionally constrained versions of (\ref{eq:m2def}). There are two\footnote{Recall that throughout this paper we are 
already making the assumption that the daughters are the same.} additional assumptions one can make:
that the parents $P_i$ are the same (or, more generally, that they have the same mass)
\beq
M_{P_1} = M_{P_2},
\label{eq:parents}
\eeq
or that the relatives have the same mass
\beq
M_{R_1} = M_{R_2}.
\label{eq:relatives}
\eeq
Of course, one could also impose (\ref{eq:parents}) and (\ref{eq:relatives}) simultaneously, 
giving us a total of 4 possibilities. We choose to enumerate these 4 cases by adding two additional 
subscripts on the $M_2$ variable to indicate whether the constraints 
(\ref{eq:parents}) and (\ref{eq:relatives}) were imposed during the minimization or not.
The first subscript always refers to the parents and their constraint, (\ref{eq:parents}),
while the second subscript always refers to the relatives and their constraint, (\ref{eq:relatives}). 
The value of the subscript will be ``C" if the corresponding constraint is imposed and ``X" otherwise. 
Altogether, we have the following four types of variables:
\bea
M_{2XX} &\equiv& \min_{\vec{q}_{1},\vec{q}_{2}}\left\{\max\left[M_{P_1}(\vec{q}_{1},\tilde m),\;M_{P_2} (\vec{q}_{2},\tilde m)\right] \right\},  
\label{eq:m2XXdef}
\\
\vec{q}_{1T}+\vec{q}_{2T} &=& \mpt   \nonumber
\eea 
\bea
M_{2CX} &\equiv& \min_{\vec{q}_{1},\vec{q}_{2}}\left\{\max\left[M_{P_1}(\vec{q}_{1},\tilde m),\;M_{P_2} (\vec{q}_{2},\tilde m)\right] \right\},  
\label{eq:m2CXdef}\\
\vec{q}_{1T}+\vec{q}_{2T} &=& \mpt   \nonumber \\
M_{P_1}&=& M_{P_2} \nonumber 
\eea 
\bea
M_{2XC} &\equiv& \min_{\vec{q}_{1},\vec{q}_{2}}\left\{\max\left[M_{P_1}(\vec{q}_{1},\tilde m),\;M_{P_2} (\vec{q}_{2},\tilde m)\right] \right\},  
\label{eq:m2XCdef}\\
\vec{q}_{1T}+\vec{q}_{2T} &=& \mpt   \nonumber \\
M_{R_1}^2&=& M_{R_2}^2 \nonumber 
\eea 
\bea
M_{2CC} &\equiv& \min_{\vec{q}_{1},\vec{q}_{2}}\left\{\max\left[M_{P_1}(\vec{q}_{1},\tilde m),\;M_{P_2} (\vec{q}_{2},\tilde m)\right] \right\}.
\label{eq:m2CCdef}\\
\vec{q}_{1T}+\vec{q}_{2T} &=& \mpt   \nonumber \\
M_{P_1}&=& M_{P_2} \nonumber  \\
M_{R_1}^2&=& M_{R_2}^2 \nonumber 
\eea 

A few comments are in order. In the equations above, the masses of the parents, $M_{P_i}$, 
and the masses of the relatives, $M_{R_i}$, are always understood to be functions of the 
invisible 3-momenta $\vec{q}_i$. Thus the constraints $M_{P_1}= M_{P_2} $ and 
$M_{R_1} = M_{R_2}$ simply further restrict the allowed values 
for those momenta (in addition to the missing transverse momentum constraint, (\ref{eq:mpt})). 
Obviously, the unrestricted variable $M_{2XX}$ is nothing but the variable defined in 
(\ref{eq:m2def}), so in this sense the pair of indices ``XX" may seem redundant.
Nevertheless, given the existence of the other three choices (\ref{eq:m2CXdef}-\ref{eq:m2CCdef}),
it seems wise to indicate explicitly the absence of any on-shell constraints in that case.

We note that while a parent mass squared is always positive, there is one case when the 
mass squared of a {\em relative} can be negative --- for subsystem $(a)$, the relative particle
is $C_i$ and its mass squared is $M_{R_i}^2=(p_{B_i}-p_{b_i})^2$ (see Fig.~\ref{fig:process}(a)).
Each of the 4-momenta $p_{B_i}^\mu$ and $p_{b_i}^\mu$ is time-like, but their difference 
may be time-like or space-like. Thus, in that situation, one has the option of additionally requiring positivity of the 
masses squared of relative particles. In this paper we shall not do that;
we shall allow the relative masses squared obtained after the minimization to have either sign\footnote{The 
reason is that the momenta obtained in the minimization do not necessarily have to correspond to the 
momenta of any physical particles; as our reconstruction ansatz may not reflect the actual process.
A similar dilemma arises in the case of $M_{T2}$, when some invisible momenta found by the minimization
may turn out to be anomalously large, well beyond the scale of the collider energy.}.
This is why in Eqs.~(\ref{eq:m2XCdef}) and (\ref{eq:m2CCdef}), the constraint for the relatives 
is written as $M_{R_1}^2 = M_{R_2}^2$ instead of simply as $M_{R_1}= M_{R_2}$.

\begin{table}[t]
\centering
\begin{tabular}{||c|c||c|c||c|c||}
\hline
\multicolumn{2}{||c||}{Subsystem $(ab)$} & \multicolumn{2}{c||}{Subsystem $(a)$} & \multicolumn{2}{c||}{Subsystem $(b)$} \\ \hline
variable & constraints &variable & constraints &variable & constraints \\ 
\hline \hline
$M_{2XX}(ab)$ & --                                        & $M_{2XX}(a)$ & -- & $M_{2XX}(b)$ & -- \\ \hline
$M_{2CX}(ab)$ & $M_{A_1}^2=M_{A_2}^2$ & $M_{2CX}(a)$ & $M_{A_1}^2=M_{A_2}^2$ & $M_{2CX}(b)$ & $M_{B_1}^2=M_{B_2}^2$ \\ \hline
$M_{2XC}(ab)$ & $M_{B_1}^2=M_{B_2}^2$ & $M_{2XC}(a)$ & $M_{C_1}^2=M_{C_2}^2$ & $M_{2XC}(b)$ & $M_{A_1}^2=M_{A_2}^2$ \\ \hline
\multirow{2}{*}{$M_{2CC}(ab)$} & $M_{A_1}^2=M_{A_2}^2$ & \multirow{2}{*}{$M_{2CC}(a)$} & $M_{A_1}^2=M_{A_2}^2$ & \multirow{2}{*}{$M_{2CC}(b)$} & $M_{B_1}^2=M_{B_2}^2$ \\
 &$M_{B_1}^2=M_{B_2}^2$ & &$M_{C_1}^2=M_{C_2}^2$ & & $M_{A_1}^2=M_{A_2}^2$\\
\hline
\end{tabular}
\caption{\label{tab:M2variables} A summary of the twelve $M_2$ variables defined in the text.
For each of the three subsystems $(ab)$, $(a)$, and $(b)$, one may choose to apply neither, one, 
or both of the constraints (\ref{eq:parents}) and (\ref{eq:relatives}).
In each case, the trial daughter masses are assumed to be the same, $\tilde{m}$.}
\end{table}

Applying (\ref{eq:m2XXdef}-\ref{eq:m2CCdef}) to the three possible subsystems of 
Fig.~\ref{fig:DecaySubsystem}, we obtain a total of 12 
on-shell constrained $M_2$ variables which are listed in Table~\ref{tab:M2variables}.
Some of these variables ($M_{2XX}$ and $M_{2CX}$) are simply 3+1 dimensional versions of $M_{T2}$ 
\cite{Barr:2011xt,Mahbubani:2012kx}, while  $M_{2XC}(ab)$ and $M_{2CC}(ab)$ were
mentioned in~\cite{Mahbubani:2012kx}. The remaining 4 variables 
 $M_{2XC}(a)$,  $M_{2CC}(a)$,  $M_{2XC}(b)$, and $M_{2CC}(b)$ are new.  
Notice that the meaning of a ``C" index depends on both its position (first or second)
and on the chosen subsystem. For example, a ``C" index sitting in first position, $M_{2CX}(ab)$, 
implies equality of the parents: $M_{A_1}^2=M_{A_2}^2$, while when sitting in second position, $M_{2XC}(ab)$, it
indicates equality of the relatives: $M_{B_1}^2=M_{B_2}^2$. Similarly, 
contrast analogous variables in the three subsystems:
$M_{2XC}(ab)$ is calculated assuming $M_{B_1}^2=M_{B_2}^2$;
$M_{2XC}(a)$ is obtained with $M_{C_1}^2=M_{C_2}^2$; while
$M_{2XC}(b)$ implies $M_{A_1}^2=M_{A_2}^2$.

At this point, it is instructive to consider a couple of specific examples, in order to better familiarize the reader with our notation.
Consider, for example, $M_{2CC}(ab)$. It applies to the $(ab)$ subsystem, where $A_i$ are the parents, 
$C_i$ are the daughters (with test masses $\tilde m$) and $B_i$ are the relatives. Both indices are ``on", 
so the constraints (\ref{eq:parents}) and (\ref{eq:relatives}) are applied. Explicitly, we have
\bea
M_{2CC}^2(ab) &\equiv& \min_{\vec{q}_{1},\vec{q}_{2}}\left\{\max\left[(p_{a_1}+p_{b_1}+q_1)^2,\;(p_{a_2}+p_{b_2}+q_2)^2 \right] \right\}.
\label{eq:m2CCab}\\
q_1^2&=& \tilde{m}^2 \nonumber \\
q_2^2&=& \tilde{m}^2 \nonumber \\
\vec{q}_{1T}+\vec{q}_{2T} &=& \mpt   \nonumber \\
(p_{a_1}+p_{b_1}+q_1)^2&=&(p_{a_2}+p_{b_2}+q_2)^2 \nonumber \\
(p_{b_1}+q_1)^2&=&(p_{b_2}+q_2)^2 \nonumber
\eea 

As another example, consider $M_{2XC}(a)$. It applies to the $(a)$ subsystem with 
$A_i$ as parents, $B_i$ as daughters, and $C_i$ as relatives. 
Note that the test mass, $\tilde m$, now refers to $m_{B_i}$.
The parents are not assumed to have equal masses, but the relatives are, thus 
\bea
M_{2XC}^2(a) &\equiv& \min_{\vec{q}_{1},\vec{q}_{2}}\left\{\max\left[(p_{a_1}+p_{b_1}+q_1)^2,\;(p_{a_2}+p_{b_2}+q_2)^2 \right] \right\}.
\label{eq:m2XCa}\\
(q_1+p_{b_1})^2&=& \tilde{m}^2 \nonumber \\
(q_2+p_{b_2})^2&=& \tilde{m}^2 \nonumber \\
\vec{q}_{1T}+\vec{q}_{2T} &=& \mpt   \nonumber \\
q_1^2&=&q_2^2 \nonumber
\eea 

Our final example is $M_{2XC}(b)$, which reads
\bea
M_{2XC}^2(b) &\equiv& \min_{\vec{q}_{1},\vec{q}_{2}}\left\{\max\left[(p_{b_1}+q_1)^2,\;(p_{b_2}+q_2)^2 \right] \right\}.
\label{eq:m2XCb}\\
q_1^2&=& \tilde{m}^2 \nonumber \\
q_2^2&=& \tilde{m}^2 \nonumber \\
\vec{q}_{1T}+\vec{q}_{2T} &=& \mpt   \nonumber \\
(p_{a_1}+p_{b_1}+q_1)^2&=&(p_{a_2}+p_{b_2}+q_2)^2 \nonumber 
\eea 
If it wasn't for the very last constraint, this would have been simply $M_{T2}(b)$, i.e.,~the $M_{T2}$ variable 
for the $(b)$ subsystem, in the presence of upstream momentum $p_{a_1}+p_{a_2}$. However, the 
constraint for the relatives $M_{A_1}=M_{A_2}$ is non-trivial and leads to a qualitatively new result.

\section{Relations among the $M_2$ type variables and $M_{T2}$}
\label{sec:relation}

In this section, we examine the relations among the four $M_2$ type variables defined in the preceding section
and compare them to the conventional $M_{T2}$ variable. 
For concreteness, we shall focus on the $(ab)$ subsystem\footnote{However,
our results will hold for the other two subsystems as well; see the summary in Sec.~\ref{sec:summary}.}
and consider the set
\bea
M_{T2}(ab),\;\;\;M_{2XX}(ab),\;\;\;M_{2CX}(ab),\;\;\;M_{2XC}(ab),\;\;\;M_{2CC}(ab).
\label{eq:variables}
\eea
We shall perform our study under the assumption that the intermediate particles, $B_i$, are 
{\it on-shell} as in Fig.~\ref{fig:process}(a). The {\it off-shell} scenario of Fig.~\ref{fig:process}(b)
will be discussed in Sec.~\ref{sec:topology} in the context of applications. 
In Sec.~\ref{sec:equivalence}, we first show that the three variables, $M_{2XX}(ab)$, $M_{2CX}(ab)$,
and $M_{T2}(ab)$, have the same value event-by-event. Informed by this
discussion, in
Sec.~\ref{sec:uniqueness}, we shall
also discuss the question of the uniqueness of the invisible momentum configurations found in the 
process of minimization. Then, in Sec.~\ref{sec:2CC}, we shall discuss the hierarchy among the 
three distinct variables on the list (\ref{eq:variables}), namely $M_{2CX}(ab)$, $M_{2XC}(ab)$, and $M_{2CC}(ab)$.
In Sec.~\ref{sec:summary}, we summarize the main results from Sec.~\ref{sec:relation}.

\subsection{Equivalence theorem among $M_{2XX},\;M_{2CX}$, and $M_{T2}$}
\label{sec:equivalence}

Applying the general definition (\ref{eq:mt2def}) to the $(ab)$ subsystem, 
$M_{T2}(ab)$ can be expressed as follows~\cite{Lester:1999tx}:
\bea
M_{T2}^2(ab)&=&\min_{\vec{q}_{1T},\vec{q}_{2T}}
\left\{ \max \left[ M_{TA_1}^2(\vec{q}_{1T},\tilde m),\;M_{TA_2}^{2}(\vec{q}_{2T},\tilde m) \right] \right\}
\label{eq:mt2(ab)}
\\
\vec{q}_{1T}+\vec{q}_{2T} &=& \mpt   \nonumber
\eea
where
\bea
M_{TA_i}^2(\vec{q}_{iT},\tilde m)= {\tilde m}^2 + m_{v_i}^2 + 2 \left[ E_{v_iT}E_{q_iT}-\vec{p}_{v_iT}\cdot\vec{q}_{iT} \right],
\label{eq:MTAi}
\eea
$v_i$ is the visible state $a_i+b_i$ belonging to the $i$-th decay chain:
\beq
\vec{p}_{v_i}\equiv \vec{p}_{a_i}+\vec{p}_{b_i},
\eeq
and $E_T$ denotes the transverse energy: 
\bea
E_{v_i T}=\sqrt{m_{v_i}^2+\vec{p}_{v_iT}^{\,2}}\, ; 
\qquad 
E_{q_i T}=\sqrt{{\tilde m}^2+\vec{q}_{iT}^{\,2}}.
\eea 
Using (\ref{eq:m2XXdef}), we can construct $M_{2XX}(ab)$ in a similar manner:
\bea
M_{2XX}^2(ab) &=& \min_{\stackrel{\vec{q}_{1T},\vec{q}_{2T}}{q_{1z},q_{2z}} }
\left\{ \max \left [ M_{A_1}^2(\vec{q}_{1T},q_{1z},\tilde m),\;M_{A_2}^2(\vec{q}_{2T},q_{2z},\tilde m) \right] \right\}.
\label{eq:m2XX(ab)}
\\
\vec{q}_{1T}+\vec{q}_{2T} &=& \mpt   \nonumber
\eea
The invariant masses of $A_1$ and $A_2$ can be written as
\bea
M_{A_i}^2(\vec{q}_{iT},q_{iz},\tilde m)= {\tilde m}^2 + m_{v_i}^2 
+ 2\left[ E_{v_iT}E_{q_iT}\cosh(\Delta \eta_i)-\vec{p}_{v_iT}\cdot\vec{q}_{iT}\right],\label{eq:MAi}
\eea
where $\Delta \eta_i$ is the rapidity difference between the visible state $v_i$ and particle $C_i$. 
The minimization of (\ref{eq:m2XX(ab)}) over the transverse momenta, $\vec{q}_{iT}$, and the 
longitudinal momenta, $q_{iz}$, can in principle be done in any order, but it is much easier to minimize over
$q_{iz}$ first, since they do not enter the $\mpt$ constraint. Furthermore, the longitudinal momenta 
are decoupled from each other, and thus the two minimizations can be performed independently. 
We can therefore rewrite (\ref{eq:m2XX(ab)}) as
\bea
M_{2XX}^2(ab) &=& \min_{\vec{q}_{1T},\vec{q}_{2T}} 
\left\{ \max \left [ \min_{q_{1z}} 
\left\{M_{A_1}^2(\vec{q}_{1T},q_{1z},\tilde m)\right\},\;\min_{q_{2z}} 
\left\{M_{A_2}^2(\vec{q}_{2T},q_{2z},\tilde m)\right\}\right] \right\}.~~~
\label{eq:m2XX(ab)_switched}
\\
\vec{q}_{1T}+\vec{q}_{2T} &=& \mpt   \nonumber
\eea
where we have switched the order of the $\min_{q_{iz}}\{\}$ and $\max\{\}$ operations.
The minimization over $q_{iz}$ is equivalent to minimization over $\Delta\eta_i$.
From (\ref{eq:MAi}) it is easy to see that the minimum is obtained for $\Delta\eta_i=0$,
which reduces (\ref{eq:MAi}) to (\ref{eq:MTAi}), so that (\ref{eq:m2XX(ab)_switched}) becomes simply
\bea
M_{2XX}^2(ab) &=& \min_{\vec{q}_{1T},\vec{q}_{2T}} 
\left\{ \max \left [ 
M_{TA_1}^2(\vec{q}_{1T},\tilde m),\;
M_{TA_2}^2(\vec{q}_{2T},\tilde m)\right] \right\}.~~~
\label{eq:m2XX(ab)_final}
\\
\vec{q}_{1T}+\vec{q}_{2T} &=& \mpt   \nonumber
\eea
Comparing (\ref{eq:m2XX(ab)_final}) with (\ref{eq:mt2(ab)}), we see that~\cite{Barr:2011xt}
\bea
M_{2XX}^2 (ab)= M_{T2}^2(ab). 
\label{eq:m2xxmt2}
\eea

Moving our attention to $M_{2CX}(ab)$, we see that the proof of its equivalence to $M_{T2}(ab)$ 
is not difficult either. A formal proof based on the method of Lagrange multipliers is presented in
Appendix~\ref{app:proofMT2M2CX}, so here we shall give just the heuristic argument.

Starting from Eq.~(\ref{eq:m2xxmt2}), without any loss of generality we can assume that 
$M_{2XX}^2(ab)$ is obtained by minimizing $M_{A_1}^2$, i.e.,~that in the 
neighborhood of the minimum, we have $M_{A_2}^2 < M_{A_1}^2$, and thus the $\max$
function in the definition (\ref{eq:m2XX(ab)}) picks up $M_{A_1}^2$ for the minimization.
The parent constraint $M_{A_1}=M_{A_2}$ is clearly not satisfied, but this can be fixed
without changing the value obtained in Eq.~(\ref{eq:m2xxmt2}).
Keeping $\vec{q}_{1T}$, $q_{1z}$, and $\vec{q}_{2T}$ fixed to their values at the 
$M_{2XX}^2(ab)$ minimum, we start varying $q_{2z}$ in the direction of increasing 
$M_{A_2}$. Eventually, we will find a value for $q_{2z}$ for which $M_{A_2}$
will reach $M_{A_1}$ and the parent constraint $M_{A_1}=M_{A_2}$ will be satisfied. 
In the meantime, nothing has changed regarding the $M_{A_1}^2$ function:
since $\vec{q}_{1T}$ and $q_{1z}$ were kept the same as before, its
value is still given by (\ref{eq:m2xxmt2}). 

This simple exercise shows that by adjusting the longitudinal invisible momenta, one can
always turn $M_{2XX}$ into $M_{2CX}$:
\bea
M_{2CX}^2(ab) =M_{2XX}^2(ab).
\label{eq:m2cxm2xx}
\eea
The main lesson is that this comes at a price --- the invisible momentum configuration
selected by the $M_{2XX}$ minimization may be different from the configuration
obtained in the $M_{2CX}$ minimization. We shall have much more to say about 
this in Sec.~\ref{sec:uniqueness} below.

Combining (\ref{eq:m2cxm2xx}) with (\ref{eq:m2xxmt2}), we also trivially obtain the relation~\cite{Mahbubani:2012kx}
\bea
M_{2CX}^2(ab)=M_{T2}^2(ab).
\label{eq:m2cxmt2}
\eea
\begin{figure}[t]
\centering
\includegraphics[width=4.9cm]{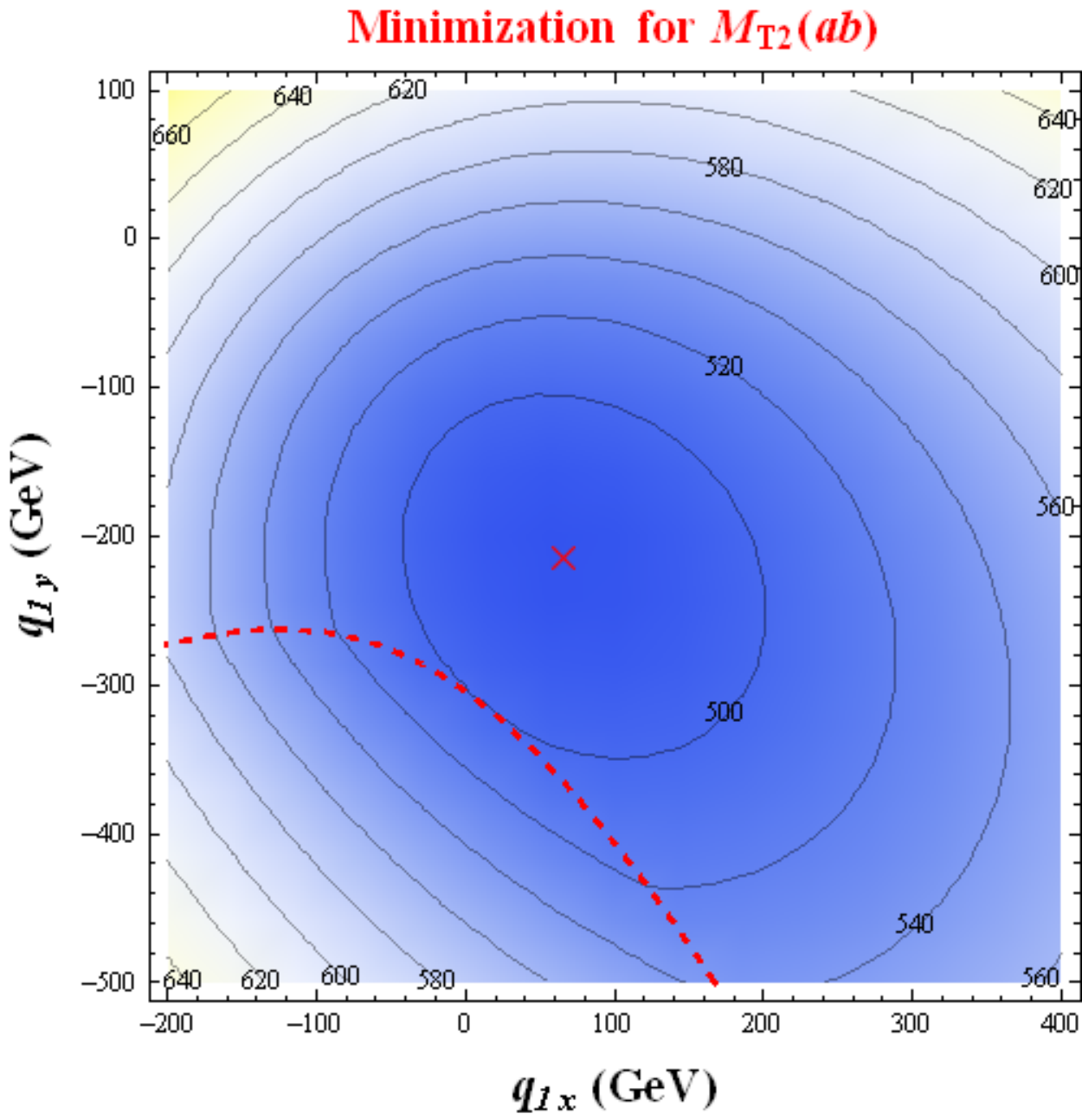}
\includegraphics[width=4.9cm]{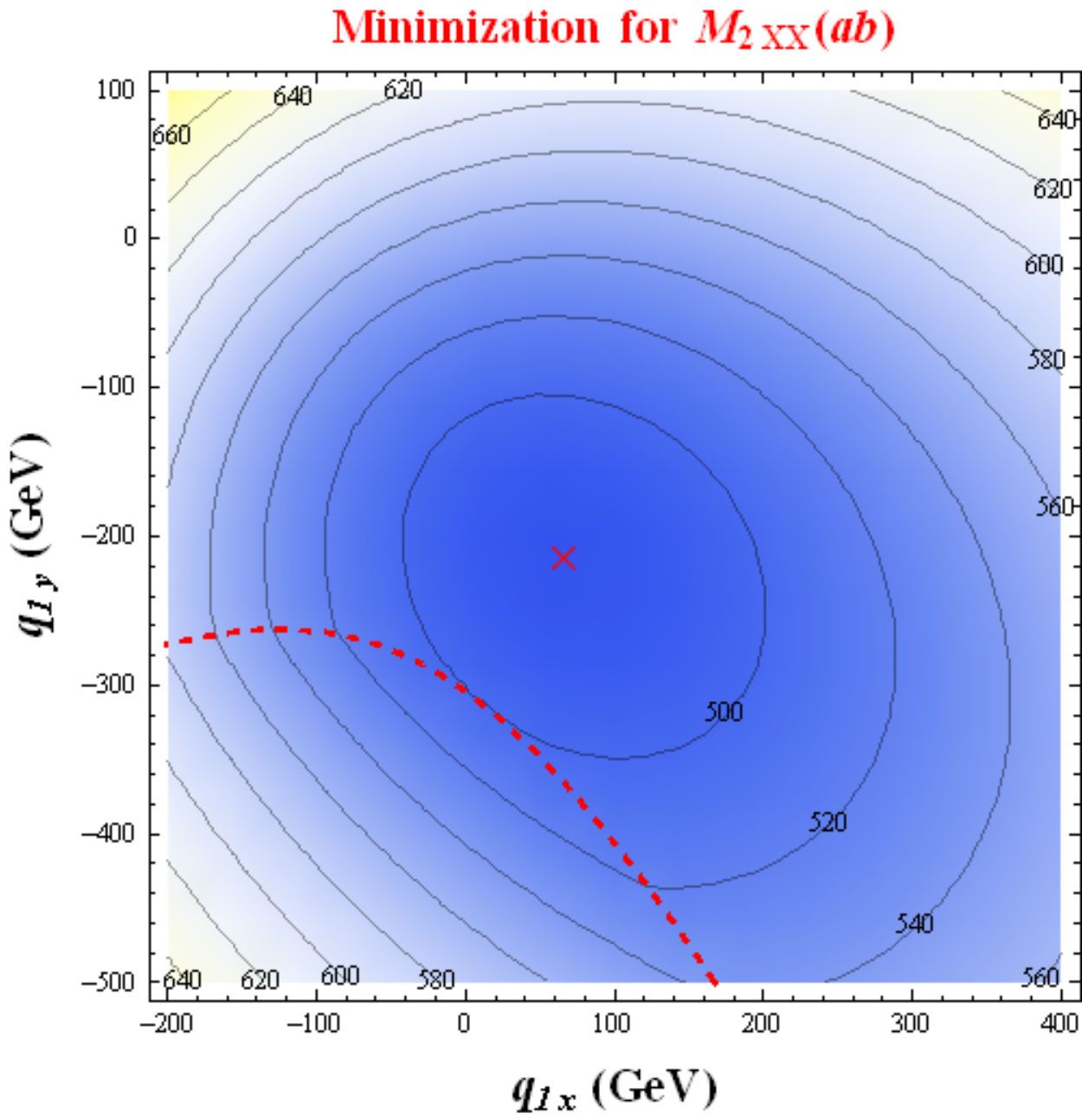}
\includegraphics[width=4.9cm]{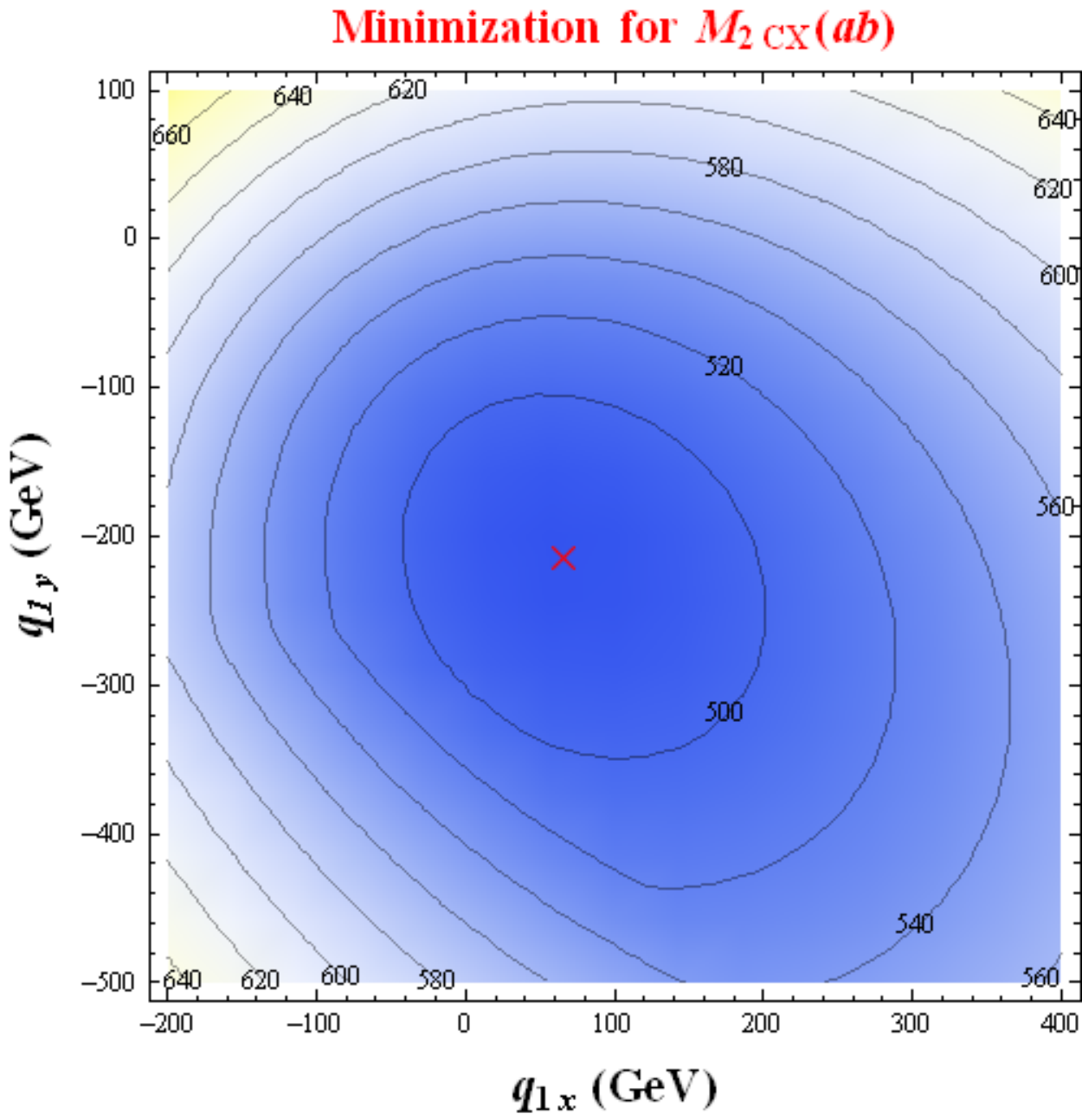}
\caption{\label{fig:MT2M2XXM2CX} 
Contour plots of the functions $f_{T2}(\vec{q}_{1T})$ (left panel),
$f_{2XX}(\vec{q}_{1T})$ (middle panel), and $f_{2CX}(\vec{q}_{1T})$ (right panel) in the plane of $\vec{q}_{1T}$. 
The chosen event leads to an unbalanced solution for $M_{T2}(ab)$. 
(The red dashed curve delineates the points with $M_{TA_1}=M_{TA_2}$.) 
The red $\times$ symbol marks the global minimum of the function in each case.
At the minimum, $M_{T2}(ab)=M_{2XX}(ab)=M_{2CX}(ab)=483.71$ GeV, 
and the corresponding solution for $\vec{q}_{1T}$ is given by $\vec{q}_{1T}^{\, \times}=(66.09,-212.90)$ GeV. } 
\end{figure}
In order to illustrate (\ref{eq:m2xxmt2}-\ref{eq:m2cxmt2}) pictorially, in 
Fig.~\ref{fig:MT2M2XXM2CX} we plot the three functions
\bea
f_{T2}(\vec{q}_{1T}) &\equiv & \max \left[ M_{TA_1}(\vec{q}_{1T},\tilde m),\;M_{TA_2}(\mpt-\vec{q}_{1T},\tilde m) \right],
\label{eq:ft2}  \\ [2mm]
f_{2XX}(\vec{q}_{1T})  &\equiv & \min_{q_{1z},q_{2z} }
\left\{ \max \left [ M_{A_1}(\vec{q}_{1T},q_{1z},\tilde m),\;M_{A_2}(\mpt-\vec{q}_{1T},q_{2z},\tilde m) \right] \right\},
\label{eq:f2XX}  \\ [2mm]
f_{2CX}(\vec{q}_{1T})  &\equiv & \min_{\stackrel{q_{1z},q_{2z}}{M_{A_1}=M_{A_2}} }
\left\{ \max \left [ M_{A_1}(\vec{q}_{1T},q_{1z},\tilde m),\;M_{A_2}(\mpt-\vec{q}_{1T},q_{2z},\tilde m) \right] \right\}
\label{eq:f2CX}
\eea
in the $\vec{q}_{1T}$ plane.  ($\vec{q}_{2T}$ is then determined from the $\mpt$ constraint as $\vec{q}_{2T}=\mpt-\vec{q}_{1T}$.)
These are precisely the functions which need to be minimized over $\vec{q}_{1T}$ in order to obtain the variables
$M_{T2}(ab)$, $M_{2XX}(ab)$, and $M_{2CX}(ab)$, respectively.
Note that these functions already contain different number of minimizations over longitudinal momenta:
$f_{2XX}(\vec{q}_{1T})$ has two, $f_{2CX}(\vec{q}_{1T})$ has one (the other longitudinal degree of freedom is fixed by the 
$M_{A_1}=M_{A_2}$ constraint), while $f_{T2}(\vec{q}_{1T})$ has none. 
The event chosen for Fig.~\ref{fig:MT2M2XXM2CX} was selected such that the associated 
$M_{T2}(ab)$ value comes from an unbalanced situation, i.e., the 
minimum of $f_{T2}(\vec{q}_{1T})$, marked with the red $\times$ symbol, 
is at $M_{TA_1}\ne M_{TA_2}$. 

Fig.~\ref{fig:MT2M2XXM2CX} demonstrates that the three functions 
(\ref{eq:ft2}-\ref{eq:f2CX}) are identical, thus justifying the identities (\ref{eq:m2xxmt2}-\ref{eq:m2cxmt2}).
In other words, once the minimization over the longitudinal components is done for the $M_{2XX}(ab)$ and $M_{2CX}(ab)$
variables, the remaining functions $f_{2XX}(\vec{q}_{1T})$ and $f_{2CX}(\vec{q}_{1T})$
become identical to $f_{T2}(\vec{q}_{1T})$, so the remaining minimization over $\vec{q}_{1T}$ 
will converge to the common point marked with the $\times$ symbol.
This means that all three variables $M_{2XX}(ab)$, $M_{2CX}(ab)$, and $M_{T2}(ab)$
not only have a common value, but also select the same {\em transverse} components
$\vec{q}_{iT}$ for the invisible momenta at their respective minima.
However, this is not the case for the {\em longitudinal} invisible momenta, $q_{iz}$,
which will be the subject of the next subsection.

\subsection{Uniqueness of the longitudinal momenta found by $M_{2XX}$ and $M_{2CX}$}
\label{sec:uniqueness}

As already mentioned in the Introduction, one of the main advantages of the $M_2$-type variables
over purely transverse analogues like $M_{T2}$, $M_{CT2}$ etc., is that they supply values for
not just the transverse, but also the longitudinal components of the invisible particle momenta.
The knowledge of the full 4-momentum of each invisible particle enables us to reconstruct the 
mass of each particle along the decay chain, and in particular the relative particles; see Sec.~\ref{sec:peak}. 
One should keep in mind that the momenta found by the $M_2$ minimization are {\em not}
the actual momenta of the invisible particles in the event. Nevertheless, the MAOS approach 
demonstrates that they can be successfully used for reconstruction~\cite{Cho:2008tj,Park:2011uz,Guadagnoli:2013xia}.
 
Let us now investigate the solutions for $q_{1z}$ and $q_{2z}$ more closely. Consider the starting point of
the $M_{2XX}$ calculation, the function
\beq
G_{2XX}(\vec{q}_{1T},q_{1z},q_{2z})  \equiv 
\max \left [ M_{A_1}(\vec{q}_{1T},q_{1z},\tilde m),\;M_{A_2}(\mpt-\vec{q}_{1T},q_{2z},\tilde m) \right].
\label{eq:g2XX}  
\eeq
As we saw in
Sec.~\ref{sec:equivalence}, its minimization along the transverse directions $\vec{q}_{1T}$
results in unique solutions; we call them $\vec{q}_{1T}^{\,(\times)}$. 
(See the red $\times$ symbols in Fig.~\ref{fig:MT2M2XXM2CX}).
Therefore, for the purposes of discussing the minimization over the longitudinal momentum components,
we can fix the transverse momenta, $\vec{q}_{1T} = \vec{q}_{1T}^{\,(\times)}$,  
and investigate the $q_{iz}$ dependence of the function
\bea
g_{2XX}(q_{1z},q_{2z})  &\equiv & 
\max \left [ M_{A_1}(\vec{q}_{1T}^{\,(\times)},q_{1z},\tilde m),\;M_{A_2}(\mpt-\vec{q}_{1T}^{\,(\times)},q_{2z},\tilde m) \right].
\label{eq:g2XX_x}  
\eea
The unconstrained minimization of $g_{2XX}(q_{1z},q_{2z})$ over $q_{1z}$ and $q_{2z}$ yields the value of $M_{2XX}$,
while minimizing (\ref{eq:g2XX_x}) subject to the parent constraint, $M_{A_1}=M_{A_2}$,  
gives the value of $M_{2CX}$.

Let us first study the effect of the parent constraint\footnote{Recall that throughout this section 
we have in mind the $(ab)$ subsystem.}
\bea
(p_{a_1}+p_{b_1}+q_1)^2=(p_{a_2}+p_{b_2}+q_2)^2, 
\label{eq:balancednessofM2CX}
\eea
which can be solved for $q_{2z}$ in terms of $q_{1z}$:
\bea
q_{2z}=\frac{p_{v_2z}K\pm E_{v_2}\sqrt{K^2-E_{q_2T}^2(E_{v_2}^2-p_{v_2z}^2 ) }}{E_{v_2}^2-p_{v_2z}^2},
\label{eq:q2zinq1z}
\eea
where
\bea
K\equiv\frac{m_{v_1}^2-m_{v_2}^2}{2}+E_{q_1}E_{v_1}-\vec{q}_{1T}\cdot\vec{p}_{v_1T}-q_{1z}p_{v_1z}
+\vec{q}_{2T}\cdot\vec{p}_{v_2T}.
\label{eq:expA}
\eea
One can obtain an analogous expression for $q_{1z}$ in terms of $q_{2z}$, 
by substituting $v_1 \leftrightarrow v_2$ and $q_1 \leftrightarrow q_2$ in 
Eqs.~(\ref{eq:q2zinq1z}) and~(\ref{eq:expA}). 

A couple of observations can be made from these equations. First, 
one can easily see from (\ref{eq:q2zinq1z}) that the $q_{2z}$ solution is 
not uniquely determined, i.e.,~$q_{2z}$ has a twofold ambiguity for a fixed $q_{1z}$,
unless the expression inside the square root, the {\it discriminant}, vanishes. 
The same argument can be made regarding the analogous expression giving 
$q_{1z}$ in terms of $q_{2z}$. Then the question becomes whether both 
$q_{1z}$ and $q_{2z}$ have double roots for some $\vec{q}_{1T}$. 
This is where the second observation comes into play. It turns out that
for the value of $q_{1z}$ which minimizes the function (\ref{eq:g2XX}),
$q_{1z}^{\,(min)}$,  
the discriminant in Eq.~(\ref{eq:q2zinq1z}) is proportional to the difference 
between the transverse masses of $A_1$ and $A_2$:
\bea
\left.K^2-E_{q_2T}^2(E_{v_2}^2-p_{v_2z}^2)\right|_{q_{1z}^{\,(min)}} 
\propto\left(M_{TA_1}^2-M_{TA_2}^2\right).
\eea
On the other hand, the discriminant that would appear in the expression 
analogous to (\ref{eq:q2zinq1z}) giving $q_{1z}$ in terms of $q_{2z}$,
will be proportional to $M_{TA_2}^2-M_{TA_1}^2$, 
i.e.,~the difference of the same squared transverse masses, only taken in opposite order. 
This suggests an interesting complementarity, in which $q_{1z}$ and $q_{2z}$
do not suffer from twofold ambiguities simultaneously, i.e.,~if $q_{2z}$ has two solutions
in Eq.~(\ref{eq:q2zinq1z}), then $q_{1z}$ is uniquely determined, and vice versa. 
This observation also reveals the necessary condition for both $q_{1z}$ and $q_{2z}$ to be
uniquely determined simultaneously: the transverse masses of $A_1$ and $A_2$ 
must be the same, $M_{TA_1}=M_{TA_2}$, see the red dashed curves in Fig.~\ref{fig:MT2M2XXM2CX}.

\begin{figure}[t]
\centering
\includegraphics[width=6.5cm]{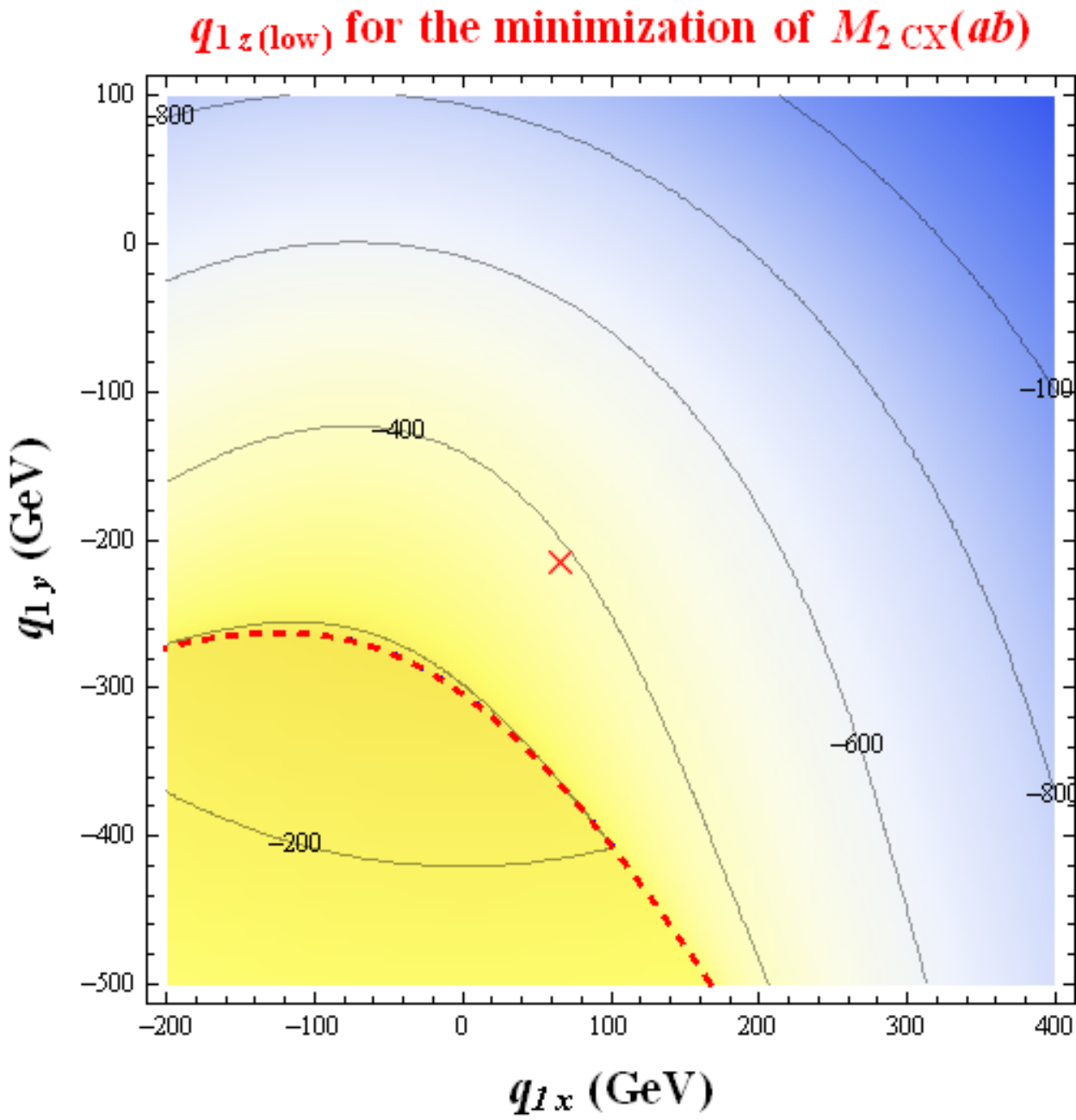}
\includegraphics[width=6.5cm]{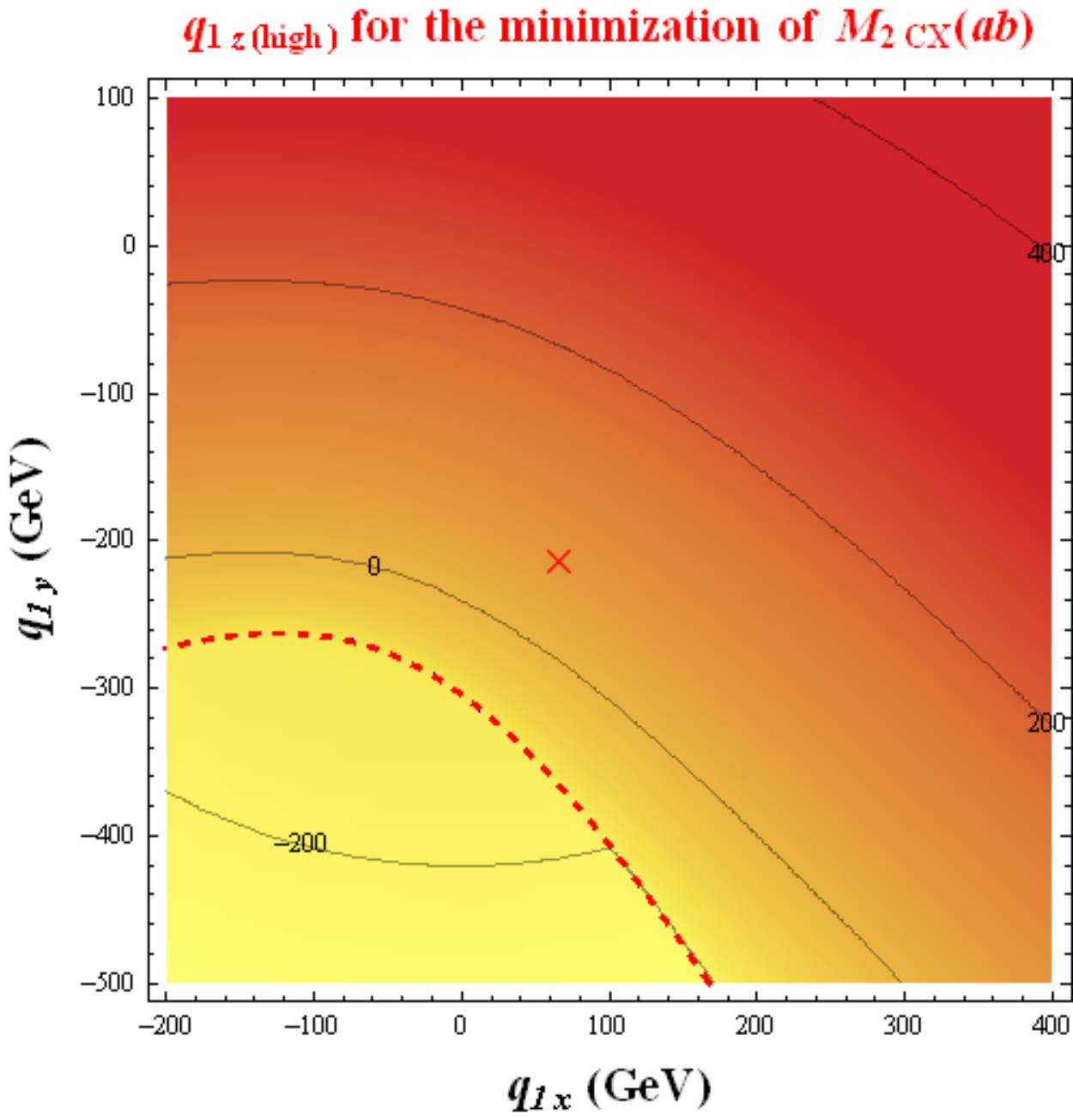} \\
\includegraphics[width=6.5cm]{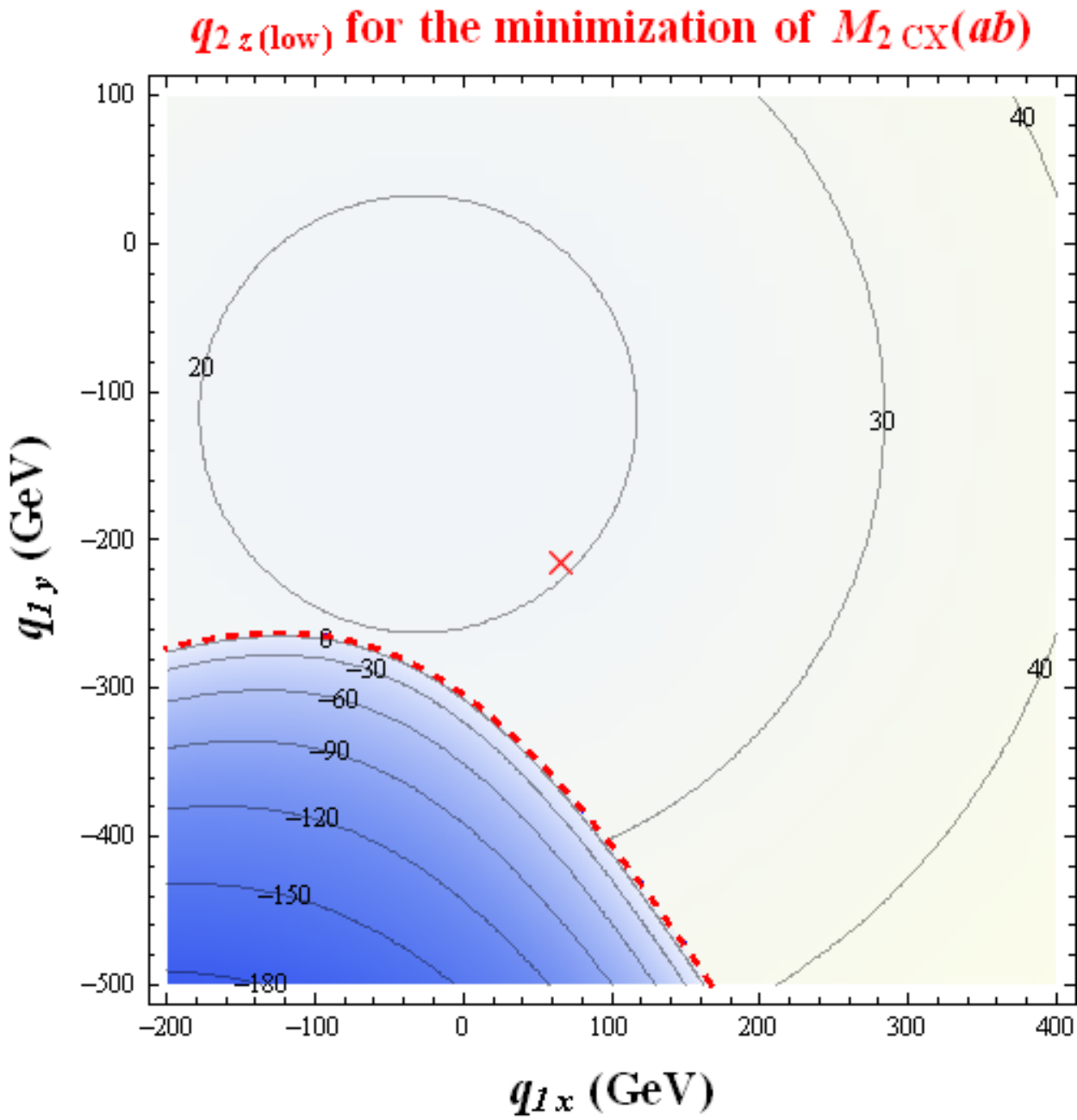}
\includegraphics[width=6.5cm]{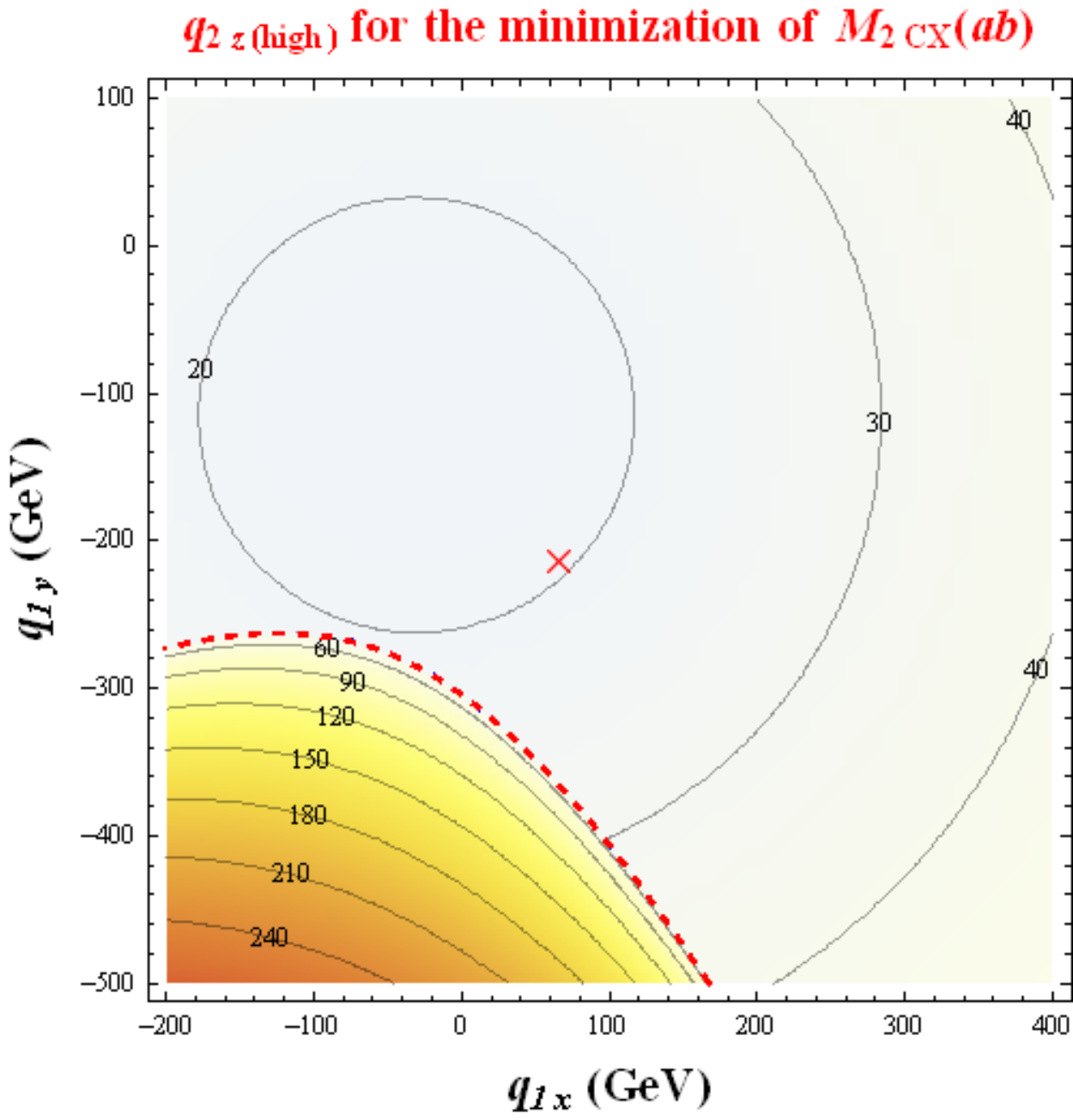}
\caption{\label{fig:M2CXqz} The values of the longitudinal invisible momenta $q_{iz(low)}$ (left panels)
and $q_{iz(high)}$ (right panels) which solve the parent constraint, (\ref{eq:balancednessofM2CX}), 
in the $\vec{q}_{1T}$ plane ($\vec{q}_{2T}$ is then given by the $\mpt$ condition (\ref{eq:mpt})),
for the same event shown in Fig.~\ref{fig:MT2M2XXM2CX}.
The upper row shows $q_{1z(low)}$ and $q_{1z(high)}$ for $q_{2z}=q_{2z}^{\,(min)}$,  
while the lower row shows $q_{2z(low)}$ and $q_{2z(high)}$ for $q_{1z}=q_{1z}^{\,(min)}$.
($q_{iz}^{\,(min)}$ is always found by minimizing (\ref{eq:g2XX}).) 
The red dashed curves denote the contours where the solutions to both $q_{1z}$ and $q_{2z}$ are unique. }
\end{figure}

Fig.~\ref{fig:M2CXqz}, which was made for the same unbalanced event used in Fig.~\ref{fig:MT2M2XXM2CX},
pictorially illustrates the above discussion. Let us call the two solutions of (\ref{eq:q2zinq1z}) 
$q_{2z(low)}$ (corresponding to the ``$-$" sign) and 
$q_{2z(high)}$ (corresponding to the ``$+$" sign).
They are plotted in the lower two panels of Fig.~\ref{fig:M2CXqz} 
in the $\vec{q}_{1T}$ plane. The remaining momenta are fixed as follows:
at each point of the plane, $\vec{q}_{2T}$ is given by the $\mpt$ condition (\ref{eq:mpt}),
while $q_{1z}$ is chosen so that it minimizes the function (\ref{eq:g2XX}):
$q_{1z}=q_{1z}^{\,(min)}$.
The upper two panels of Fig.~\ref{fig:M2CXqz} show the analogous plots where 
the roles of $q_{1z}$ and $q_{2z}$ are reversed --- we find 
$q_{2z}$ by minimizing (\ref{eq:g2XX}), $q_{2z}=q_{2z}^{\,(min)}$, and then plot the two solutions
for $q_{1z}$,  $q_{1z(low)}$ and $q_{1z(high)}$.
The red dashed lines delineate the points with balanced solutions for $M_{T2}$,
$M_{TA_1}=M_{TA_2}$. 

Fig.~\ref{fig:M2CXqz} confirms that the red dashed line is a watershed boundary ---
in the region above and to the right of that line we always find two possible values for 
$q_{1z}$, but a single value for $q_{2z}$. Conversely, in the area below and to the left of that line
there is always a unique solution for $q_{1z}$, but two solutions for $q_{2z}$ instead.
Now recall that the event depicted in Figs.~\ref{fig:MT2M2XXM2CX} and \ref{fig:M2CXqz}
was unbalanced, i.e., the true global minimum was obtained at the red $\times$ point, at which
$M_{TA_1}\ne M_{TA_2}$. This point also happens to be located in the region where the
solution for $q_{2z}$ is unique, but the solution for $q_{1z}$ has a twofold ambiguity.
On the other hand, if we had chosen a balanced event, the global minimum would fall 
somewhere on the red dashed line, and both $q_{1z}$  and $q_{2z}$ will be uniquely determined.

\begin{figure}[t]
\centering
\includegraphics[width=6.5cm]{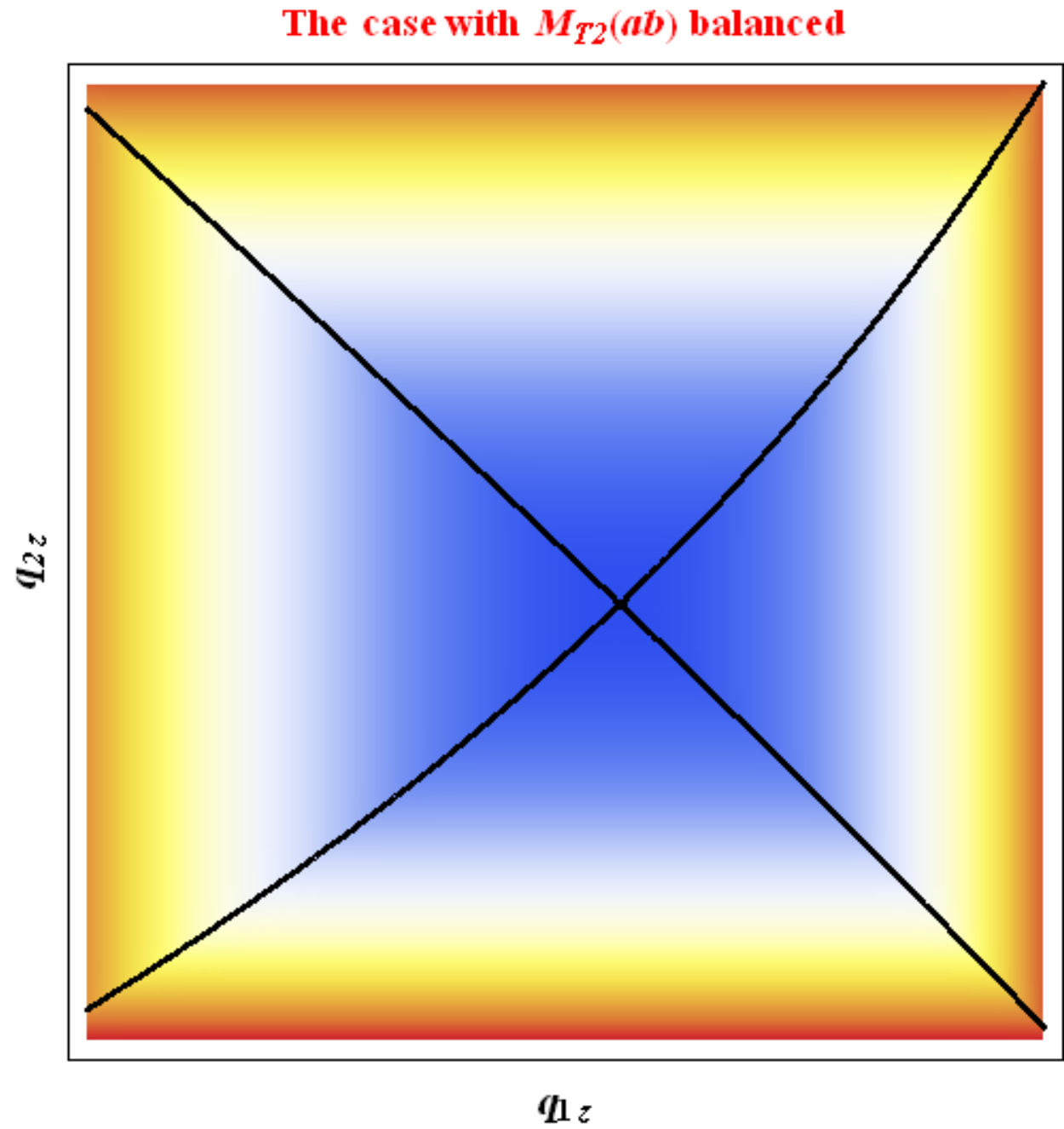}
\includegraphics[width=6.5cm]{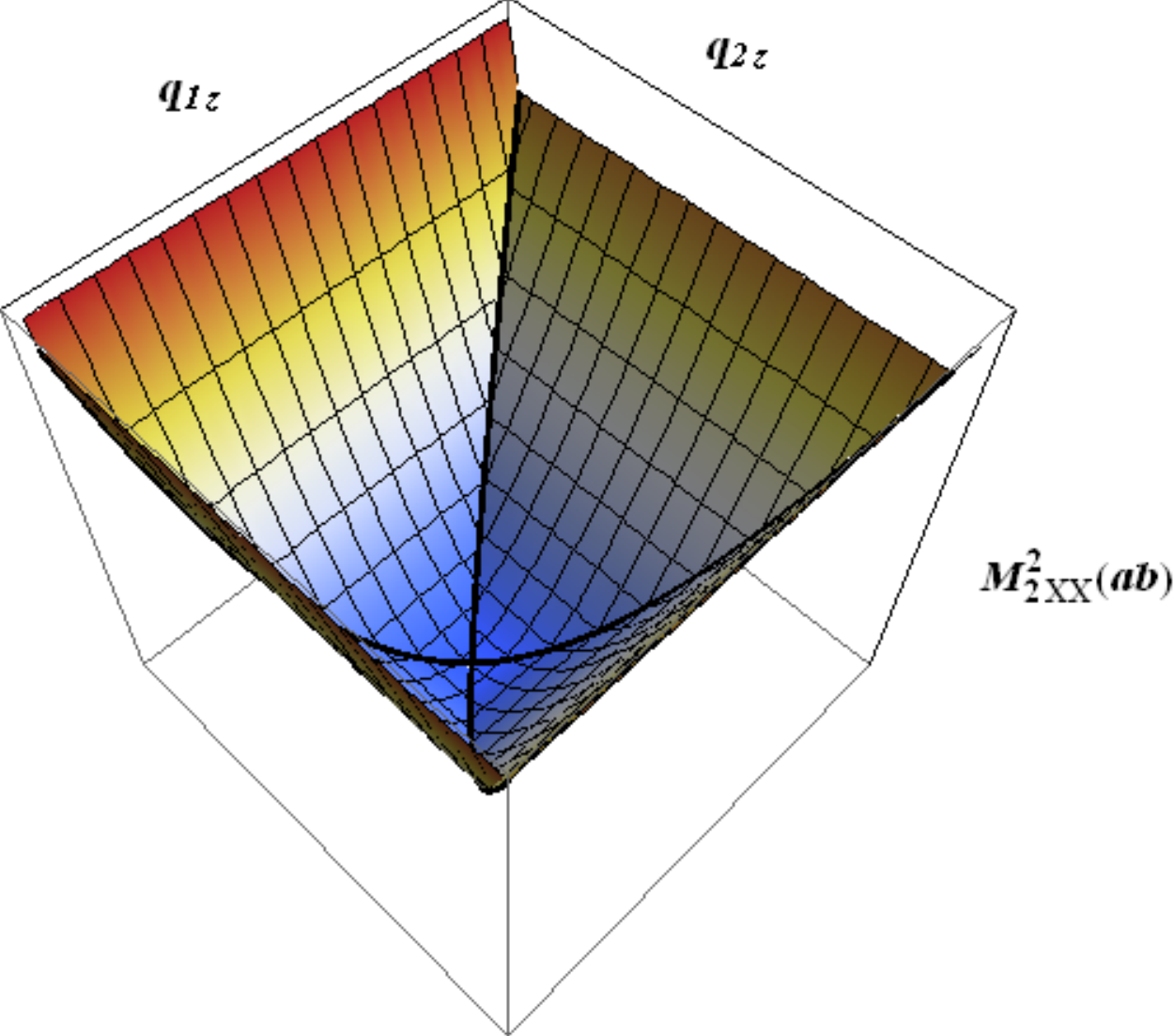} \\
\includegraphics[width=6.5cm]{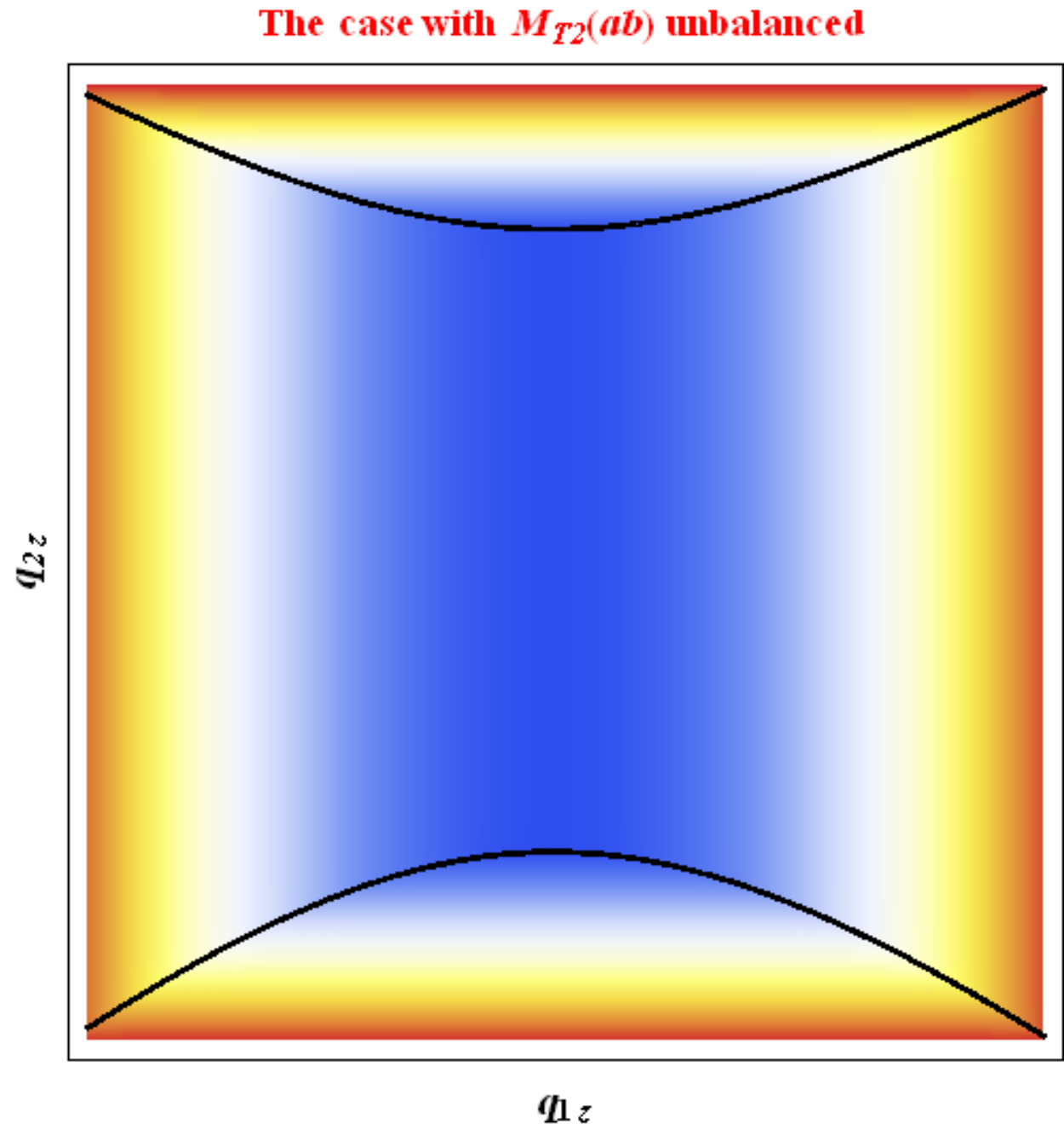}
\includegraphics[width=6.5cm]{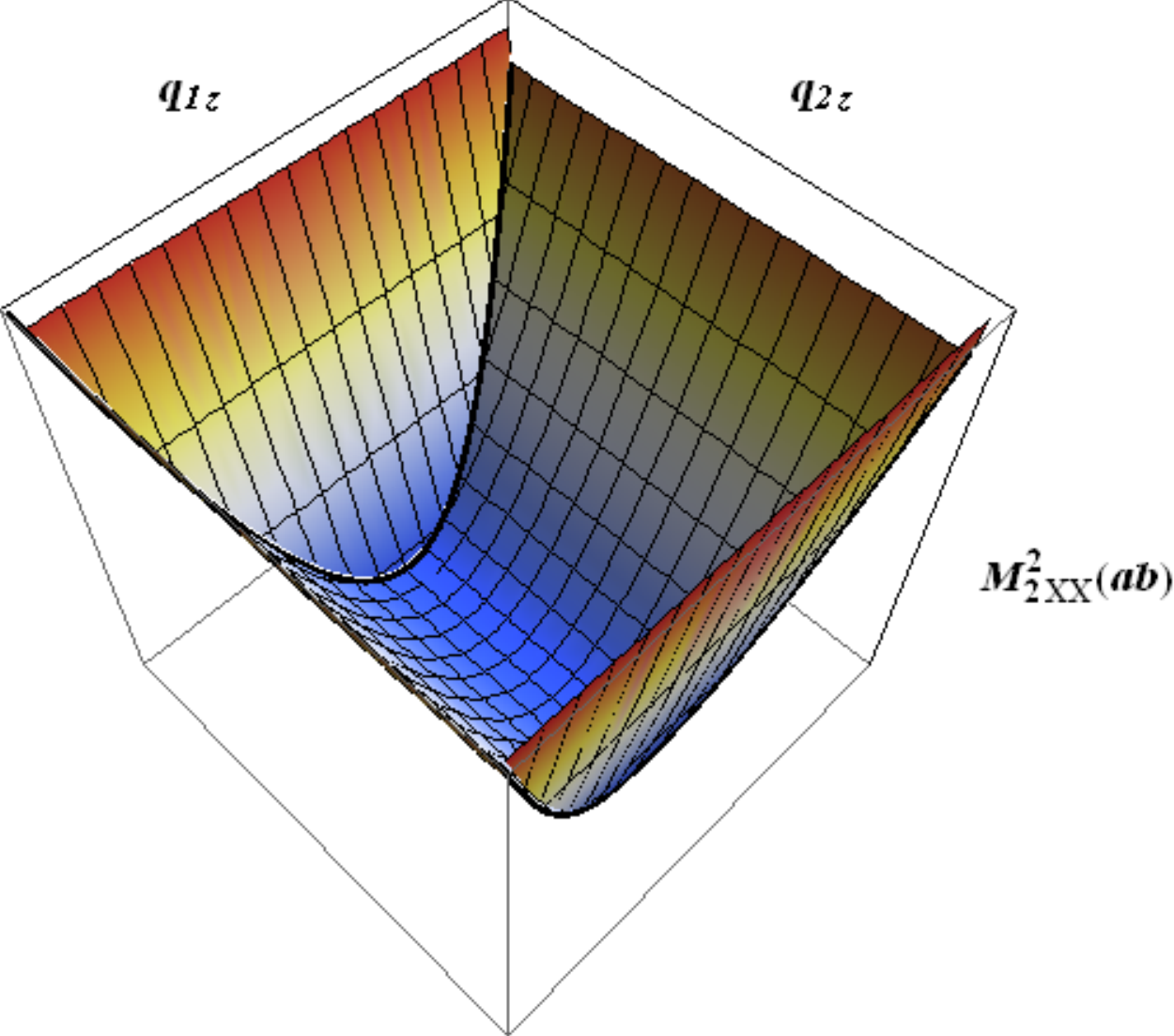}
\caption{\label{fig:balunbalspace} 
Plot of the function (\ref{eq:g2XX_x}) in the $(q_{1z},q_{2z})$ plane, 
for a balanced event with $M_{TA_1}=M_{TA_2}$ at the minimum (top row)
and an unbalanced event with $M_{TA_1}\ne M_{TA_2}$ at the minimum (bottom row).
The left panels are contour plots, while the right panels show the corresponding 
3-dimensional view. The black solid curves mark the points satisfying the parent constraint,
$M_{A_1}=M_{A_2}$. }
\end{figure}

Having understood the minimization of $M_{2CX}(ab)$, it is easy to infer the corresponding 
solutions for $q_{1z}$ and $q_{2z}$ in the case of $M_{2XX}(ab)$. The ambiguity problem 
is now even more serious, because whenever $q_{iz(low)}\ne q_{iz(high)}$, {\em any} value
of $q_{iz}\in \left(q_{iz(low)}, q_{iz(high)}\right)$ is also allowed, i.e.,~the ambiguity is not just twofold,
instead there is a flat direction. However, these ambiguities are present only for unbalanced events ---
for balanced events, $q_{iz(low)}= q_{iz(high)}$, and the solution for both $q_{1z}$ and $q_{2z}$ 
is unique. This is pictorially illustrated in Fig.~\ref{fig:balunbalspace}, which shows 
the function (\ref{eq:g2XX_x}) as a function of $q_{1z}$ and $q_{2z}$.
Since $\vec{q}_{1T}$ is already fixed to its correct value, $\vec{q}_{1T}^{\,(\times)}$,
at the global minimum of (\ref{eq:g2XX}), the global unconstrained minimum of the function
(\ref{eq:g2XX_x}) seen in Fig.~\ref{fig:balunbalspace} corresponds to $M_{2XX}$,
while the constrained minimization along the black solid lines with $M_{A_1}=M_{A_2}$
yields the value of $M_{2CX}$. The two plots in the top row of Fig.~\ref{fig:balunbalspace}
correspond to a balanced event, in which there is a single global minimum, 
and thus the longitudinal momentum configuration at the minimum is unique. 
Furthermore, the global minimum is at the intersection of the two black solid lines, 
implying that the parent constraint, $M_{A_1}=M_{A_2}$, is satisfied and therefore
$M_{2XX}=M_{2CX}$, in agreement with the theorem from Sec.~\ref{sec:equivalence}.
On the other hand, the bottom two plots show an unbalanced event, 
in which the unconstrained minimization reveals a flat direction along $q_{2z}$.
Any value of $q_{2z}$ along the bottom of that valley is acceptable and will give the 
correct value of $M_{2XX}$. If we now consider the constrained minimization along 
the black solid lines to obtain $M_{2CX}$, we find two degenerate global minima --- one
on the upper black solid curve and one on the lower black solid curve.
Thus, as expected, there is a twofold ambiguity --- in this case in the value of $q_{2z}$,
while $q_{1z}$ is unique. Again, the values of $M_{2XX}$ and $M_{2CX}$ are the same, since 
the function (\ref{eq:g2XX_x}) is constant along the flat direction.
 
Since later on we shall be using the momenta found by the minimization for 
reconstruction purposes, the results from this subsection raise the
question of how one should deal with unbalanced events, for which
(some of) the momentum components are not uniquely determined. There
can be several approaches:
\begin{itemize}
\item Restrict one's attention to balanced events only, incurring some
  (minor) loss in statistical significance.
\item Sum over all possible kinematic solutions (i.e.,~integrate over the flat direction in 
Fig.~\ref{fig:balunbalspace}), and enter the results 
in histograms with correspondingly reduced weights.
\item Instead of obtaining the momenta from $M_{2XX}$ and $M_{2CX}$, use the 
variables with relative constraints, $M_{2XC}$ and $M_{2CC}$, for which these ambiguities
generally do not arise, see Sec.~\ref{sec:2CC}.
\end{itemize} 

\subsection{The variables $M_{2XC}$ and $M_{2CC}$}
\label{sec:2CC}

Having seen in Sec.~\ref{sec:equivalence} that $M_{2XX}(ab)$ and
$M_{2CX}(ab)$ are equivalent to $M_{T2}(ab)$, we now shift our focus
to the new variables $M_{2XC}(ab)$ and $M_{2CC}(ab)$ and investigate
their relationship with the other variables. 

First we argue that the result from minimization with respect to
$\vec{q}_{1T}$ will be different in general when obtaining these new
variables.
For this purpose, let us assume the opposite, i.e.,~consider the
function (\ref{eq:g2XX_x})
in which $\vec{q}_{1T}$ has been fixed to the result
$\vec{q}_{1T}^{\,(\times)}$
found in the $\vec{q}_{1T}$ minimization in
Fig.~\ref{fig:MT2M2XXM2CX}.
We then discuss its minimization in the $(q_{1z},q_{2z})$ plane as in
Fig.~\ref{fig:balunbalspace}. 
The new element here is the presence of the relative constraint,
$M_{B_1}^2=M_{B_2}^2$,
which can be written as
\bea
(p_{b_1}+q_1)^2=(p_{b_2}+q_2)^2,
\label{eq:rel}
\eea
and which can be solved for $q_{2z}$ in analogy to (\ref{eq:q2zinq1z}):
\bea
q_{2z}=\frac{p_{b_2z}K'\pm E_{b_2}\sqrt{K^{'2}-E_{q_2T}^2p_{b_2T}^2}}{p_{b_2T}^2},
\label{eq:q2zinq1zM2XC}
\eea
where
\bea
K'\equiv E_{q_1}E_{b_1}-\vec{q}_{1T}\cdot\vec{p}_{b_1T}-q_{1z}p_{b_1z}+\vec{q}_{2T}\cdot\vec{p}_{b_2T}.
\label{eq:Kprime}
\eea
A similar expression can be obtained for $q_{1z}$ in terms of $q_{2z}$, with the replacements
$q_{2z}\leftrightarrow q_{1z}$ and $b_1\leftrightarrow b_2$ in (\ref{eq:q2zinq1zM2XC},\ref{eq:Kprime}). 
Due to the ``$\pm$'' sign in (\ref{eq:q2zinq1zM2XC}), the relative constraint, (\ref{eq:rel}), again
implies two branches in the $(q_{1z},q_{2z})$ plane, analogous to the black solid curves in Fig.~\ref{fig:balunbalspace}.
If at least one of these two curves passes through the global minimum point\footnote{Recall from Fig.~\ref{fig:balunbalspace}
that balanced events lead to a unique global minimum as shown in the top panels while unbalanced events 
lead to a flat direction along a finite line segment as shown in the bottom panels.} found previously 
for the case of $M_{2XX}$, then $M_{2XC}$ will turn out to be the same as $M_{2XX}(ab)$. 
However, the chances of a plane curve passing through a given point (or even a given finite line segment)
are minimal, therefore we expect that, in general, the solution found previously for $M_{2XX}$ 
will not obey the relative constraint, (\ref{eq:rel}). This means that our choice of 
$\vec{q}_{1T}=\vec{q}_{1T}^{\,(\times)}$ was wrong, and that the minimum for
$M_{2XC}$ is obtained at a different value for $\vec{q}_{1T}$ than the one found in
Fig.~\ref{fig:MT2M2XXM2CX}. In particular, the constrained global minimum found by $M_{2XC}$
will be higher than the corresponding unconstrained global minimum $M_{2XX}$:
\bea
M_{2XC}(ab)\geq M_{2XX}(ab)=M_{2CX}(ab)=M_{T2}(ab).
\eea

\begin{figure}[t]
\centering
\includegraphics[width=4.9cm]{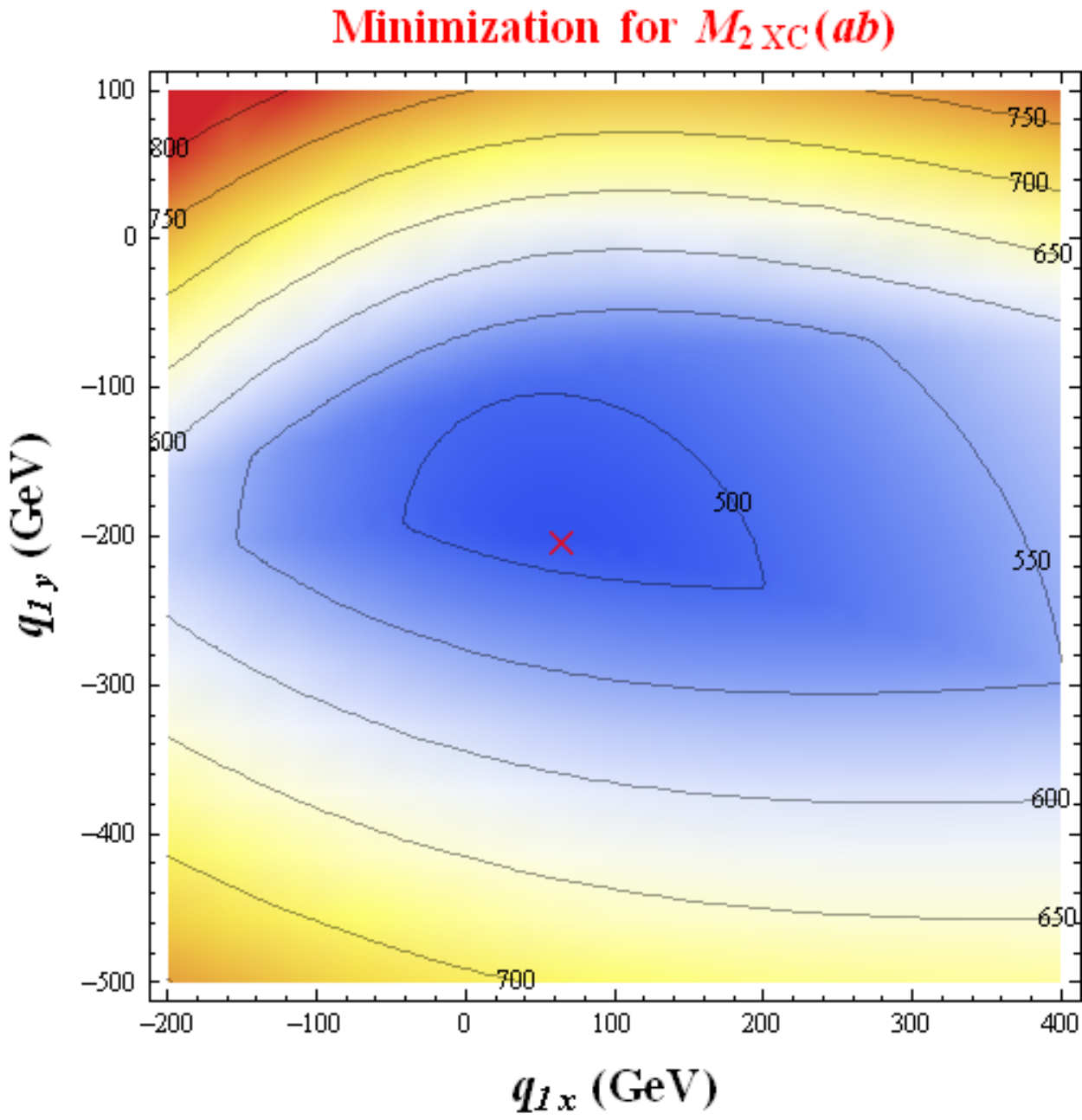}
\includegraphics[width=4.9cm]{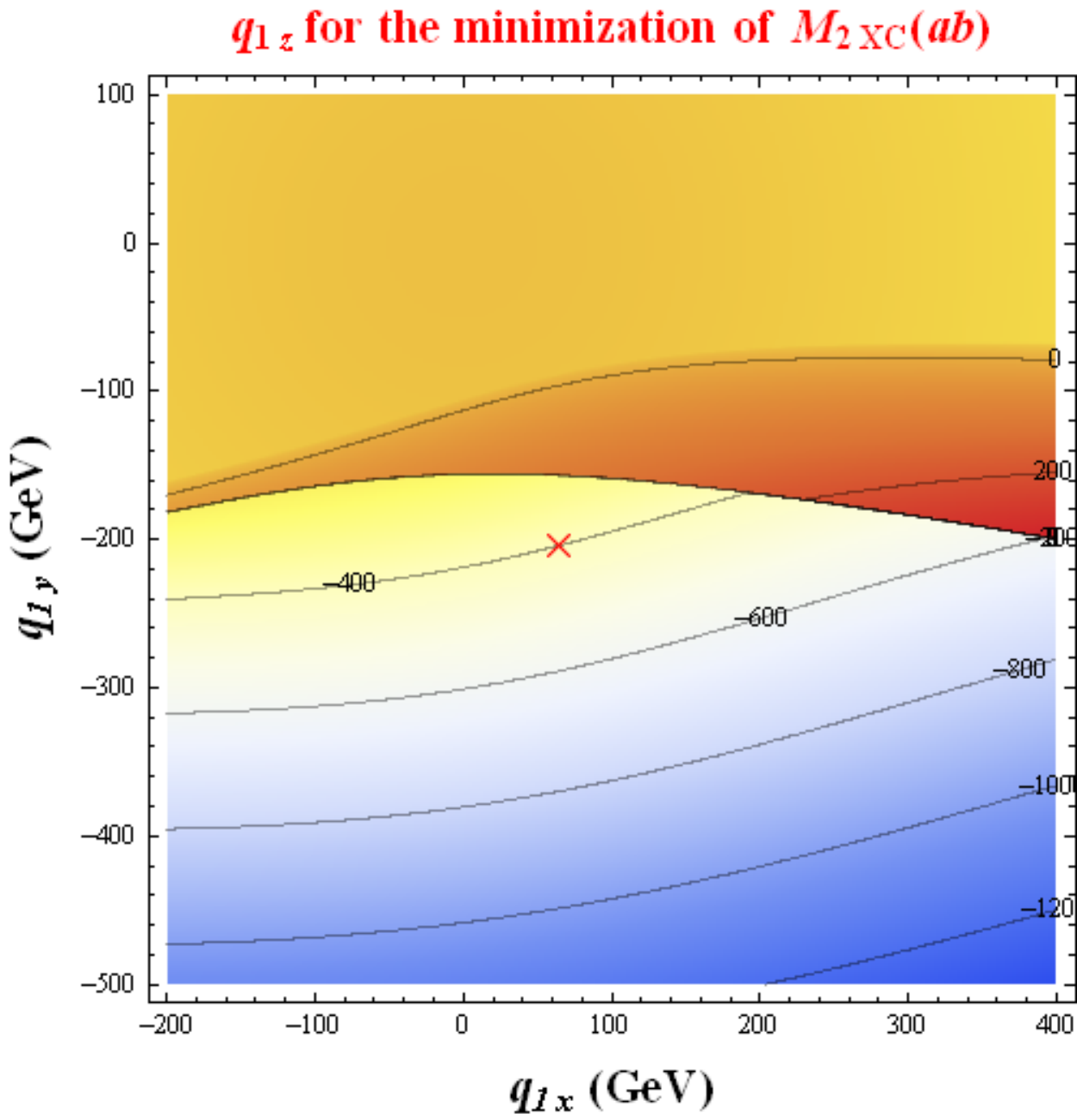}
\includegraphics[width=4.9cm]{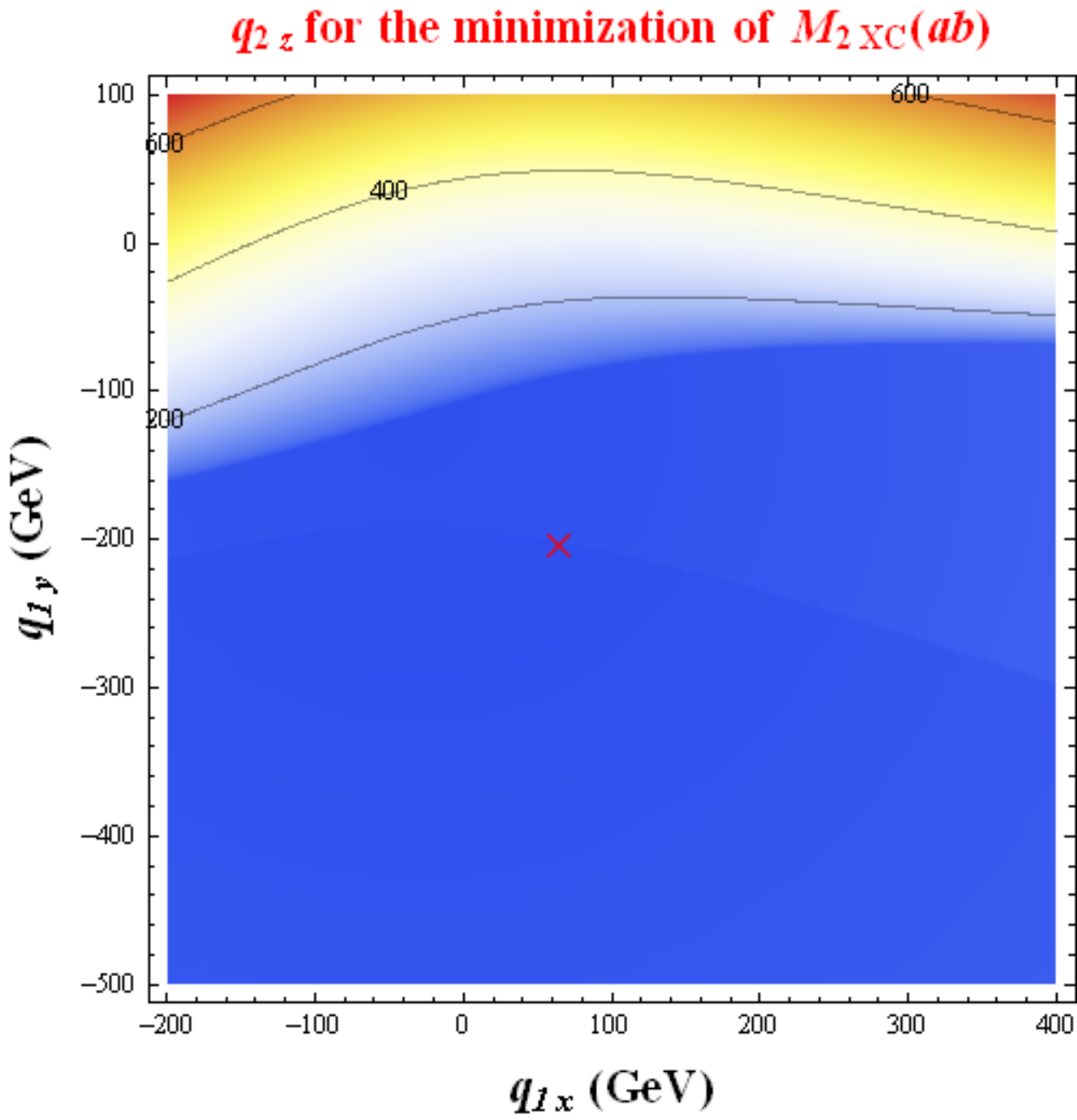}
\caption{\label{fig:M2XC} 
The analogues of Fig.~\ref{fig:MT2M2XXM2CX} (left panel)
and Fig.~\ref{fig:M2CXqz} (middle and right panels) for the case of $M_{2XC}$.
The cross symbols show the location of the global minimum for $M_{XC}$, at which
$M_{2XC}(ab)=483.85$ GeV, and the invisible momenta are given by 
$(\vec{q}_{1T},\;q_{1z},\;q_{2z})=(64.61,\;-202.37,\;-395.75,\;19.07)$ GeV. 
}
\end{figure}

Fig.~\ref{fig:M2XC} pictorially illustrates the above discussion.
The left panel shows the function to be minimized when calculating $M_{2XC}$.
As compared with the analogous Fig.~\ref{fig:MT2M2XXM2CX} for the case of $M_{2XX}$,
we see that the shape of the function is completely different, and as a result the 
global minimum (marked with a red $\times$ symbol) is obtained at a different 
point in $\vec{q}_T$ space, $\vec{q}_{1T}=(64.61,-202.37)$ GeV
(as opposed to $\vec{q}_{1T}^{\, \times}=(66.09,-212.90)$ GeV, which was found in Fig.~\ref{fig:MT2M2XXM2CX}).

Another important lesson from the middle and right panels in Fig.~\ref{fig:M2XC} is that the solutions for $q_{1z}$ and $q_{2z}$ 
are now unique, unlike in the case of $M_{2XX}$ and $M_{2CX}$ exhibited in Fig.~\ref{fig:M2CXqz}. 
We shall use this fact later on when reconstructing the mass of relative particles and studying the 
event topology.

\begin{figure}[ht]
\centering
\includegraphics[width=4.9cm]{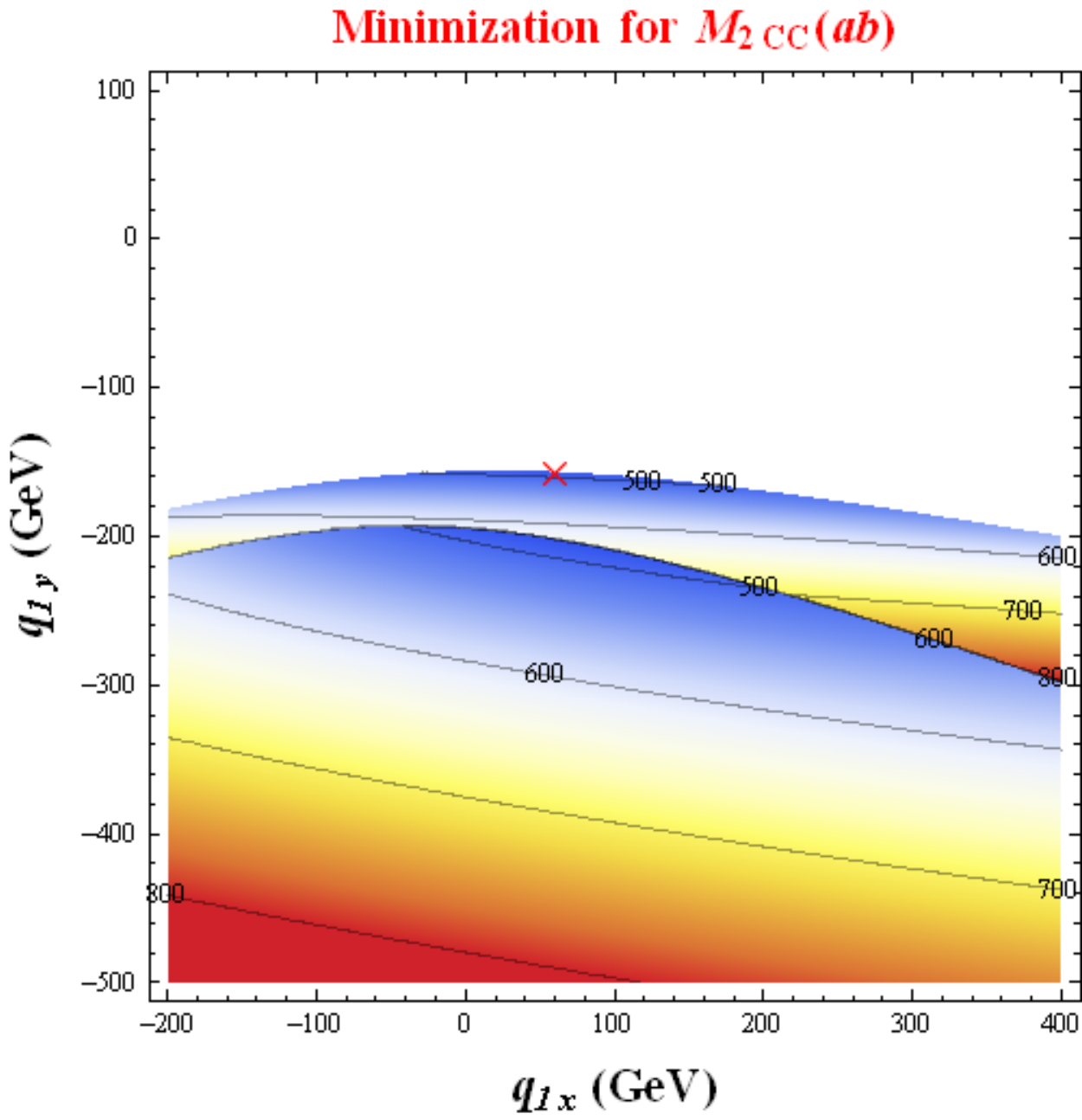}
\includegraphics[width=4.9cm]{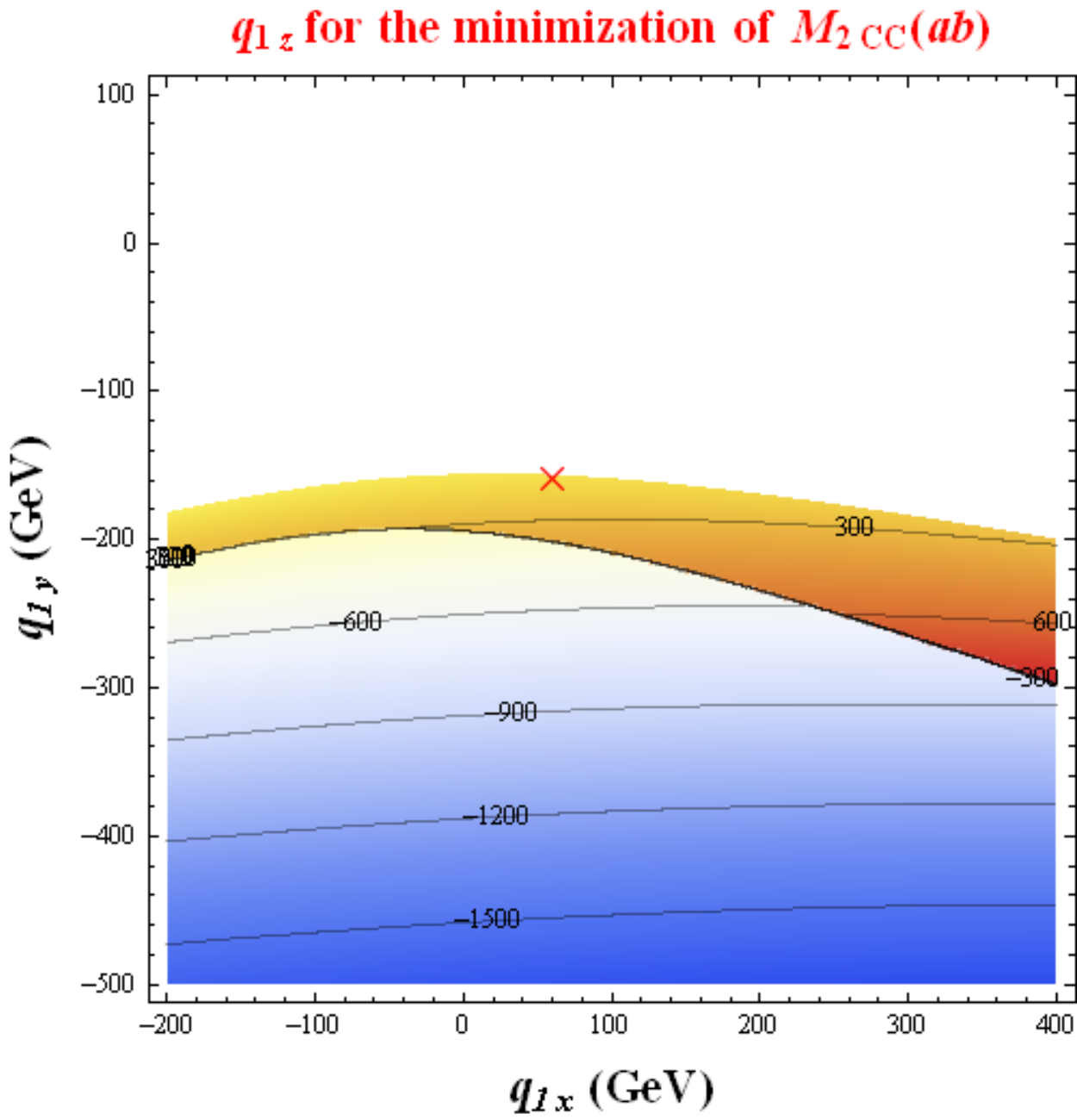}
\includegraphics[width=4.9cm]{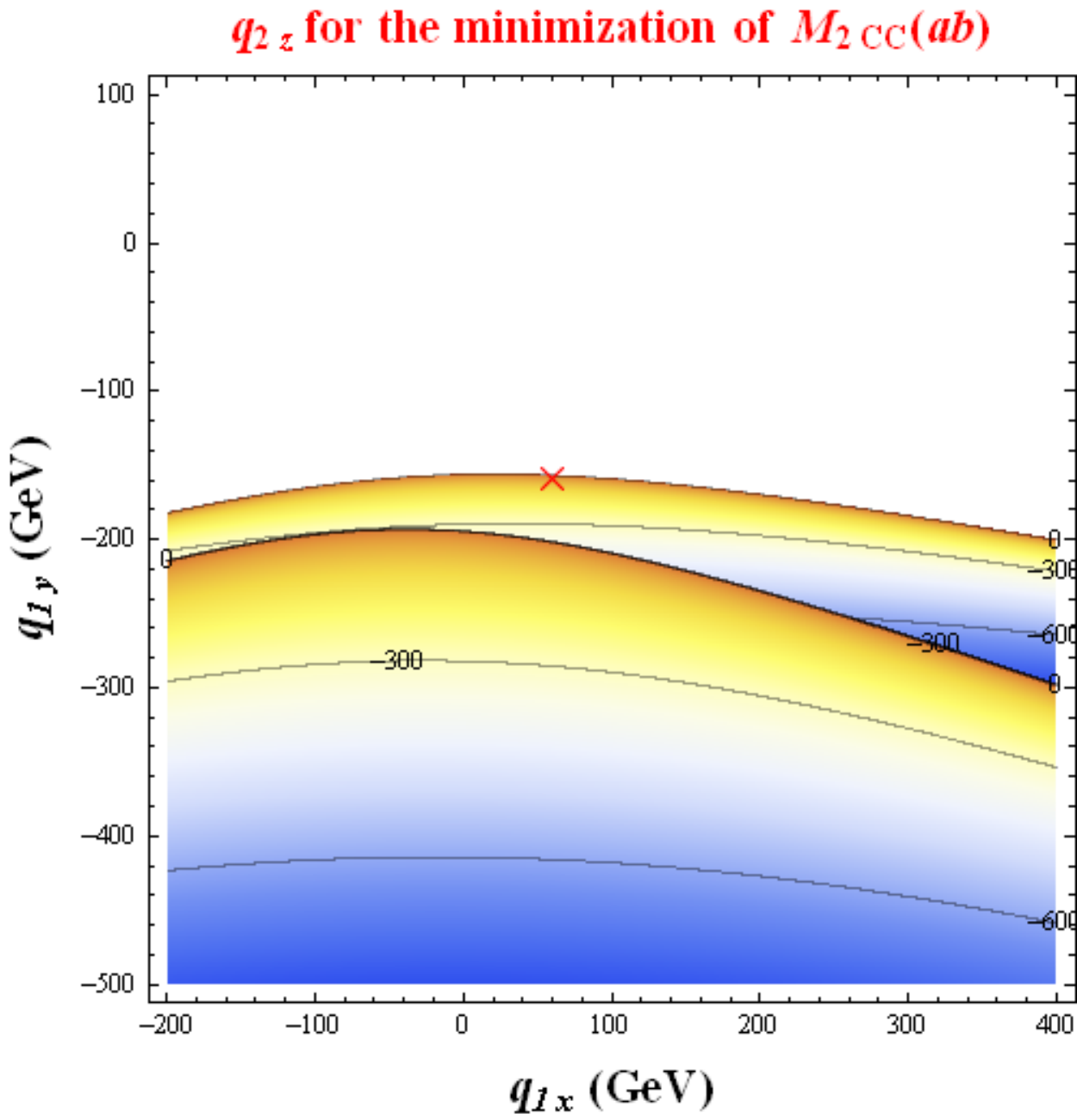}
\caption{\label{fig:M2CC} The same as Fig.~\ref{fig:M2XC}, but for the case of $M_{2CC}(ab)$.
At the global minimum (marked with the red $\times$ symbol), $M_{2CC}(ab)=487.86$ GeV, and 
the solution for the invisible momenta is given by $(\vec{q}_{1T},\;q_{1z},\;q_{2z})=(60.11,\;-156.62,\;121.35,\;17.44)$ GeV. 
Within the white region, the constraints (\ref{eq:balancednessofM2CX}) and (\ref{eq:rel})
cannot be simultaneously satisfied.}
\end{figure}

Finally, it remains to discuss the variable $M_{2CC}(ab)$, where the parent and relative constraints, 
(\ref{eq:balancednessofM2CX}) and (\ref{eq:rel}), are simultaneously applied. The analysis proceeds 
very similarly to the case of $M_{2XC}(ab)$ and the corresponding results are displayed in Fig.~\ref{fig:M2CC}.
Because of the additional constraint, the true global minimum for
$M_{2CC}(ab)$ is now even greater than
$M_{2XC}(ab)$.  We thus arrive at our final result relating the variables (\ref{eq:variables}):
\bea
M_{2CC}(ab)\geq M_{2XC}(ab) \geq M_{2XX}(ab)=M_{2CX}(ab)=M_{T2}(ab). 
\label{eq:hierarchy}
\eea
Note that there are large regions in invisible momentum parameter space (the white areas in Fig.~\ref{fig:M2CC}),
for which the on-shell kinematic constraints (\ref{eq:balancednessofM2CX}) and (\ref{eq:rel}) cannot be simultaneously satisfied.
As before, the red $\times$ symbol marks the solution for $\vec{q}_{1T}$, which is found at a new location,
$\vec{q}_{1T}=(60.11,-156.62)$ GeV. The corresponding $M_{2CC}(ab)$ value is $487.86$ GeV, which is slightly larger than 
$M_{2XC}(ab)=483.85$ GeV, in agreement with (\ref{eq:hierarchy}). The solutions for the longitudinal
momenta are also unique (just as in the case of $M_{2XC}(ab)$ in Fig.~\ref{fig:M2XC}), and
are found at $(q_{1z},q_{2z})=(121.35,17.44)$ GeV.

\subsection{Summary of the properties of the on-shell constrained $M_2$ variables}
\label{sec:summary}

We now collect our main results from Sec.~\ref{sec:notation} and Sec.~\ref{sec:relation} before moving on to 
the practical applications of the $M_2$ variables in the next few sections.

In Sec.~\ref{sec:notation}, we defined five different types of variables 
for each of the three subsystems in Fig.~\ref{fig:DecaySubsystem} (see Table~\ref{tab:M2variables}).
The hierarchy among those variables is\footnote{Strictly speaking, in this section we only discussed 
the $(ab)$ subsystem and the relations (\ref{eq:abrelations}), but the analysis leading to 
(\ref{eq:arelations}) and (\ref{eq:brelations}) is very similar.} 
\bea
&M_{2CC}(ab)\geq M_{2XC}(ab) \geq M_{2XX}(ab)=M_{2CX}(ab)=M_{T2}(ab);& 
\label{eq:abrelations}
\\ [2mm]
&M_{2CC}(a)\geq M_{2XC}(a) \geq M_{2XX}(a)=M_{2CX}(a)=M_{T2}(a);& 
\label{eq:arelations}
\\ [2mm]
&M_{2CC}(b)\geq M_{2XC}(b) \geq M_{2XX}(b)=M_{2CX}(b)=M_{T2}(b).&
\label{eq:brelations}
\eea
Thus, out of the fifteen variables seen in (\ref{eq:abrelations}-\ref{eq:brelations}), 
there are only nine which are quantitatively different.

\begin{table}[t]
\centering
\begin{tabular}{||c||c|c||c|c||}
\hline
                        &  \multicolumn{2}{c||}{Balanced events} &  \multicolumn{2}{c||}{Unbalanced events} \\
\cline{2-5}
Variable           &  $\vec{q}_{iT}$  & $q_{iz}$ & $\vec{q}_{iT}$ & $q_{iz}$\\ 
\hline \hline
$M_{T2}(ab)$    &   unique   &    NA       &     unique   &    NA  \\
$M_{2XX}(ab)$  &   unique   & unique    &     unique   &  flat direction \\
$M_{2CX}(ab)$  &   unique   &  unique   &     unique   &  twofold ambiguity  \\ 
$M_{2XC}(ab)$  &   unique   & unique    &     unique   &  unique  \\
$M_{2CC}(ab)$  &   unique   &  unique   &     unique   &  unique \\
\hline
\end{tabular}
\caption{\label{tab:unique} Table summarizing the uniqueness of the invisible momentum configurations 
corresponding to the global minimum. }
\end{table}

Each of the $M_2$ variables in Table~\ref{tab:M2variables} is calculated by minimizing 
a suitably defined mass function in terms of the invisible momenta, see (\ref{eq:m2XXdef}-\ref{eq:m2CCdef}).
The global minimum thus selects a special configuration of the invisible momenta which can be used for
kinematical studies. In this section, we also investigated the uniqueness of the global minimum and 
consequently, the uniqueness of the associated invisible momenta. Our results are summarized in
Table~\ref{tab:unique}. For completeness, in the table we also include the $M_{T2}$ variable, 
which, however, cannot determine the longitudinal components of the invisible momenta.
In the case of balanced events, all four $M_2$ variables uniquely determine the invisible 3-momenta, 
while for unbalanced events, only $M_{2XC}$ and $M_{2CC}$ do so. Note that the twofold ambiguity 
in the case of $M_{2CX}$ and the flat direction in the case of $M_{2XX}$ are only with respect to 
{\em one} of the $q_{iz}$ components, while the {\em other} $q_{iz}$ component is uniquely determined.

\section{Mass measurements}

We now discuss several physics examples illustrating the potential uses and advantages of the 
$M_2$ variables. In this section, we first consider the simpler scenario 
where we have made the correct hypothesis about the true physics model and show how the 
use of $M_2$ variables can improve the precision of the mass measurements (in Sec.~\ref{sec:mass})
and provide a generalization of the MAOS technique~\cite{Cho:2008tj} (in Sec.~\ref{sec:peak}).
Then in Sec.~\ref{sec:topology}, we move to the case where we are
uncertain about which new physics model is correct. We show how one can then use the $M_2$ variables to rule out
the incorrect model assumptions and hone in on the correct event topology.

The model to be studied in this section is the one depicted in Fig.~\ref{fig:process}(a), 
where the two decay chains are assumed to be identical:
\bea
m_{A_1}=m_{A_2}\equiv m_{A},\;\;\;m_{B_1}=m_{B_2}\equiv m_{B},\;\;\;m_{C_1}=m_{C_2}\equiv m_{C}.
\label{eq:truemass}
\eea
In order to avoid confusion, from here on we shall use lowercase letters as in (\ref{eq:truemass})
to denote the true physical masses of the particles, reserving the corresponding uppercase letters 
$M_A$, $M_B$, etc., for masses which are {\em reconstructed} using kinematic information 
from the visible decay products in the event. Where necessary, input {\em test} masses (i.e.,~mass ans\"atze) 
will be denoted with a tilde. Throughout the paper, for our simulations we shall use event samples of $\sim 100,000$ events 
each, generated at threshold ($\sqrt{\hat{s}}=m_{A_1}+m_{A_2}$) without any spin correlations
(i.e.~we use pure ``phase space" distributions)\footnote{In general, depending on the details of the 
new physics model, the parents $A_i$ will be produced with some non-zero boost, i.e., 
$\sqrt{\hat{s}}>m_{A_1}+m_{A_2}$. However, given the current LHC bounds,
the parents $A_i$ are expected to be heavy, so that they should be predominantly produced near threshold. 
We have also tested our methods below with more realistic event samples, including the effects 
from initial state radiation and proton structure, and found that our conclusions remain unchanged.}.

\subsection{$M_2$ kinematic endpoints and parent mass measurements}
 \label{sec:mass}
 
The relations (\ref{eq:abrelations}-\ref{eq:brelations}) imply that the on-shell constrained variables 
$M_{2XC}$ and $M_{2CC}$ can provide progressively better measurements of an upper kinematic 
endpoint, as compared with the conventional variables $M_{T2}$, $M_{2XX}$, and $M_{2CX}$.
The reason is that the shapes of the $M_{2XC}$ and $M_{2CC}$ distributions will be skewed to the right, 
thus better populating the bins in the vicinity of the endpoint. 
\begin{figure}[t]
\centering
\includegraphics[trim=0.5cm 0.5cm 1cm 0.3cm, width=7.5cm]{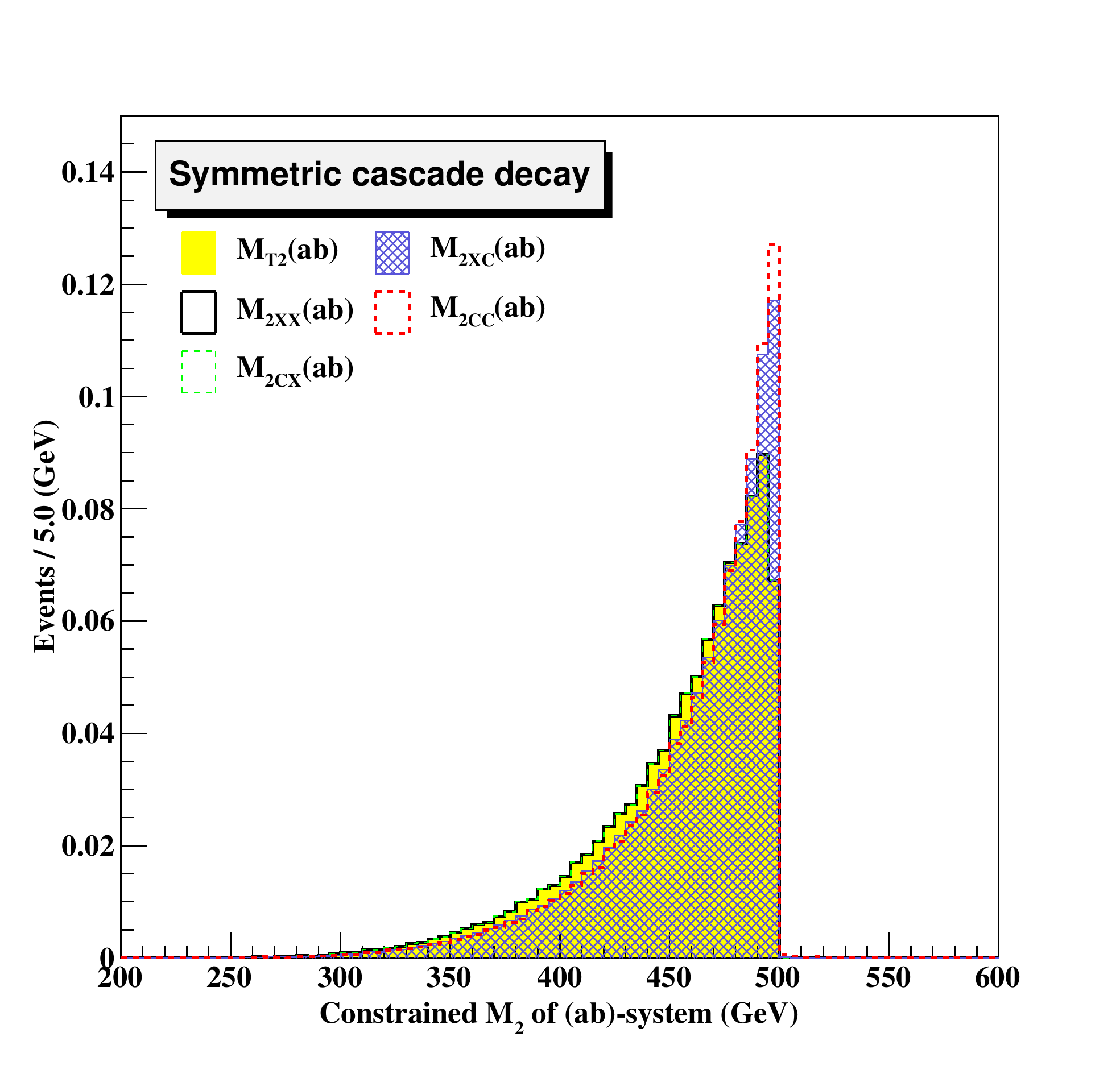}
\includegraphics[trim=0.5cm 0.5cm 1cm 0.3cm, width=7.5cm]{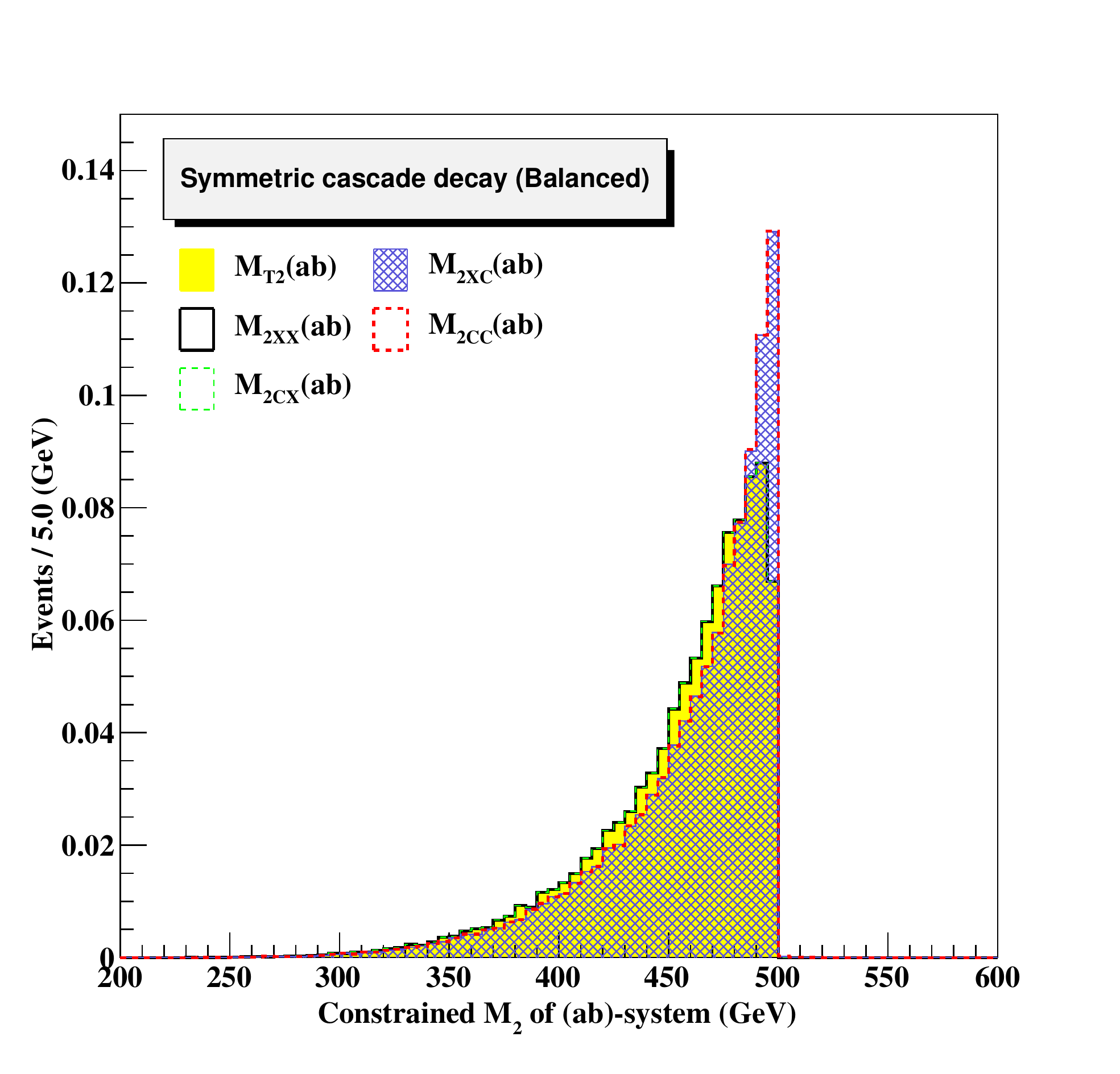} \\
\includegraphics[trim=0.5cm 0.5cm 1cm 0.3cm, width=7.5cm]{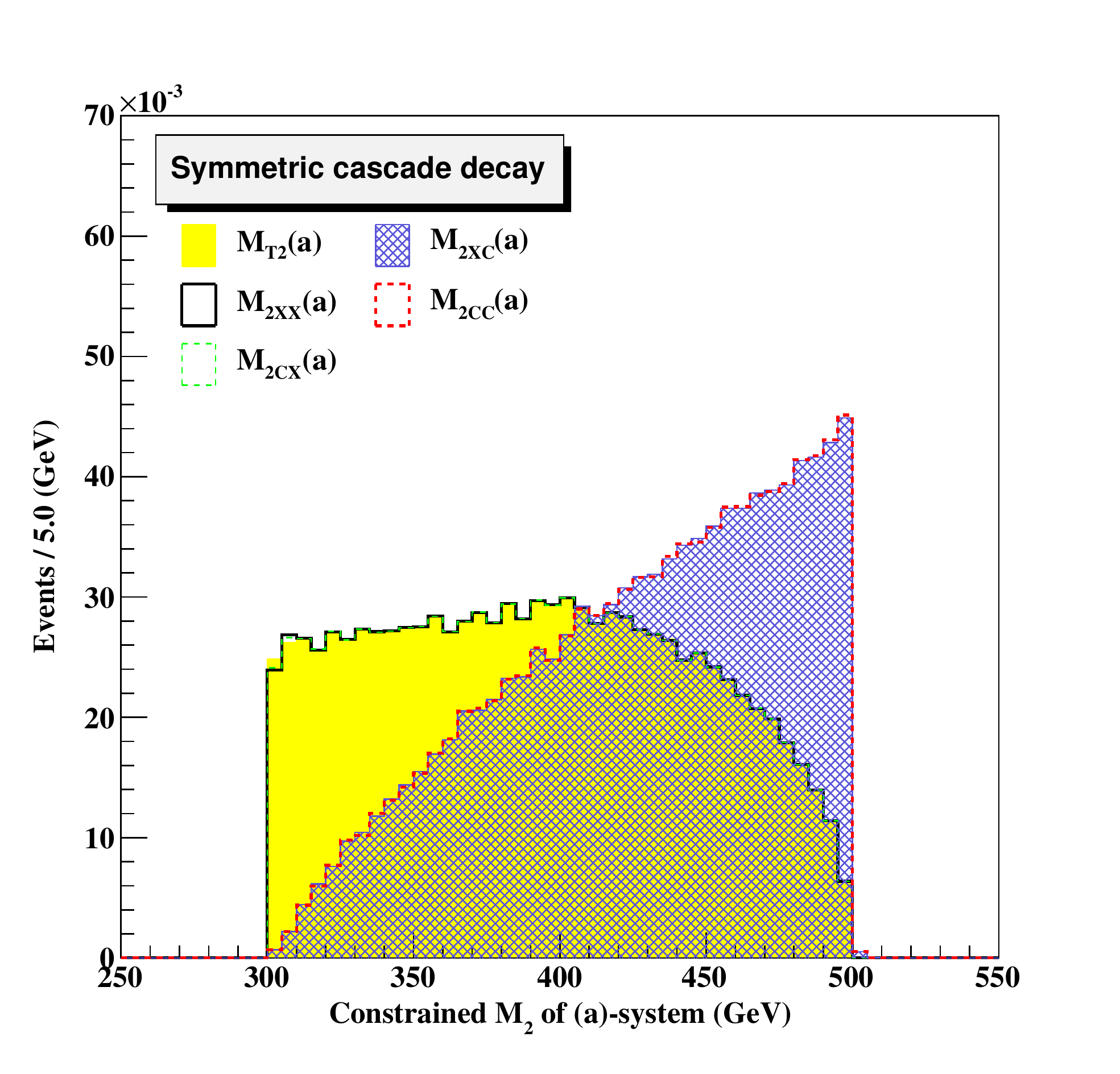}
\includegraphics[trim=0.5cm 0.5cm 1cm 0.3cm, width=7.5cm]{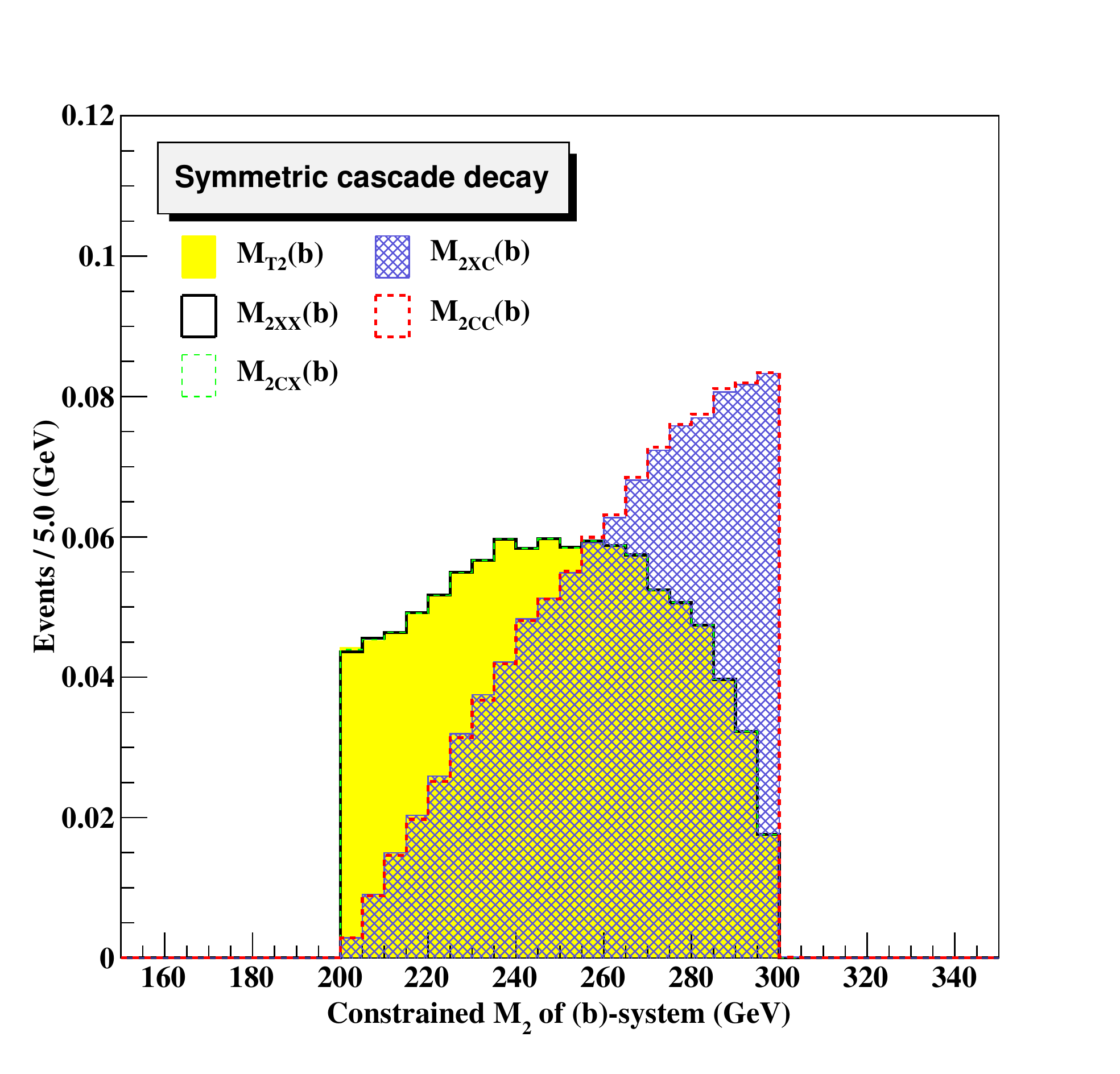}
\caption{\label{fig:plotcomparison} Unit-normalized differential distributions of the variables 
$M_{T2}$ (yellow shaded histograms), $M_{2XX}$ (black solid line),
$M_{2CX}$ (green dashed line), $M_{2XC}$ (blue hatched histograms), and
$M_{2CC}$ (red dashed line) for the process of Fig.~\ref{fig:process}(a)
with mass spectrum $(m_A,\;m_B,\;m_C)=(500,\;300,\;200)$ GeV.
Results are shown for subsystem $(ab)$ (upper left panel), 
subsystem $(ab)$ with balanced events only (upper right panel), 
subsystem $(a)$ (lower left panel), and subsystem $(b)$ (lower right panel). 
The input trial mass is chosen to be the same as the true mass of the relevant daughter particle. }
\end{figure}
This expectation is confirmed in Fig.~\ref{fig:plotcomparison}, where we compare the distributions of 
these five variables for the example of (\ref{eq:truemass}) with mass spectrum 
$(m_A,\;m_B,\;m_C)=(500,\;300,\;200)$ GeV.
For concreteness and simplicity, we choose the input trial mass, $\tilde{m}$, to be the same 
as the actual daughter mass in each case.
Comparisons are made for each subsystem of Fig.~\ref{fig:DecaySubsystem}:
subsystem $(ab)$ (upper left panel), 
subsystem $(ab)$ but using only balanced events (upper right panel), 
subsystem $(a)$ (lower left panel), and subsystem $(b)$ (lower right panel). 
Although each panel shows results for five variables, only three 
distributions (at most) can be seen, because the distributions of $M_{T2}$, $M_{2XX}$, and $M_{2CX}$
are identical, in accordance with the equivalence theorem from Sec.~\ref{sec:equivalence}.
An interesting observation is that $M_{2CC}$ and $M_{2CX}$ also turn out to be the same
for balanced events (i.e., events in which the transverse masses of the parents end up being 
equal for the momentum configuration obtained when minimizing the respective mass function). 
This observation is supported by the upper right plot in Fig.~\ref{fig:plotcomparison}, which 
uses only events in which $M_{T2}(ab)$ is obtained from a balanced configuration\footnote{In our sample, 
64\% (36\%) of the events have balanced (unbalanced) solutions for $M_{T2}(ab)$.},
and by the two lower plots in Fig.~\ref{fig:plotcomparison}, in which $M_{T2}(a)$ and $M_{T2}(b)$
always come from balanced configurations.

In the case of subsystem $(ab)$, the $M_{T2}$ distribution is already very sharp near the 
kinematic endpoint, and the improvement from replacing $M_{T2}$ with $M_{2XC}$ or $M_{2CC}$ 
appears marginal. However, the effect is very drastic in the case of subsystem $(a)$
or subsystem $(b)$ (the lower two plots in Fig.~\ref{fig:plotcomparison}), where 
the $M_{T2}$ distribution (the yellow-shaded histogram) has very few events near 
the kinematic endpoint. Now, using $M_{2XC}$ or $M_{2CC}$ in place of $M_{T2}$ 
completely changes the character of the distribution, and the bins near the endpoint 
become the most populated ones. Notice the extremely sharp drop-off at the endpoint 
of the $M_{2XC}(a)$ and $M_{2XC}(b)$ distributions (the blue-shaded histograms).
This feature should be easily observable over the background and would lead to more 
accurate endpoint measurements and extraction of masses.

\subsection{$M_2$-assisted mass reconstruction of relative peaks}
\label{sec:peak}

As explained in the introduction, an attractive feature of the $M_{T2}$ variable 
is that it provides an ansatz for the transverse momenta of the invisible particles.
The $M_2$ variables, being 3+1 dimensional extensions of $M_{T2}$,
take this one step further and extend the ansatz to the full 4-momenta 
of the invisible particles. This allows us to apply the MAOS method for 
mass reconstruction~\cite{Cho:2008tj,Choi:2009hn,Choi:2010dw} in a pure form, i.e.,~without the need for 
additional assumptions in order to solve for the longitudinal momenta of the 
invisible particles --- since those are already provided by the $M_2$ minimization 
itself\footnote{Thus in our case, the MAOS abbreviation should perhaps be thought of 
as ``$M_2$-assisted on-shell" reconstruction.}.
As shown in Sec.~\ref{sec:relation}, the variables $M_{2CC}$ and $M_{2XC}$
are somewhat better suited for our purpose (in comparison to $M_{2XX}$ and $M_{2CX}$), 
since they provide a unique ansatz for the invisible
particle momenta in the case of unbalanced events. Of course, for balanced events, 
any of our four types of $M_2$ variables can be used.

\begin{figure}[t]
\centering
\includegraphics[trim=0.6cm 0.5cm 1cm 1cm ,width=4.9cm]{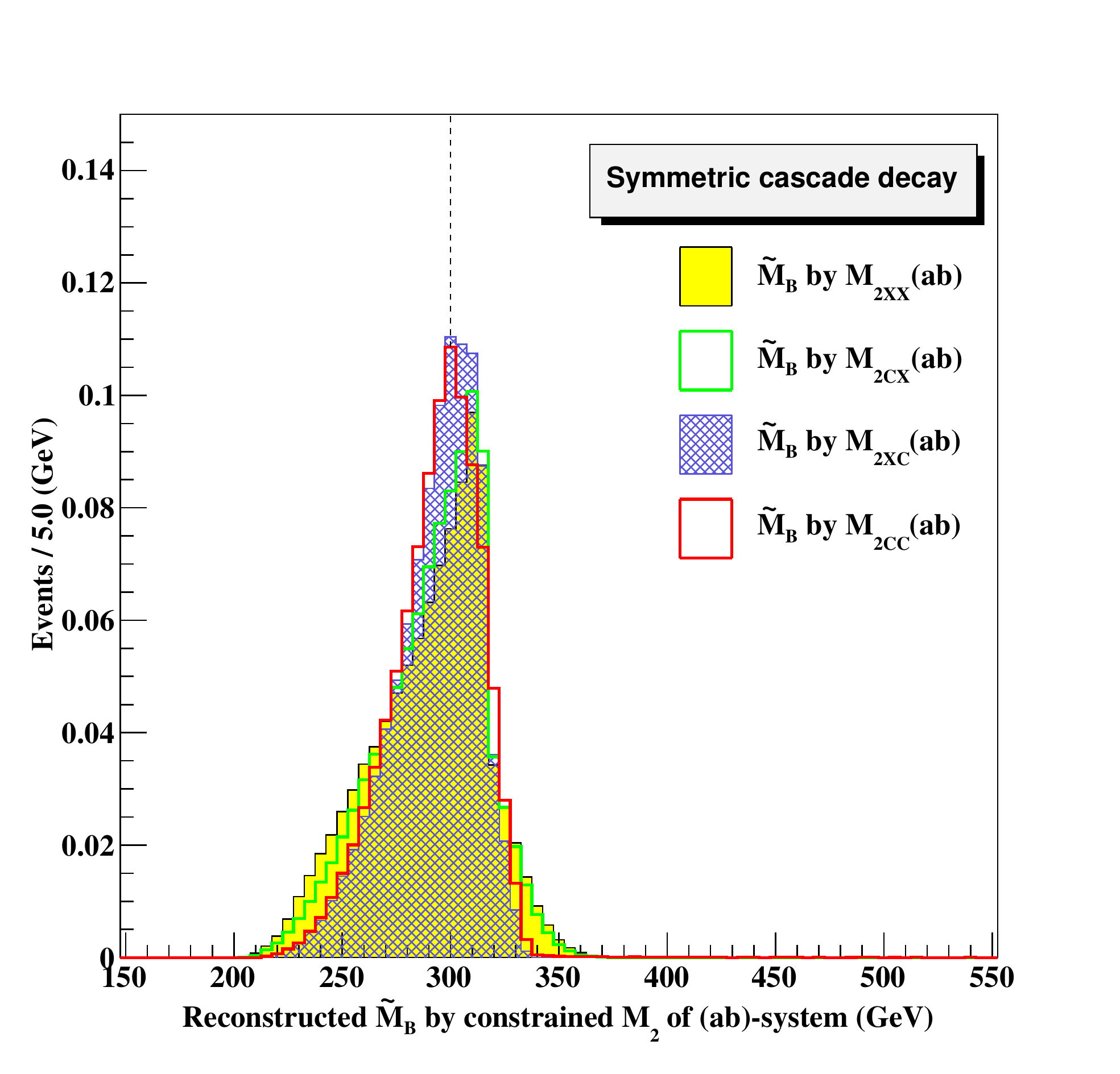}
\includegraphics[trim=0.6cm 0.5cm 1cm 1cm ,width=4.9cm]{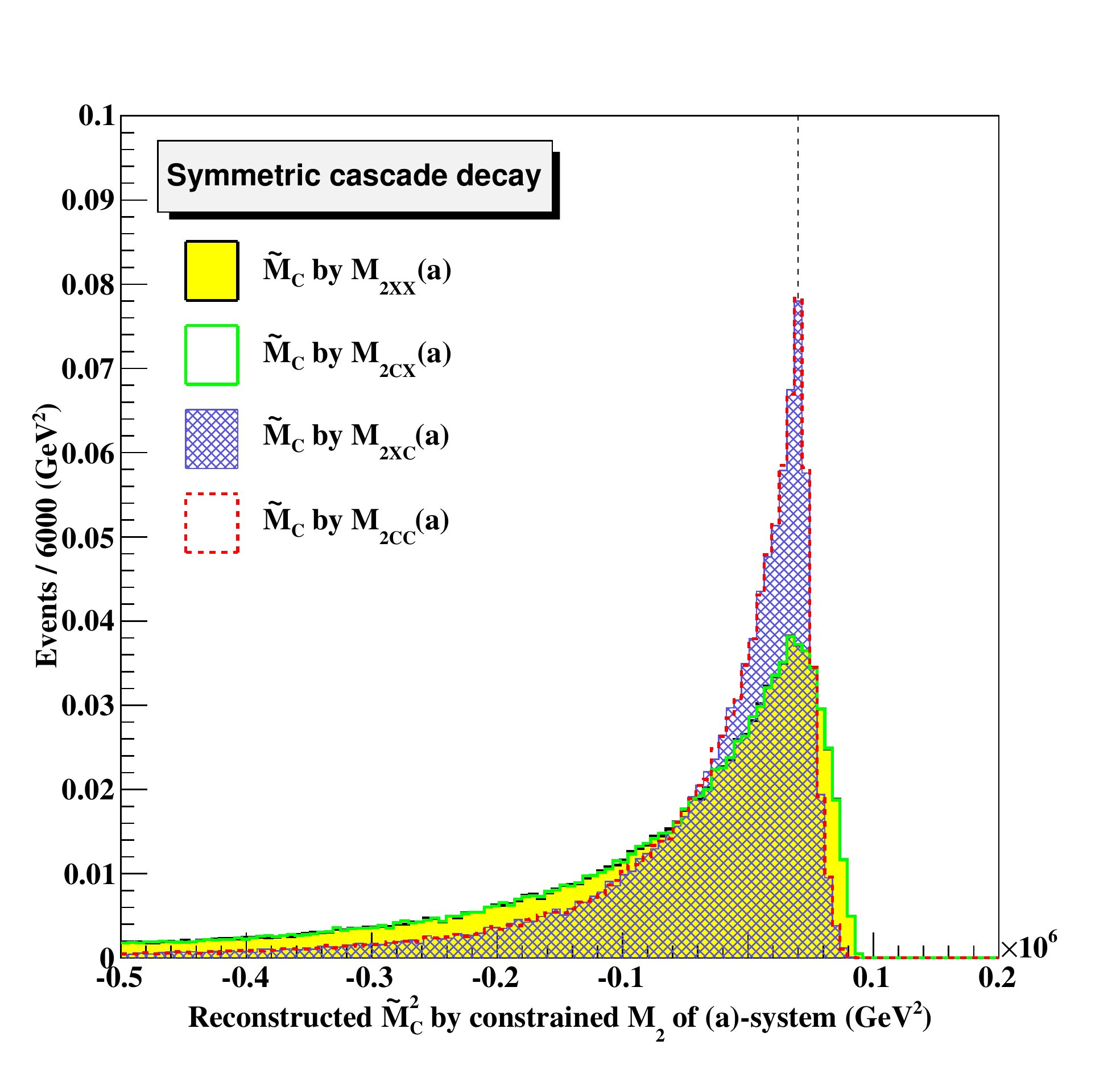}
\includegraphics[trim=0.6cm 0.5cm 1cm 1cm ,width=4.9cm]{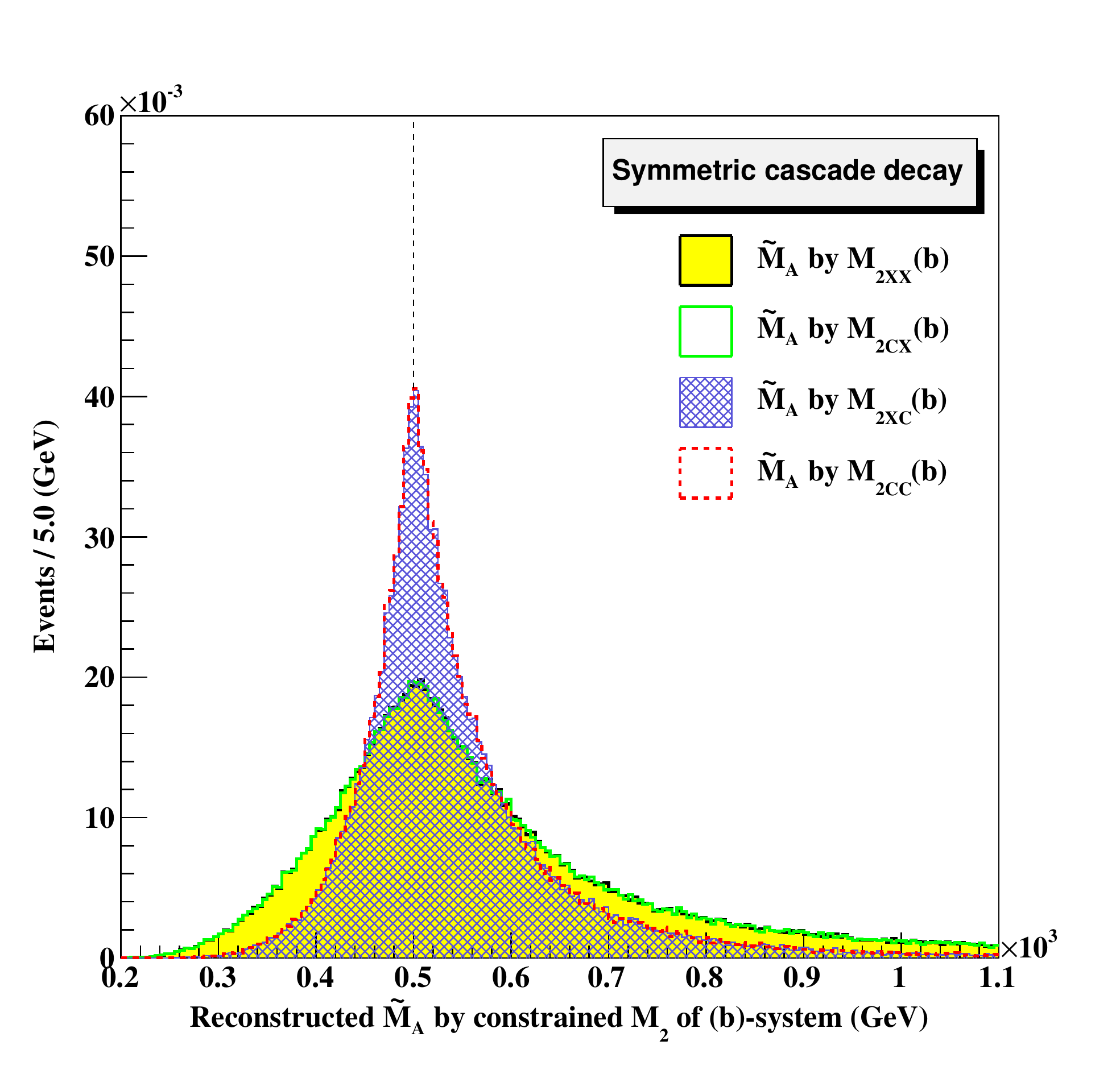}
\caption{\label{fig:plotreconstruction} Reconstruction of the mass of the relative particle
in the case of subsystem $(ab)$ (left panel), subsystem $(a)$ (middle panel), and subsystem $(b)$ (right panel). 
The ansatz for the invisible particle momenta can be taken from the corresponding 
$M_{2XX}$ variable (yellow-shaded histogram), $M_{2CX}$ variable (green histogram),
$M_{2XC}$ variable (blue-shaded histogram), or $M_{2CC}$ variable (red histogram).
The true mass spectrum and trial masses are chosen as in Fig.~\ref{fig:plotcomparison}. 
The vertical black dashed line in each plot denotes the true mass of the associated relative 
particle. The middle panel (for subsystem $(a)$) shows the mass {\em squared} of the relative particle,
which can be negative at times.}
\end{figure}

Fig.~\ref{fig:plotreconstruction} shows the results for the reconstruction of the masses of the relative 
particles in each of the three subsystems from Fig.~\ref{fig:DecaySubsystem}.
In the left panel of Fig.~\ref{fig:plotreconstruction}, we use the invisible momenta obtained from various $M_{2}(ab)$-type variables
to reconstruct the mass\footnote{From here on, a tilde over a quantity implies that it is a function of the test mass $\tilde m$.}, 
$\tilde M_B$, of the relative particle, $B$, in subsystem $(ab)$;
in the middle panel we use the momenta obtained from $M_{2}(a)$-type variables
to find the mass {\em squared}, $\tilde M_C^2$, of the relative particle, $C$, in subsystem $(a)$; 
and finally, in the right panel, we use the momenta from $M_{2}(b)$-type variables
to reconstruct the mass, $\tilde M_A$, of the relative particle, $A$, in subsystem $(b)$.
Each distribution in Fig.~\ref{fig:plotreconstruction} is color coded according to the 
type of $M_2$ variable supplying the invisible momenta: yellow-shaded histograms 
for the case of $M_{2XX}$, green histograms for $M_{2CX}$,
blue-shaded histogram for $M_{2XC}$, and red histograms for $M_{2CC}$.
The events are generated with the mass spectrum from Eq.~(\ref{eq:truemass})
and the test mass was always chosen to be the true mass of the relevant daughter particle:
$\tilde m=m_C$ for subsystem $(ab)$ (left panel),
$\tilde m=m_B$ for subsystem $(a)$ (middle panel), and
$\tilde m=m_C$ for subsystem $(b)$ (right panel).

The most interesting feature of the plots in Fig.~\ref{fig:plotreconstruction} is that
the distributions always peak close to the true mass of the relative particle
(denoted by the vertical black dashed line in each plot). This suggests a new technique 
for measuring the mass of a relative particle --- by using the location of the peak 
of the reconstructed relative mass distribution as shown in Fig.~\ref{fig:plotreconstruction}.
A closer inspection of Fig.~\ref{fig:plotreconstruction} reveals another advantage of the 
$M_2$ variables that incorporate on-shell kinematic constraints for relative particles 
in their definition. Note that in each panel, all four distributions peak 
near the true relative mass, but in the case of $M_{2XC}$ and (especially) $M_{2CC}$,
the peak is much more narrow, and, more importantly, the peak location is very close
to the true value of the mass of the respective relative particle. We therefore anticipate that the 
precision of the new technique will be much better when using $M_{2CC}$ (and $M_{2XC}$)
as opposed to $M_{2CX}$ or $M_{2XX}$.

This technique is in principle independent of (and complementary to) the previous methods 
in which masses are measured from upper kinematic endpoints. For example, consider 
particle $B$ (the intermediate particle in the decay chains of Fig.~\ref{fig:process}).
It is known that its mass can be measured  (as a function of $\tilde m \equiv \tilde m_C$) 
from the upper kinematic endpoint $M_{T2}^{max}(b)$ of the $M_{T2}(b)$ distribution in subsystem $b$, 
where $B_i$ is treated as a parent~\cite{Barr:2003rg,Burns:2008va}
\beq
\tilde m_B(\tilde m_C) = M_{T2}^{max}(b)(\tilde m_C).
\label{mBfromMT2b}
\eeq
Using the correct value for the daughter particle mass, $m_C$, in (\ref{mBfromMT2b})
yields the correct value of the parent mass, $m_B$:
\beq
m_B = M_{T2}^{max}(b)(m_C).
\label{mBfromMT2b_correct}
\eeq

We now propose to consider subsystem $(ab)$ instead, where $B_i$ is treated as a relative,
and extract $\tilde m_B(\tilde m_C)$ from the location of the peak $\tilde M_B^{peak}$ of one of the $\tilde M_B$ distributions 
in the left panel of Fig.~\ref{fig:plotreconstruction}, e.g.,~the one where the invisible momenta are fixed by $M_{2CC}(ab)$:
\beq
\tilde m_B(\tilde m_C) = \tilde M_{B}^{peak}(ab)(\tilde m_C).
\label{mBfromM2CCb}
\eeq

\begin{figure}[t]
\centering
\includegraphics[width=13cm]{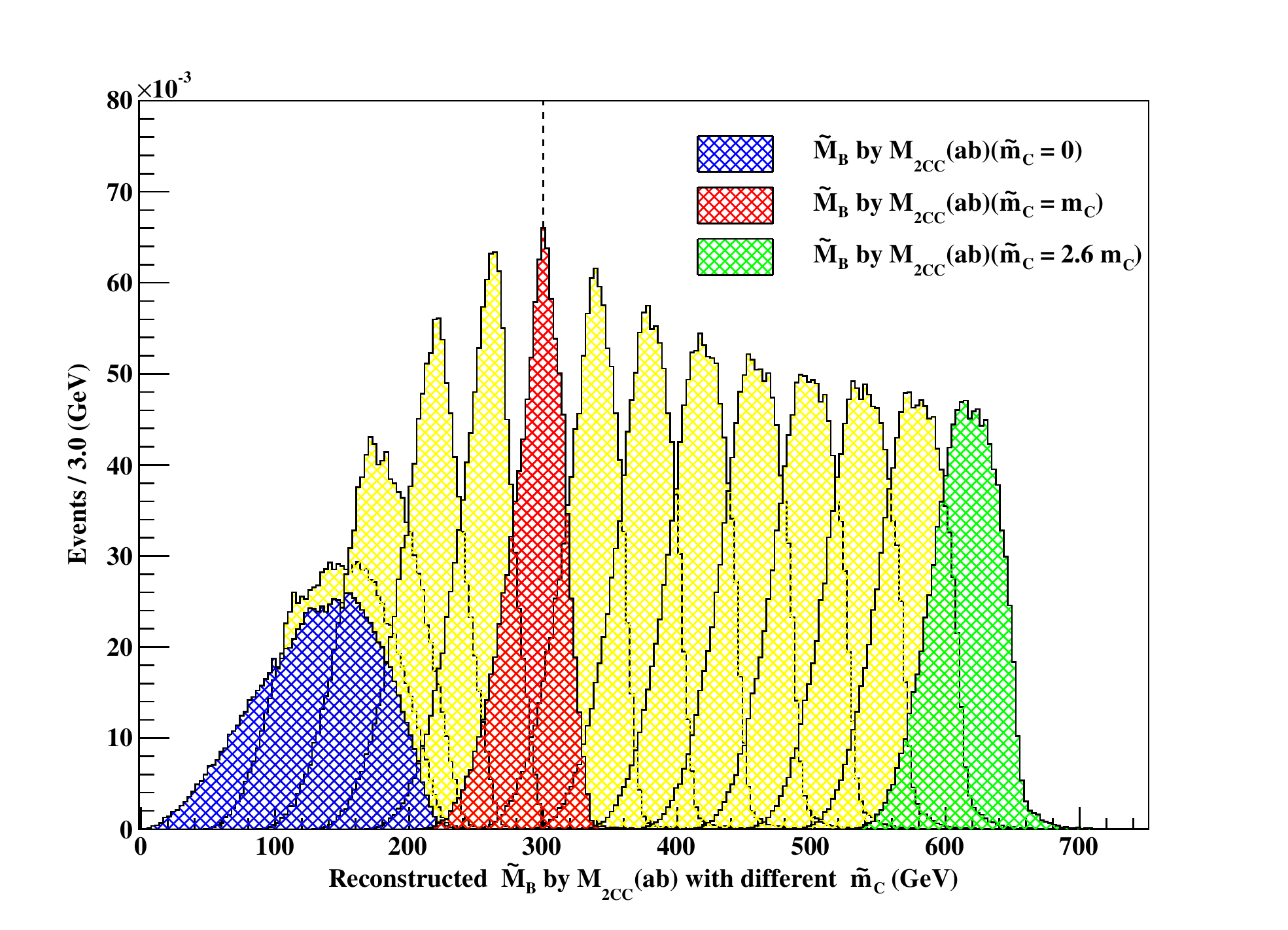}
\caption{\label{fig:Bpeak} Unit-normalized distributions of the reconstructed mass, $\tilde M_B$, of the relative particle
in subsystem $(ab)$, using invisible momenta from $M_{2CC}(ab)$, and picking a series of different values for the
input test mass, $\tilde m_C$, from $\tilde m_C=0$ (blue histogram) to $\tilde m_C=2.6 m_C$
(green histogram). The red shaded distribution corresponds 
to the true value, $\tilde m_C = m_C$, and is the same as the 
red histogram in the left panel of Fig.~\ref{fig:plotreconstruction}.
The vertical black dashed line marks the true mass, $m_B=300$ GeV, in our example. }
\end{figure}

\begin{figure}[t]
\centering
\includegraphics[width=13cm]{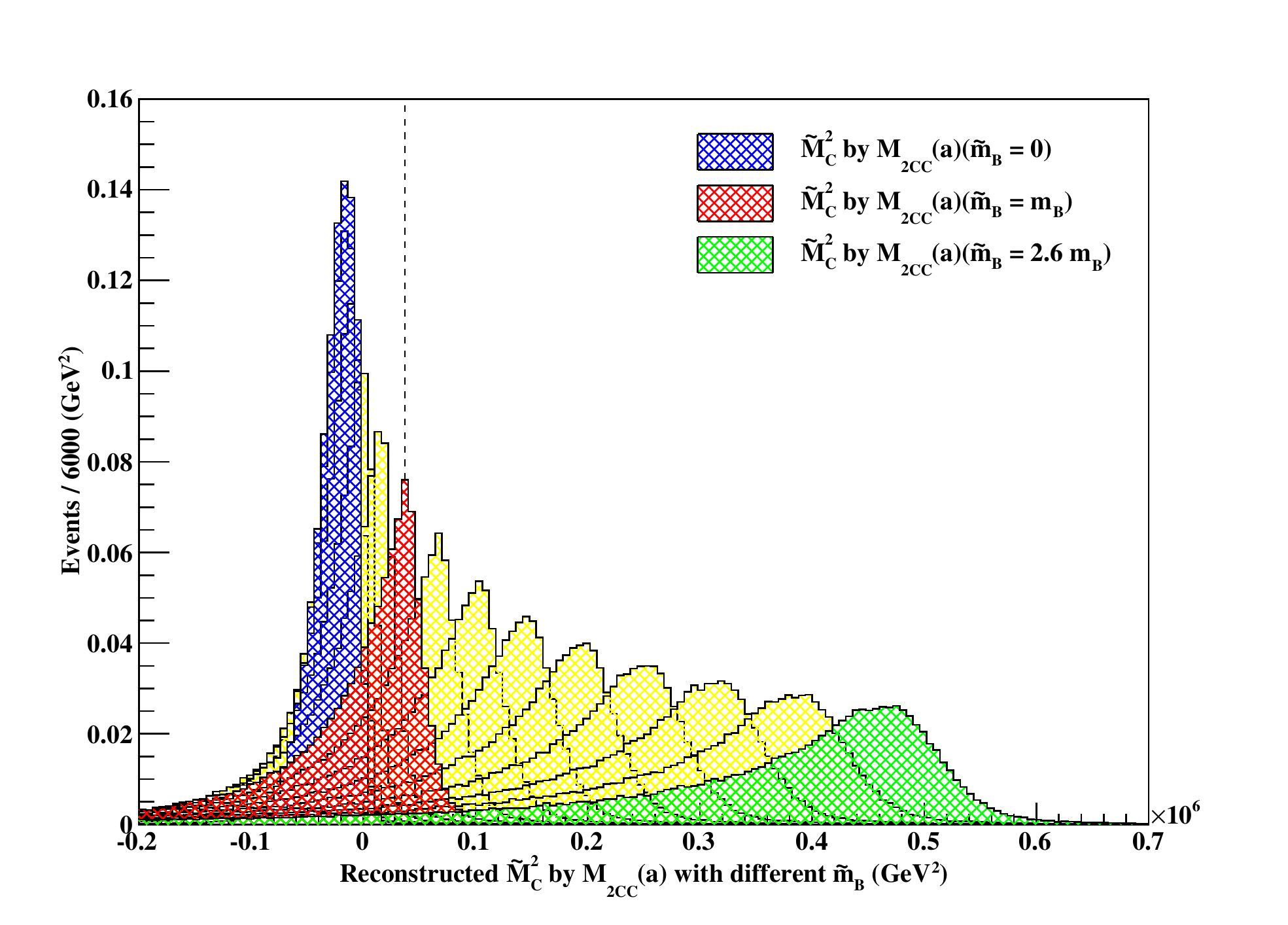}
\caption{\label{fig:Cpeak} The same as Fig.~\ref{fig:Bpeak}, this time reconstructing
the mass squared, $\tilde M_C^2$, of the relative particle in subsystem $(a)$ for several values of $\tilde m_B$, 
using invisible momenta from $M_{2CC}(a)$.
The red shaded distribution corresponds to the true value, $\tilde m_B = m_B$, and is the same as the 
red histogram in the middle panel of Fig.~\ref{fig:plotreconstruction}.
The vertical black dashed line marks the value of the true mass squared, $m_C^2=40,000\ {\rm GeV}^2$, in our example.  }
\end{figure}

The procedure is pictorially illustrated in Fig.~\ref{fig:Bpeak}. The $\tilde M_B$ distribution
from Fig.~\ref{fig:plotreconstruction} can now be re-obtained without the ``cheat" of fixing 
$\tilde m = m_C$. Instead, we can now simply vary the input test mass, $\tilde m_C$, 
and read off the location of the $\tilde M_B$ peak for each $\tilde m_C$ value, 
thus experimentally determining the function (\ref{mBfromM2CCb}).
This method relies on the fact demonstrated by the red shaded histogram in Fig.~\ref{fig:Bpeak} ---
that for the correct value, $m_C$, of the test daughter mass 
the peak of the $\tilde M_B$ distribution matches the correct value, $m_B$, of the mass for 
the relative particle\footnote{The careful reader might notice some other interesting features of the 
red-shaded histogram in Fig.~\ref{fig:Bpeak} --- it appears to be the most localized distribution
and, correspondingly, has the highest peak among all distributions shown in Fig.~\ref{fig:Bpeak}.
However, we do not pursue further this observation, since Fig.~\ref{fig:Cpeak} below provides a 
counterexample in which the highest peak is obtained for the wrong value of the test mass.}:
\beq
m_B = \tilde M_{B}^{peak}(ab)(m_C).
\label{mBfromM2CCb_correct}
\eeq
Notice the analogy between the relationships (\ref{mBfromMT2b}) and (\ref{mBfromM2CCb}) --- 
they both relate the mass of particle $B_i$ with the mass of particle $C_i$.
The difference is that the correlation (\ref{mBfromMT2b}) is derived from a kinematic endpoint
in subsystem $(b)$,
while the correlation (\ref{mBfromM2CCb}) is derived from the peak of a distribution within subsystem $(ab)$.
Also one should keep in mind that while (\ref{mBfromMT2b_correct}) is a mathematical identity,
the relation (\ref{mBfromM2CCb_correct}) at this point is a conjecture supported by
the numerical results from Figs.~\ref{fig:plotreconstruction} and \ref{fig:Bpeak}.
(Compare to the similar conjecture relating the peak of the $\sqrt{\hat{s}}_{min}$
distribution to the mass of the corresponding parents~\cite{Konar:2008ei}.)

Similar logic can be applied to particle $A$. It is known that its mass can be measured 
from the upper kinematic endpoint of the $M_{T2}(ab)$ distribution in subsystem $(ab)$, 
as a function of the input test mass, $\tilde m_C$, in complete analogy to (\ref{mBfromMT2b})
\cite{Cho:2007qv,Cho:2007dh}: 
\beq
\tilde m_A(\tilde m_C) = M_{T2}^{max}(ab)(\tilde m_C).
\label{mAfromMT2ab}
\eeq
Alternatively, it can be measured from the upper kinematic endpoint of the $M_{T2}(a)$ 
distribution in subsystem $(a)$, this time as a function of the test mass, $\tilde m_B$ \cite{Barr:2003rg,Cho:2007dh,Burns:2008va}: 
\beq
\tilde m_A(\tilde m_B) = M_{T2}^{max}(a)(\tilde m_B).
\label{mAfromMT2a}
\eeq
We now propose a third way of measuring the mass of $A_i$, by treating it as a relative particle in subsystem $(b)$:
using the invisible momenta from the $M_{2CC}(b)$ calculation, we can
reconstruct the mass of the relative, $\tilde M_A$, and read off the location of the peak, $\tilde M_A^{peak}$,
in analogy to (\ref{mBfromM2CCb})
\beq
\tilde m_A(\tilde m_C) = \tilde M_{A}^{peak}(b)(\tilde m_C).
\label{mAfromM2CCb}
\eeq

\begin{figure}[t]
\centering
\includegraphics[width=13cm]{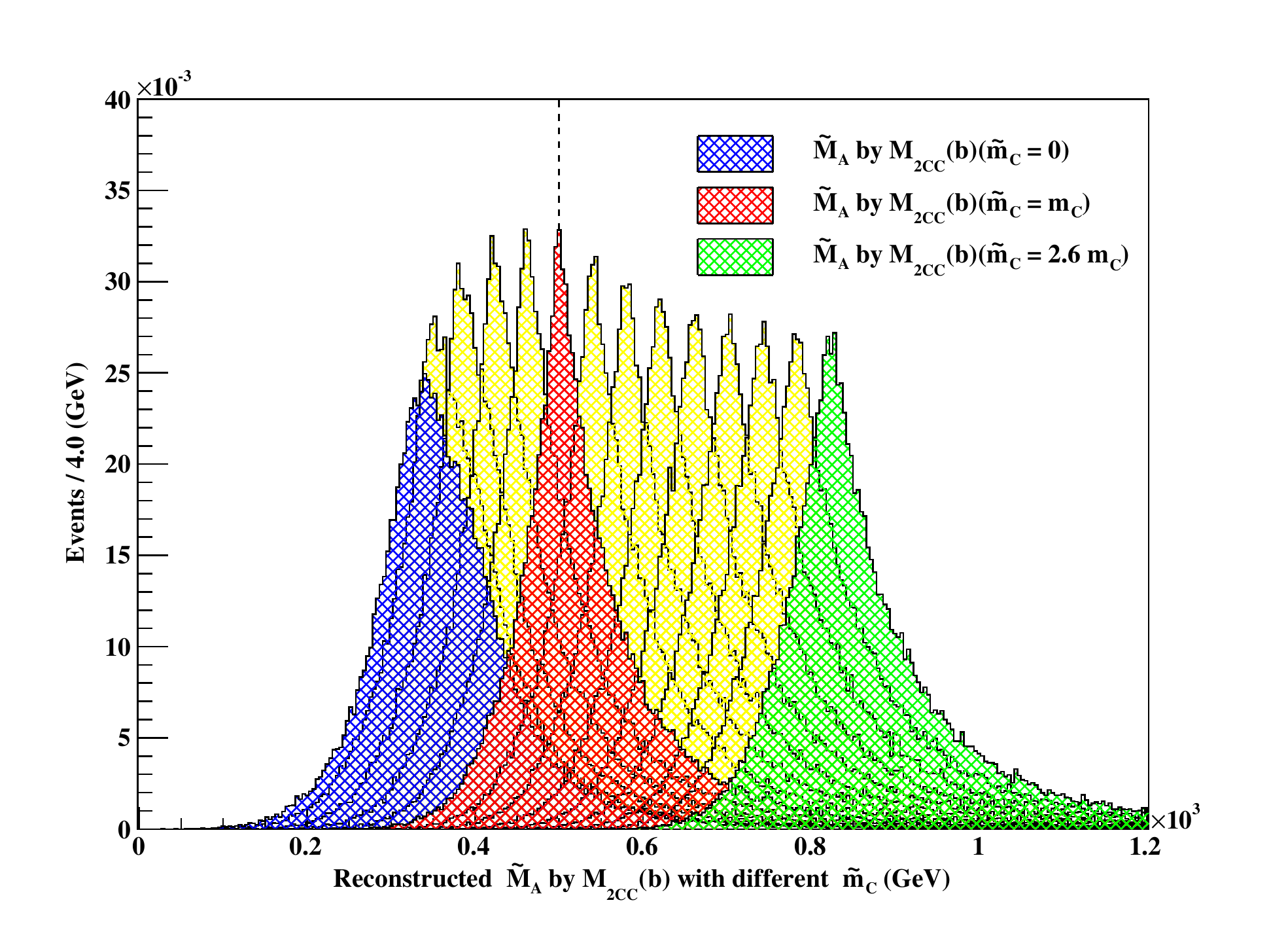}
\caption{\label{fig:Apeak} The same as Fig.~\ref{fig:Bpeak}, this time reconstructing
the mass, $\tilde M_A$, of the relative particle in subsystem $(b)$ for several values of $\tilde m_C$, 
using invisible momenta from $M_{2CC}(b)$.
The red shaded distribution corresponds to the true value, $\tilde m_C = m_C$, and is the same as the 
red histogram in the right panel of Fig.~\ref{fig:plotreconstruction}.
The vertical black dashed line marks the true mass, $m_A=500$ GeV, in our example.  }
\end{figure}

The function, (\ref{mAfromM2CCb}), can be experimentally derived as shown in Fig.~\ref{fig:Apeak} ---
one varies the test mass, $\tilde m_C$, and forms a series of $\tilde M_{A}$ distributions. 
The location of the peak of each distribution represents the value of $\tilde m_A$
for the given hypothesized value of $\tilde m_C$. The red shaded histogram in 
Fig.~\ref{fig:Apeak} corresponds to the true value of $\tilde m_C=m_C$ and again 
peaks at the correct value of the mass, $m_A$, of the relative particle:
\beq
m_A = \tilde M_{A}^{peak}(b)(m_C).
\label{mAfromM2CCb_correct}
\eeq

Finally, one may also consider the subsystem $(a)$ and study the distributions of
the reconstructed relative mass, $\tilde M_{C}$, shown in the middle panel of Fig.~\ref{fig:plotreconstruction}.
This establishes the relation
\beq
\tilde m_C(\tilde m_B) = \tilde M_{C}^{peak}(a)(\tilde m_B).
\label{mCfromM2CCa}
\eeq

The procedure is illustrated in Fig.~\ref{fig:Cpeak}, where we have used $M_{2CC}(a)$ to fix the momenta 
of the invisible particles before computing $\tilde M_{C}^2$. A peculiar feature of Fig.~\ref{fig:Cpeak} 
is that for low enough values of the test mass, $\tilde m_B$, the peak of the distribution is found at
negative values of $\tilde M_{C}^2$, which is why we do not take a
square root and instead use the mass squared in the plot. 
Nevertheless, the important feature of Fig.~\ref{fig:Cpeak} is that,
just like in Figs.~\ref{fig:Bpeak} and \ref{fig:Apeak}, for the correct choice of the test mass,
$\tilde m_B=m_B$ (see red histogram), the peak reveals the true value, $m_C$, of the relative particle
(in this case $C_i$).

\begin{figure}[t]
\centering
\includegraphics[width=12cm]{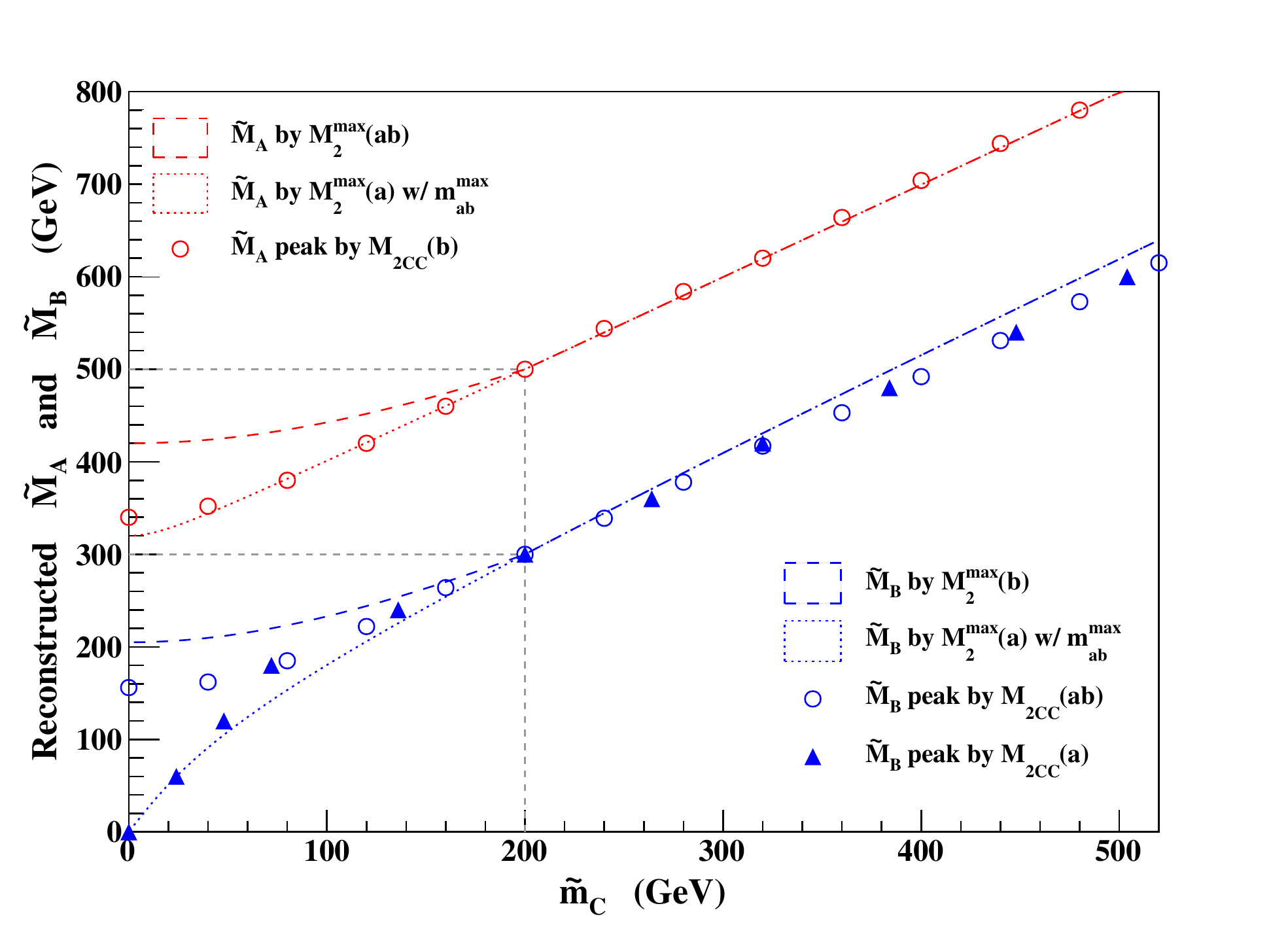}
\caption{\label{fig:masses} A summary of the different mass correlation methods discussed in the text:
(\ref{mBfromM2b}) is represented by a blue dashed line,
(\ref{mAfromM2ab}) is given by a red dashed line,
(\ref{mApeak}) is shown by the red open circles,
(\ref{mBpeak}) is denoted by the blue open circles, while
(\ref{mCpeak}) is marked by the blue triangles.
The red dotted line represents the relationship between $\tilde m_A$ and $\tilde m_C$
which is obtained by eliminating $\tilde m_B$ from (\ref{mAfromM2a}) and (\ref{Mabmax}), while 
the blue dotted line shows the orthogonal relationship among $\tilde m_B$ and $\tilde m_C$
resulting from eliminating $\tilde m_A$ from (\ref{mAfromM2a}) and (\ref{Mabmax}).
}
\end{figure}

Before concluding, in Fig.~\ref{fig:masses} we summarize the different mass determination methods 
discussed in this section. The existing method relies on measuring $M_{T2}$ kinematic endpoints
in the three subsystems of Fig.~\ref{fig:DecaySubsystem}, establishing the three relationships
(\ref{mBfromMT2b}), (\ref{mAfromMT2ab}), and (\ref{mAfromMT2a}).  In Sec.~\ref{sec:mass},
we proposed to measure the sharper $M_{2CC}$ kinematic endpoints instead, resulting in
three analogous relations
\bea
\tilde m_B(\tilde m_C) &=& M_{2}^{max}(b)(\tilde m_C),
\label{mBfromM2b} \\ [2mm]
\tilde m_A(\tilde m_C) &=& M_{2}^{max}(ab)(\tilde m_C),
\label{mAfromM2ab} \\ [2mm]
\tilde m_A(\tilde m_B) &=& M_{2}^{max}(a)(\tilde m_B).
\label{mAfromM2a}
\eea
These can be supplemented with the classic measurement of the kinematic endpoint of the 
invariant mass, $M_{ab}$, of the two visible particles, $a_i$ and $b_i$, in each decay chain 
\beq
M_{ab}^{max} = \sqrt{\frac{(\tilde m_A^2-\tilde m_B^2)(\tilde m_B^2-\tilde m_C^2)}{\tilde m_B^2}},
\label{Mabmax}
\eeq
which provides a constraint among all three masses $\tilde m_A$, $\tilde m_B$, and $\tilde m_C$.
The four measurements (\ref{mBfromM2b}-\ref{Mabmax})
are already sufficient to determine the three unknowns $\tilde m_A$, $\tilde m_B$, and $\tilde m_C$
\cite{Burns:2008va}.
The new measurements proposed in Sec.~\ref{sec:peak} are the peak determinations
\bea
\tilde m_A(\tilde m_C) &=& \tilde M_{A}^{peak}(b)(\tilde m_C),
\label{mApeak} \\ [2mm]
\tilde m_B(\tilde m_C) &=& \tilde M_{B}^{peak}(ab)(\tilde m_C),
\label{mBpeak} \\ [2mm]
\tilde m_C(\tilde m_B) &=& \tilde M_{C}^{peak}(a)(\tilde m_B).
\label{mCpeak}
\eea

The seven relations (\ref{mBfromM2b}-\ref{mCpeak}) are pictorially illustrated in Fig.~\ref{fig:masses}.
In order to display all seven relations on the same plot, we first plot (\ref{mBfromM2b}-\ref{mAfromM2ab})
and (\ref{mApeak}-\ref{mCpeak}) directly, then from the remaining two relations
(\ref{mAfromM2a}) and (\ref{Mabmax}) we either eliminate $\tilde m_B$ to obtain $\tilde m_A$ as a function of $\tilde m_C$
(red dotted line), or eliminate $\tilde m_A$ to obtain $\tilde m_B$ as a function of $\tilde m_C$ (blue dotted line).
All seven correlations (\ref{mBfromM2b}-\ref{mCpeak}) agree for the correct values for 
$m_A$, $m_B$ and $m_C$, marked with the black dotted lines in Fig.~\ref{fig:masses}.
What is more interesting is that they {\em disagree} for the wrong values of the test input mass, $\tilde m_C$.
This is particularly noticeable in the region $\tilde m_C<m_C$.
Fig.~\ref{fig:masses} suggests that by combining the results from all the different methods (\ref{mBfromM2b}-\ref{mCpeak})
one can determine the true value of $m_C$ as the location of the crossing point of the different curves shown in the figure.
Our method is complementary to other methods in the literature for determining the absolute value of $m_C$
\cite{Cho:2007qv,Gripaios:2007is,Barr:2007hy,Cho:2007dh,Barr:2009jv,Matchev:2009ad,Matchev:2009fh,Alwall:2009sv,Konar:2009wn,Cohen:2010wv,Cheng:2011ya,Serna:2008zk}.

\section{Using $M_2$ variables for topology disambiguation}
\label{sec:topology}

\begin{figure}[t]
\centering
\includegraphics[scale=0.9]{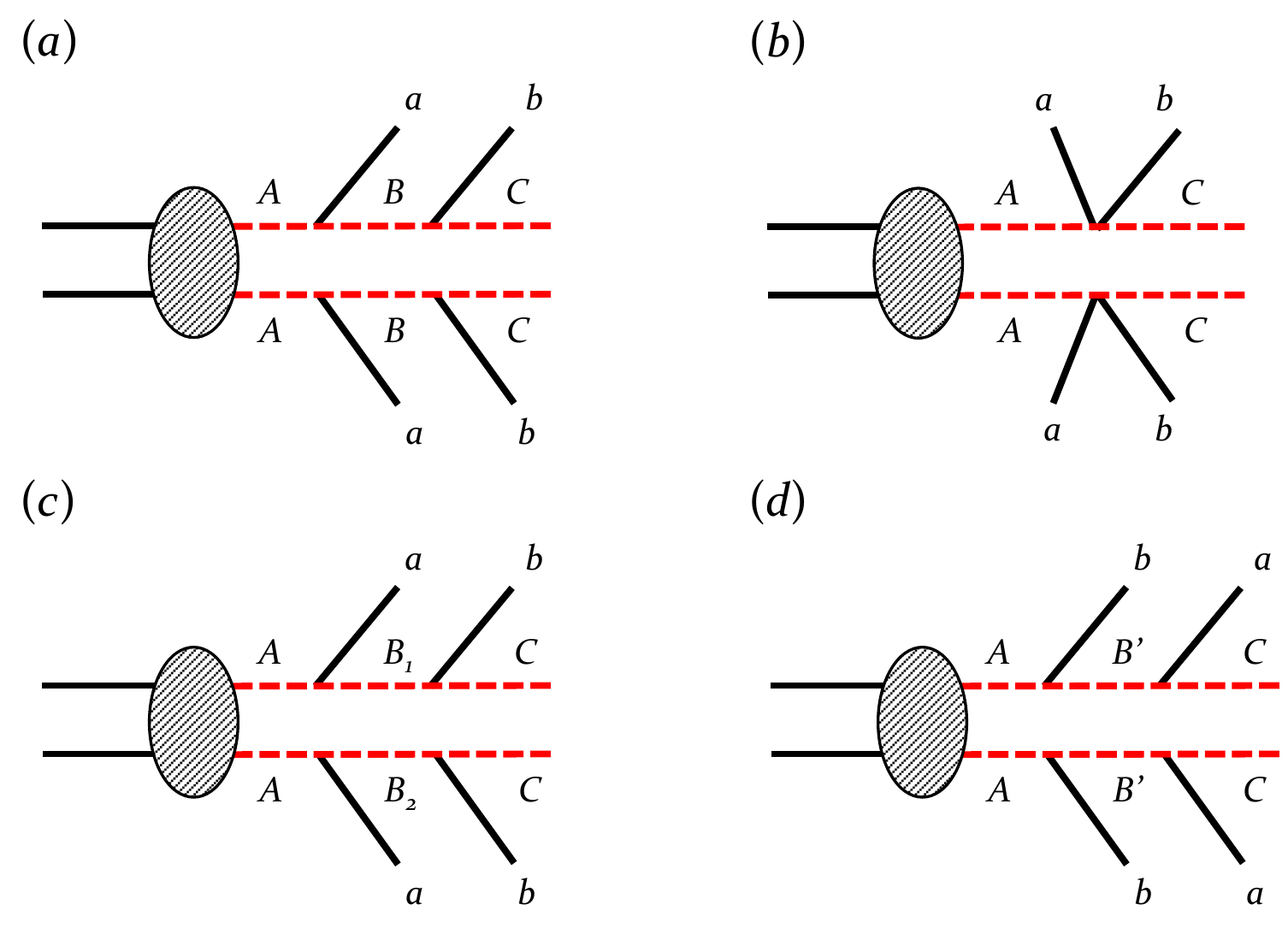}
\caption{\label{fig:DecayTopologies} The four benchmark decay topologies studied in Sec.~\ref{sec:topology}.}
\end{figure}

Up to this point, we have been studying events under the correct assumption about the event topology.
However, in a real experiment, there is no prior indication as to what
the correct event topology is
for any given observed final state, and one should consider (and test for) all possible alternatives. 
This is exactly what we set out to do in this section. Given our observed final state of two visible $a$ particles,
two visible $b$ particles and missing transverse momentum, a number of
event topologies are possible; four of which are
shown in Fig.~\ref{fig:DecayTopologies}. Fig.~\ref{fig:DecayTopologies}(a) shows our nominal scenario, (\ref{eq:truemass}),
considered so far, in which there is an on-shell $B_i$ resonance in each chain and furthermore, 
the two $B$ resonances are the same: $B_1=B_2\equiv B$. 
Fig.~\ref{fig:DecayTopologies}(b) represents the off-shell scenario in which the intermediate $B$ 
resonance is very heavy and the decays are three-body.
Fig.~\ref{fig:DecayTopologies}(c) is the same as Fig.~\ref{fig:DecayTopologies}(a), but with a slight
modification --- now the two intermediate resonances, $B_i$, are different: $B_1\ne B_2$.
Finally, Fig.~\ref{fig:DecayTopologies}(d) is the analogue of Fig.~\ref{fig:DecayTopologies}(a)
in which the visible particles, $a$ and $b$, are switched, i.e.,~the decay to $b$ takes place first,
followed by the decay to $a$.

In this section, we shall design several tests which discriminate among the alternative possibilities 
depicted in Fig.~\ref{fig:DecayTopologies}. The tests make crucial use of the constrained $M_2$ 
variables introduced in Sec.~\ref{sec:notation}.

\subsection{Endpoint test}
\label{sec:endpoint}

We first design a test to distinguish among the three event topologies shown in Fig.~\ref{fig:DecayTopologies}(a),
Fig.~\ref{fig:DecayTopologies}(b), and
Fig.~\ref{fig:DecayTopologies}(c).  (This test will not be able to discriminate 
among Fig.~\ref{fig:DecayTopologies}(a) and Fig.~\ref{fig:DecayTopologies}(d).)
The basic idea is very simple. Recall that the $M_{2XC}(ab)$ and $M_{2CC}(ab)$ variables 
from Sec.~\ref{sec:notation} were defined under the assumption of a
common relative particle.  I.e.,
\begin{itemize}
\item there is an intermediate $B_i$ resonance in each decay chain, and
\item the two $B_i$ particles are the same, so that $m_{B_1}=m_{B_2}$.
\end{itemize}
If either of these two assumptions is incorrect, the definition of $M_{2XC}(ab)$ and $M_{2CC}(ab)$ loses its
physical meaning, and as a result something will go wrong. Therefore, by testing for the consistency of 
$M_{2XC}(ab)$ and $M_{2CC}(ab)$ with another, topology-independent, variable like $M_{2XX}(ab)$,
we can verify the above two assumptions. Note that relaxing the first assumption leads to 
the event topology of Fig.~\ref{fig:DecayTopologies}(b), while dropping the second assumption
leads to the event topology of Fig.~\ref{fig:DecayTopologies}(c).

How can one test for the consistency of $M_{2XC}(ab)$ and $M_{2CC}(ab)$?
Recall that the basic property of all $M_2$ variables is that they provide a lower bound on the 
mass of the corresponding parent, and their upper kinematic endpoints saturate that bound,
revealing the mass of the parent (as a function of the test daughter mass).
Now consider the relevant variables (\ref{eq:variables}) for subsystem $(ab)$.
They bound the mass of {\em the same} parent $A$, the only difference is that they 
have various assumptions about the event topology built in. Therefore, if all those assumptions 
are correct, the kinematic endpoints of all the variables should agree as well\footnote{Of course, 
due to the equivalence theorem discussed in Sec.~\ref{sec:equivalence},
the first two equalities in Eq.~(\ref{eq:endpointtest}) are trivially satisfied, 
so that the actual test involves only the last two equalities in Eq.~(\ref{eq:endpointtest}).}:
\beq
M_{T2}^{max} = M_{2XX}^{max}= M_{2CX}^{max}= M_{2XC}^{max}= M_{2CC}^{max}.
\label{eq:endpointtest}
\eeq
Conversely, if some of the assumptions are not satisfied, (\ref{eq:endpointtest}) will be violated ---
there will be a certain number of events in which the values of $M_{2XC}(ab)$ 
and $M_{2CC}(ab)$ will violate the upper kinematic endpoint $M_{2XX}^{max}$
of the topology-independent variable $M_{2XX}(ab)$.
 
\begin{figure}[t]
\centering
\includegraphics[trim=1cm 0.5cm 1cm 1cm ,width=4.9cm]{sym_ab_M2}
\includegraphics[trim=1cm 0.5cm 1cm 1cm ,width=4.9cm]{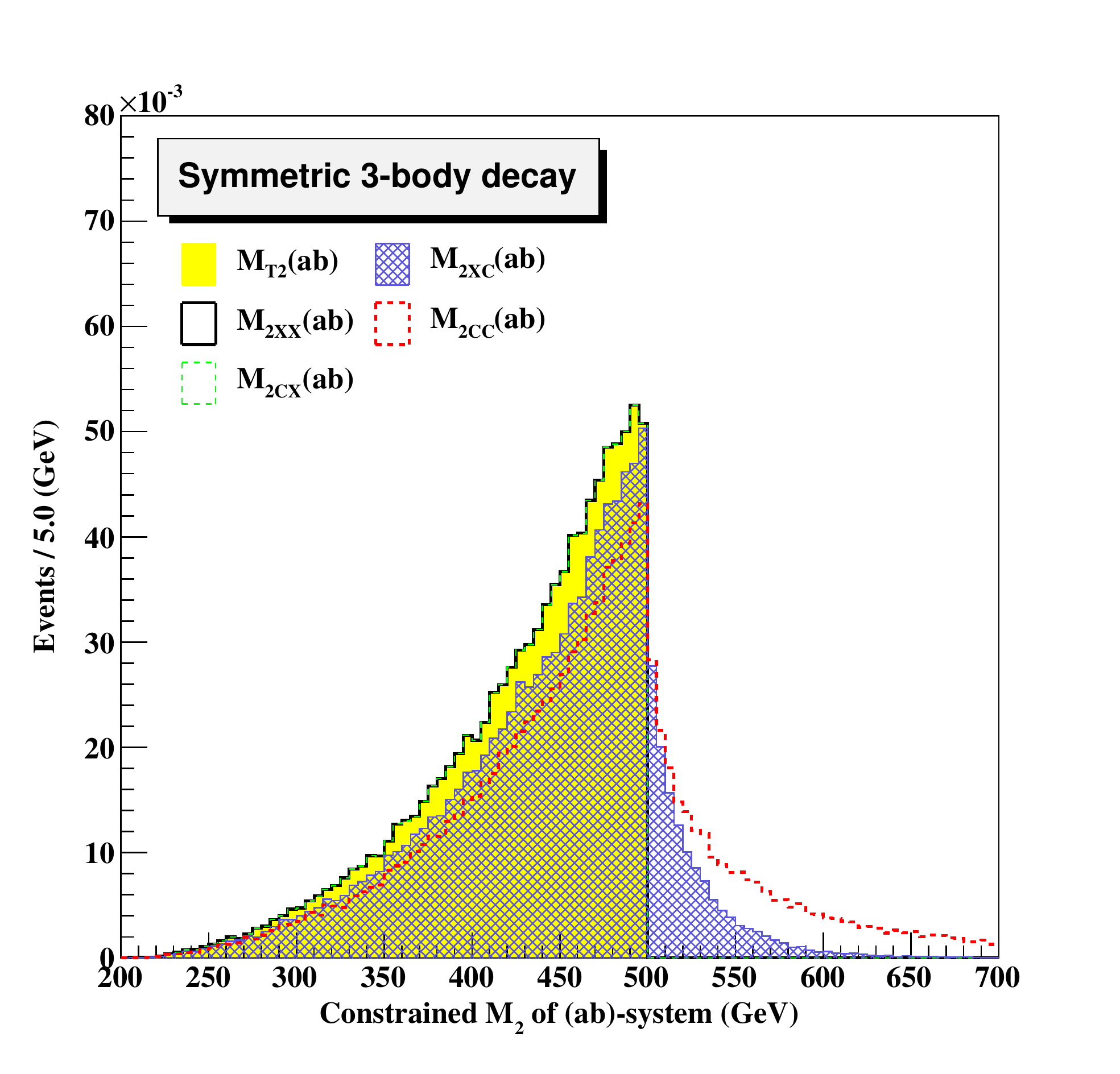}
\includegraphics[trim=1cm 0.5cm 1cm 1cm ,width=4.9cm]{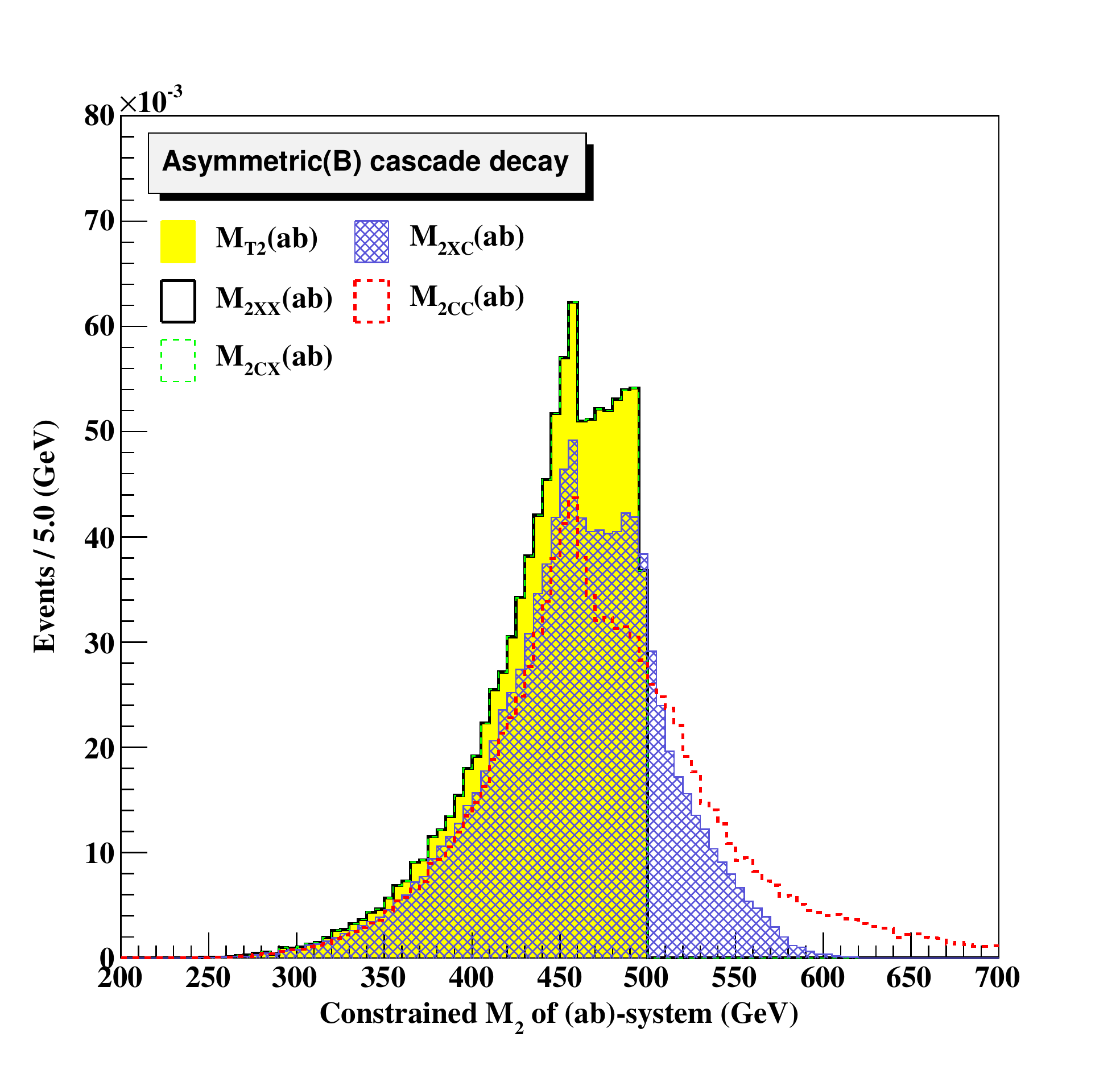}
\caption{\label{fig:3endpoints} The same as the upper left panel of Fig.~\ref{fig:plotcomparison}, 
but using events from three different scenarios. The left panel shows
the nominal event topology from Fig.~\ref{fig:DecayTopologies}(a) 
with $(m_A,\;m_B,\;m_C)=(500,\;300,\;200)$ GeV.  The middle panel 
corresponds to the off-shell case of Fig.~\ref{fig:DecayTopologies}(b) 
with $(m_A,\;m_C)=(500,\;200)$ GeV.  
The right panel represents the asymmetric 
event topology from Fig.~\ref{fig:DecayTopologies}(c) with  
$(m_A,\;m_{B_2},\;m_{B_1},\;m_C)=(500,\;400,\;300,\;200)$ GeV.
The test mass was always chosen to be $\tilde m_C=200$ GeV.}
\end{figure}

The test is performed in Fig.~\ref{fig:3endpoints}, where we compare the distributions of 
the five variables in the $(ab)$ subsystem, as in the upper left panel of Fig.~\ref{fig:plotcomparison}.
In the left panel of Fig.~\ref{fig:3endpoints}, we first consider the case of our nominal
event topology from Fig.~\ref{fig:DecayTopologies}(a) with the mass spectrum from
(\ref{eq:truemass}). As already observed in Fig.~\ref{fig:plotcomparison},
the distributions may have slightly different shapes, but their endpoints are exactly the same. 
Therefore, this case passes the endpoint test, (\ref{eq:endpointtest}), as expected.

We next consider the off-shell case of Fig.~\ref{fig:DecayTopologies}(b) 
with $(m_A,\;m_C)=(500,\;200)$ GeV and plot the results in the middle panel of 
Fig.~\ref{fig:3endpoints}. In accordance with the equivalence theorem from
Sec.~\ref{sec:equivalence}, the distributions of $M_{T2}$, $M_{2XX}$, and $M_{2CX}$
are identical, and their common endpoint provides a reference value, $M_{2XX}^{max}$,
to be compared against the endpoints of $M_{2XC}$ and $M_{2CC}$.
The plot clearly shows that the distributions of $M_{2XC}$ and $M_{2CC}$
develop long tails beyond $M_{2XX}^{max}$, thus violating (\ref{eq:endpointtest})
and failing the endpoint test. The violation is more severe in the case of $M_{2CC}$ 
(the red histograms in Fig.~\ref{fig:3endpoints}),
where a larger number of events have migrated beyond the anticipated endpoint $M_{2XX}^{max}$.
The reason for this violation is easy to understand --- in the off-shell case of Fig.~\ref{fig:DecayTopologies}(b)
there are no intermediate resonances, $B_1$ and $B_2$. Thus when we enforce
the relative constraint, $M_{B_1}=M_{B_2}$, in constructing the $M_{2XC}$ and $M_{2CC}$
variables, we unnecessarily restrict the range of allowed values of the invisible momenta
during the minimization, and thus arrive at an unphysical global minimum.
Based on the results from the middle panel of Fig.~\ref{fig:3endpoints},
we can therefore safely rule out the on-shell event topology of Fig.~\ref{fig:DecayTopologies}(a) 
as being the source of these events.


The right panel in Fig.~\ref{fig:3endpoints} shows the case of 
the asymmetric event topology from Fig.~\ref{fig:DecayTopologies}(c) with  
$(m_A,\;m_{B_2},\;m_{B_1},\;m_C)=(500,\;400,\;300,\;200)$ GeV.
This time, the intermediate resonances, $B_1$ and $B_2$, are present, 
but their masses are not equal: $m_{B_1}=300$ GeV, while $m_{B_2}=400$ GeV.
Thus applying the relative constraint, $M_{B_1}=M_{B_2}$, during the 
minimization for $M_{2XC}$ and $M_{2CC}$ once again leads to an unphysical situation.
As a result, the $M_{2XC}$ and $M_{2CC}$ distributions again develop tails
beyond $M_{2XX}^{max}$, failing the test (\ref{eq:endpointtest}) and ruling out
the on-shell event topology of Fig.~\ref{fig:DecayTopologies}(a) 
as being the source of these events.

Note that in the last two cases, when the endpoint test failed,
it simply told us which event topology is wrong, but it did not specify the correct answer.
For this, we must develop further tests as in the next two subsections.
However, notice the distinctive shape of the distributions in the right panel of Fig.~\ref{fig:3endpoints}
in comparison with the middle panel. One might hope to use this shape difference 
to further discriminate among the event topologies of Fig.~\ref{fig:DecayTopologies}(b) 
and Fig.~\ref{fig:DecayTopologies}(c). However, such detailed shape analysis is beyond 
the scope of this paper.  

Note that the ability to discriminate among two alternative event topologies suggests an 
interesting application of the constrained $M_2$ variables in discriminating the signal from 
irreducible backgrounds \cite{newpaper}. The SM backgrounds have known 
event topologies, for which the corresponding on-shell constraints can be readily applied;
the resulting distributions will still have the same endpoint. With a suitably chosen cut
above this expected SM endpoint, one would be able to remove most, if not all, background events.
On the other hand, the signal event topology is generally different, and the signal events 
will migrate to higher values of $M_2$ once the kinematic constraints are imposed,
leading to a higher signal efficiency when using $M_2$ in place of $M_{T2}$.

\subsection{Dalitz plot test}
 \label{sec:Dalitz}

In this subsection, we develop a Dalitz plot test which 
enables us to discriminate the event topology in Fig.~\ref{fig:DecayTopologies}$(a)$ 
from those in~\ref{fig:DecayTopologies}$(b)$ and~\ref{fig:DecayTopologies}$(d)$. 
The idea is to use the invisible momenta obtained in the $M_2$ minimization
to form invariant mass combinations involving the final state invisible particles, $C_i$.

\begin{figure}[t]
\centering
\includegraphics[scale=0.75]{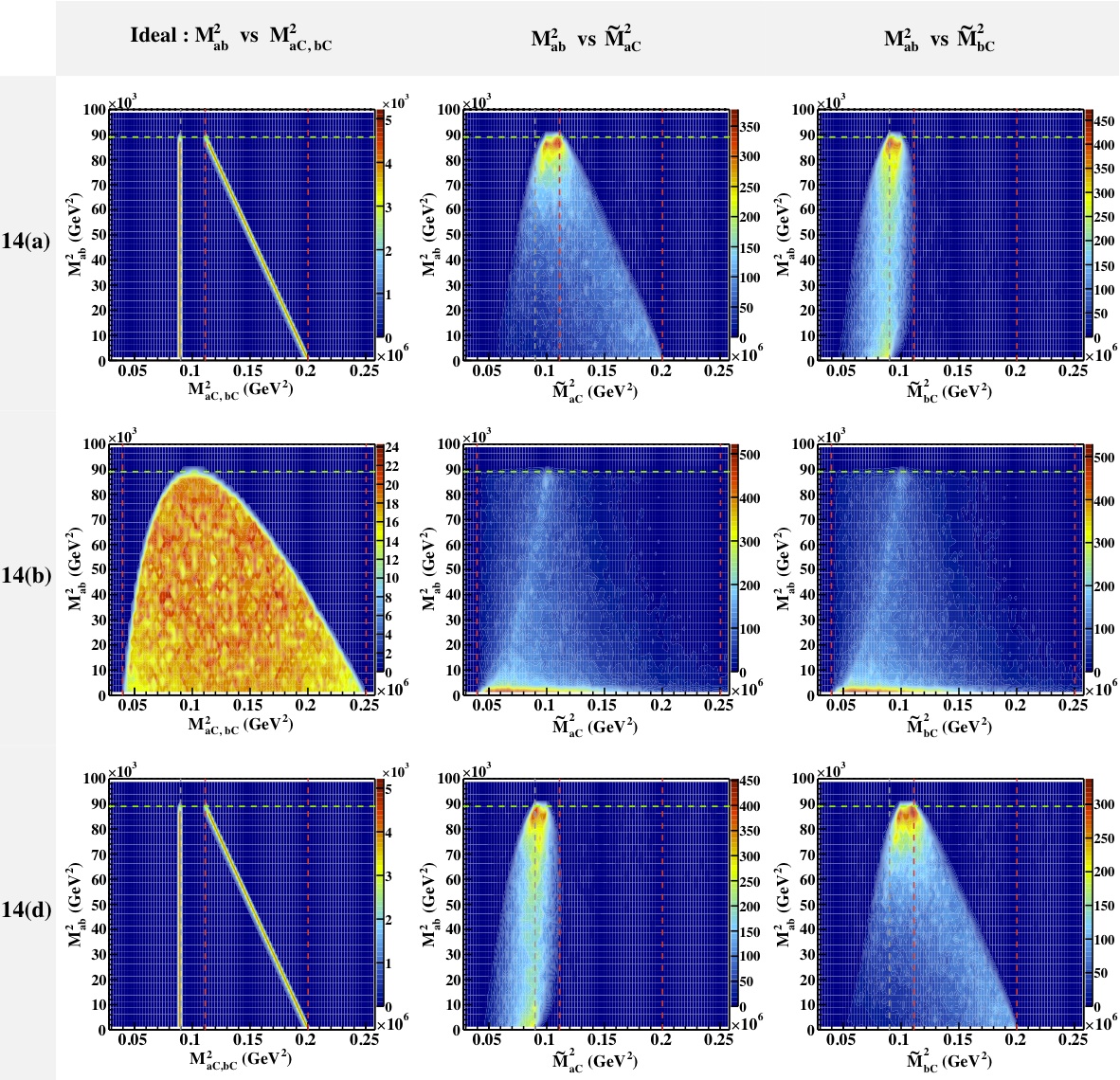}
\caption{\label{fig:3dalitzplots} Dalitz plots for the event topologies of Fig.~\ref{fig:DecayTopologies}$(a)$ (top row),
Fig.~\ref{fig:DecayTopologies}$(b)$ (middle row), and Fig.~\ref{fig:DecayTopologies}$(d)$ (bottom row). 
Using the invisible particle momenta obtained from $M_{2CC}(ab)$, 
we show scatter plots of $\tilde M_{aC}^2$ versus $M_{ab}^2$ (middle column) and
$\tilde M_{bC}^2$ versus $M_{ab}^2$ (right column). 
The left column shows the corresponding results in the ideal case when we use the {\em true} 
momenta of the invisible particles in the event.
The mass spectrum is fixed as in (\ref{eq:truemass}). }
\end{figure}

To see how the method works, let us assume that the signal comes from the 
event topology of Fig.~\ref{fig:DecayTopologies}$(a)$. First consider the ideal case 
when we have exact knowledge of the four momenta of the invisible particles, $C_i$. 
Since there are three particles in the final state of each decay chain,
$a_i$, $b_i$, and $C_i$, and we know their 4-momenta, we can form three invariant mass 
combinations, $M_{ab}$, $M_{bC}$, and $M_{aC}$. 
Since particles $b_i$ and $C_i$ originate from the same mother particle, $B_i$, 
$M_{bC}$ simply equals the mass, $m_B$, of that mother particle, regardless of the value of $M_{ab}$. 
Therefore, the Dalitz plot in the $(M_{bC}^2,M_{ab}^2)$ plane is characterized by a single vertical line:
\bea
M_{bC}^2=m_B^2 \hspace{1cm}\hbox{for any }M_{ab}^2\in[0,(M_{ab}^{\max})^2],
\label{MbC2}
\eea
with $M_{ab}^{max}$ given by (\ref{Mabmax}).
On the other hand, $M_{aC}^2$ takes values within a given range consistent with the sum rule 
\bea
M_{aC}^2 = m_A^2-m_B^2+m_C^2 -M_{ab}^2  \hspace{1cm}\hbox{for any }M_{ab}^2\in[0,(M_{ab}^{\max})^2],
\label{MaC2}
\eea
which is nothing but a straight line with a negative slope in the plane of $(M_{aC}^2,M_{ab}^2)$. 
The predictions (\ref{MbC2}) and (\ref{MaC2}) in this idealized case are illustrated 
in the upper left panel of Fig.~\ref{fig:3dalitzplots},
where the vertical line corresponds to (\ref{MbC2}), and the slanted line corresponds to (\ref{MaC2}).


We are now ready to consider the more realistic case in which we do not have exact knowledge of the 
individual momenta of the invisible particles, $C_i$, but instead obtain them from the $M_{2CC}$ ansatz.
The corresponding results are shown in the remaining two plots in the top row of Fig.~\ref{fig:3dalitzplots} ---
the middle panel shows a scatter plot in the $(M_{aC}^2,M_{ab}^2)$ plane, while the 
right panel shows a scatter plot in the $(M_{bC}^2,M_{ab}^2)$ plane.
Since the invisible momenta are only approximated, the correlations are not exactly linear, 
but nevertheless they tend to follow the general trends given by (\ref{MbC2}) and (\ref{MaC2}). 

Let us now move on to the event topology of Fig.~\ref{fig:DecayTopologies}(b).
This case is illustrated in the middle row of Fig.~\ref{fig:3dalitzplots}.
Since the intermediate $B_i$ resonance is absent, the visible particles, $a_i$ and $b_i$, 
arise from the same vertex and are on equal footing. Thus, we expect the associated 
Dalitz plots in the $(M_{aC}^2,M_{ab}^2)$, and $(M_{bC}^2,M_{ab}^2)$ planes to be very similar,
and indeed this is what we observe by comparing the middle and right panels of the middle row.
We therefore conclude that the similarity between the two Dalitz plots is an indication of an 
off-shell scenario as in Fig.~\ref{fig:DecayTopologies}(b).

Finally, the bottom row in Fig.~\ref{fig:3dalitzplots} represents the case of the event topology 
from Fig.~\ref{fig:DecayTopologies}(d), which again has a pair of identical intermediate resonances,
$B_i$, only now the visible particles, $a_i$ and $b_i$, are emitted in the opposite order --- 
$b_i$ comes first and $a_i$ comes second\footnote{In this way we are trying to resolve the combinatorial ambiguity 
associated with the assignment of visible particles within a given decay chain.}.
Comparing this to the decay topology of Fig.~\ref{fig:DecayTopologies}(a),
we see that the only difference is that the roles of the visible particles, $a_i$ and $b_i$, are reversed.
Therefore our previous analysis leading up to eqs.~(\ref{MbC2}) and (\ref{MaC2}) still applies,
only now the two trends are interchanged --- the correlation in the 
$(M_{aC}^2,M_{ab}^2)$ plane is expected to be a vertical straight line, while the 
correlation in the $(M_{bC}^2,M_{ab}^2)$ plane is expected to be a slanted straight line.
This ideal case with perfect knowledge of the invisible momenta is shown in the 
left bottom panel of Fig.~\ref{fig:3dalitzplots}. The more realistic case, in which the 
invisible momenta are taken from the $M_{2CC}(ab)$ minimization, is presented in the middle and 
right bottom panels of Fig.~\ref{fig:3dalitzplots}. As expected, the behavior is exactly the 
opposite of what we observed in the corresponding plots in the upper row of Fig.~\ref{fig:3dalitzplots}.
Our conclusion, therefore, is that whenever the two scatter plots are different,
the visible particle in the scatter plot with the vertical correlation is the one which is emitted second, 
while the visible particle in the slanted scatter plot is the one which is emitted first.

\subsection{Resonance scatter plot test }
\label{sec:scatter}

Finally, we describe a test aimed at detecting and identifying any intermediate resonances, $B_i$.
In particular, we shall revisit the event topologies from Figs.~\ref{fig:DecayTopologies}(a),
\ref{fig:DecayTopologies}(b), and \ref{fig:DecayTopologies}(c) and attempt to answer the questions:
\begin{itemize}
\item Is there an intermediate $B_i$ resonance in each decay chain?
\item If so, are the two $B_i$ particles the same or not?
\end{itemize}
Once again, the idea is to use the invisible momenta found by one of the $M_2$-type minimizations
and then reconstruct the masses of the hypothesized $B_i$ resonances. 
As already discussed in Sec.~\ref{sec:relation}, the novel advantage of the $M_2$-type variables 
(e.g.,~over transverse variables like $M_{T2}$) is that they supply the full 3-momenta 
of the invisible particles, including the longitudinal components. Thus, it becomes possible 
to carry out the direct reconstruction of any heavy particles along the decay chain.
In our case, to form the mass of particle $B_i$, we simply use the measured 4-momentum 
of $b_i$ and the momentum of $C_i$ obtained in the minimization of $M_{2CX}(ab)$\footnote{Here
we prefer to avoid any bias from using momenta from $M_{2XC}(ab)$ or $M_{2CC}(ab)$, which
assume the presence of identical intermediate resonances from the outset.}.
In order to avoid the two-fold ambiguity discussed in Sec.~\ref{sec:uniqueness},
we use only ``balanced" events, for which the invisible momentum configuration is unique.

\begin{figure}[t]
\centering
\includegraphics[trim=0cm 0.2cm 0.5cm 1cm, width=4.9cm]{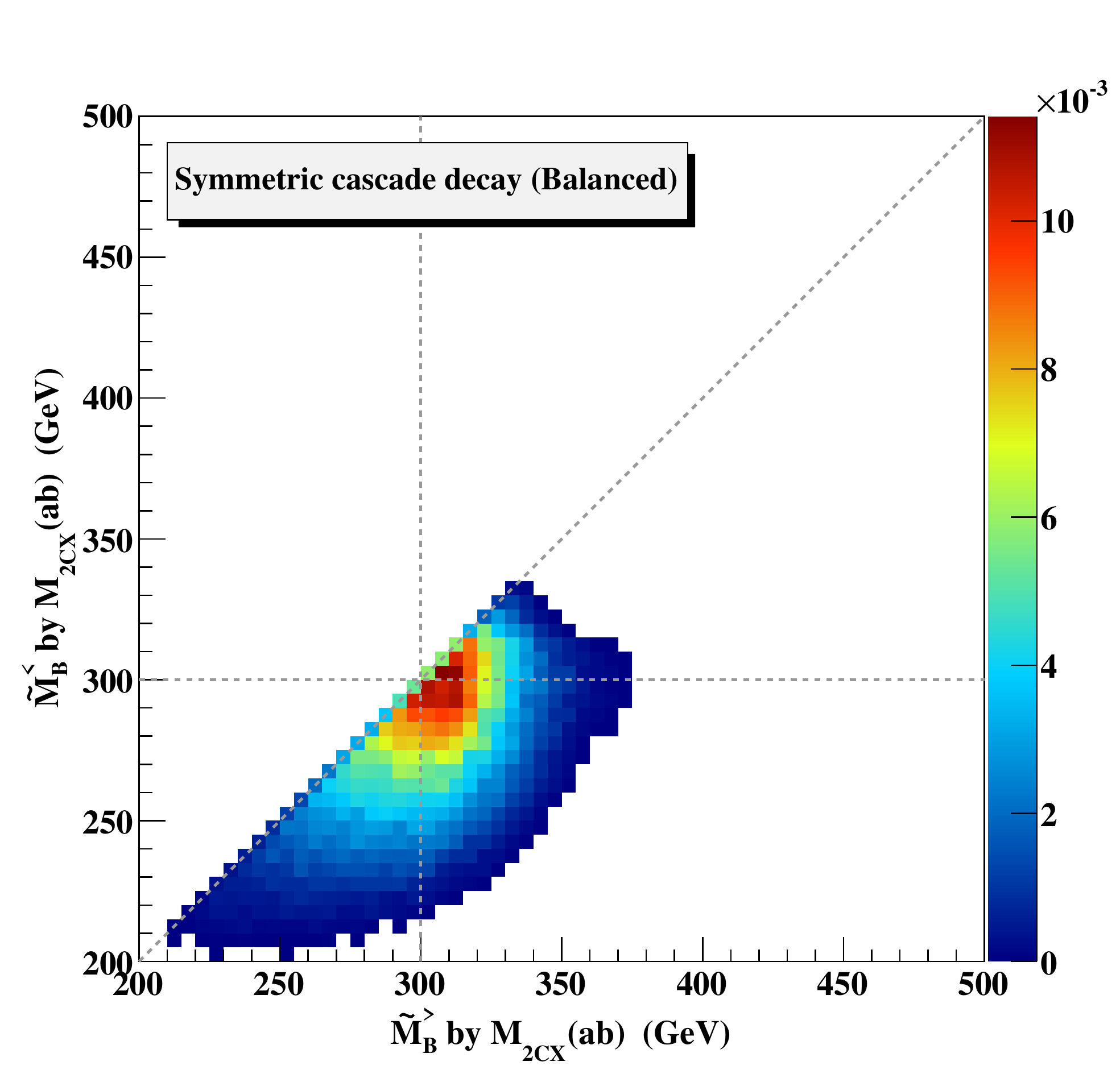}
\includegraphics[trim=0cm 0.2cm 0.5cm 1cm, width=4.9cm]{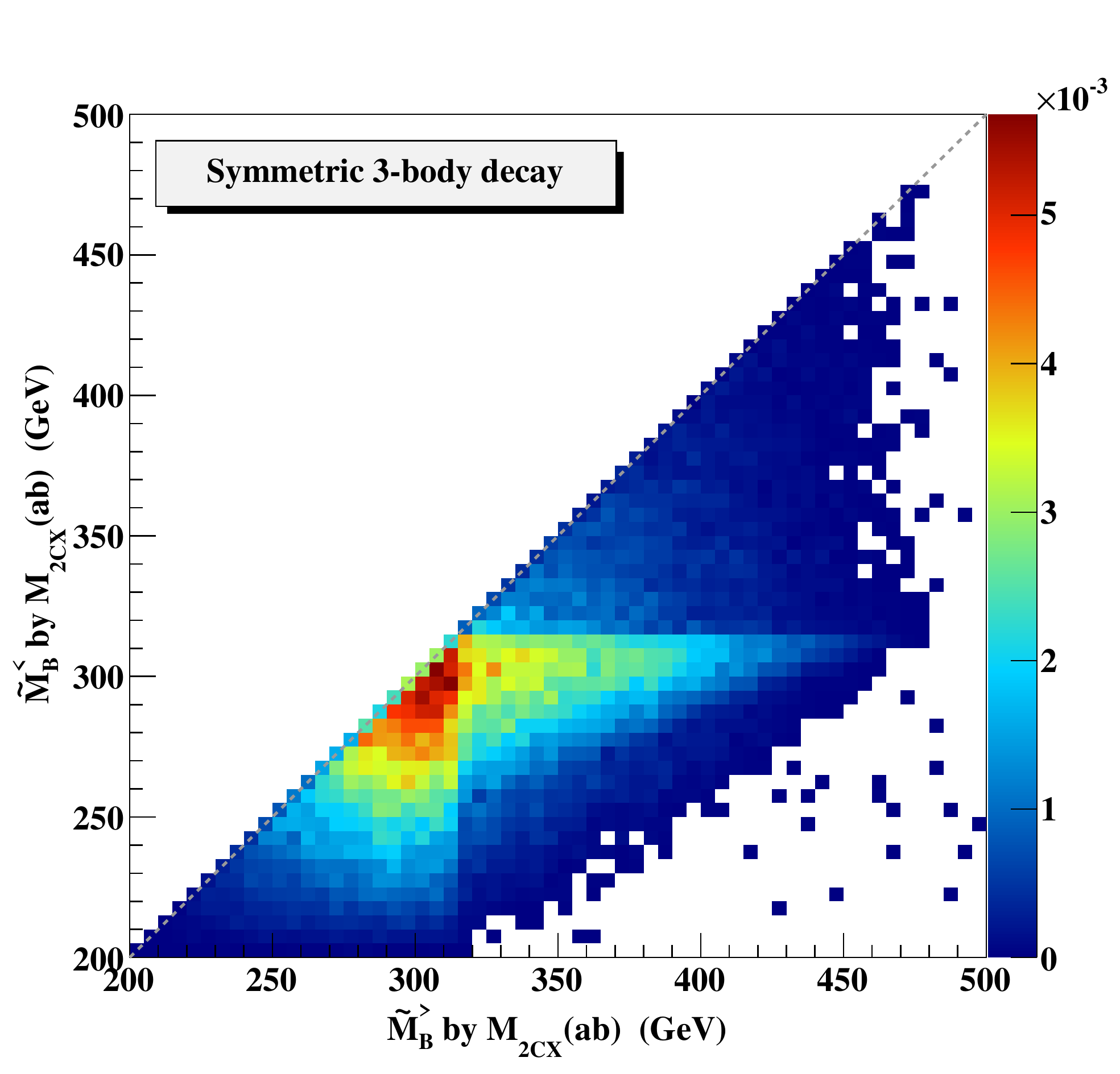}
\includegraphics[trim=0cm 0.2cm 0.5cm 1cm, width=4.9cm]{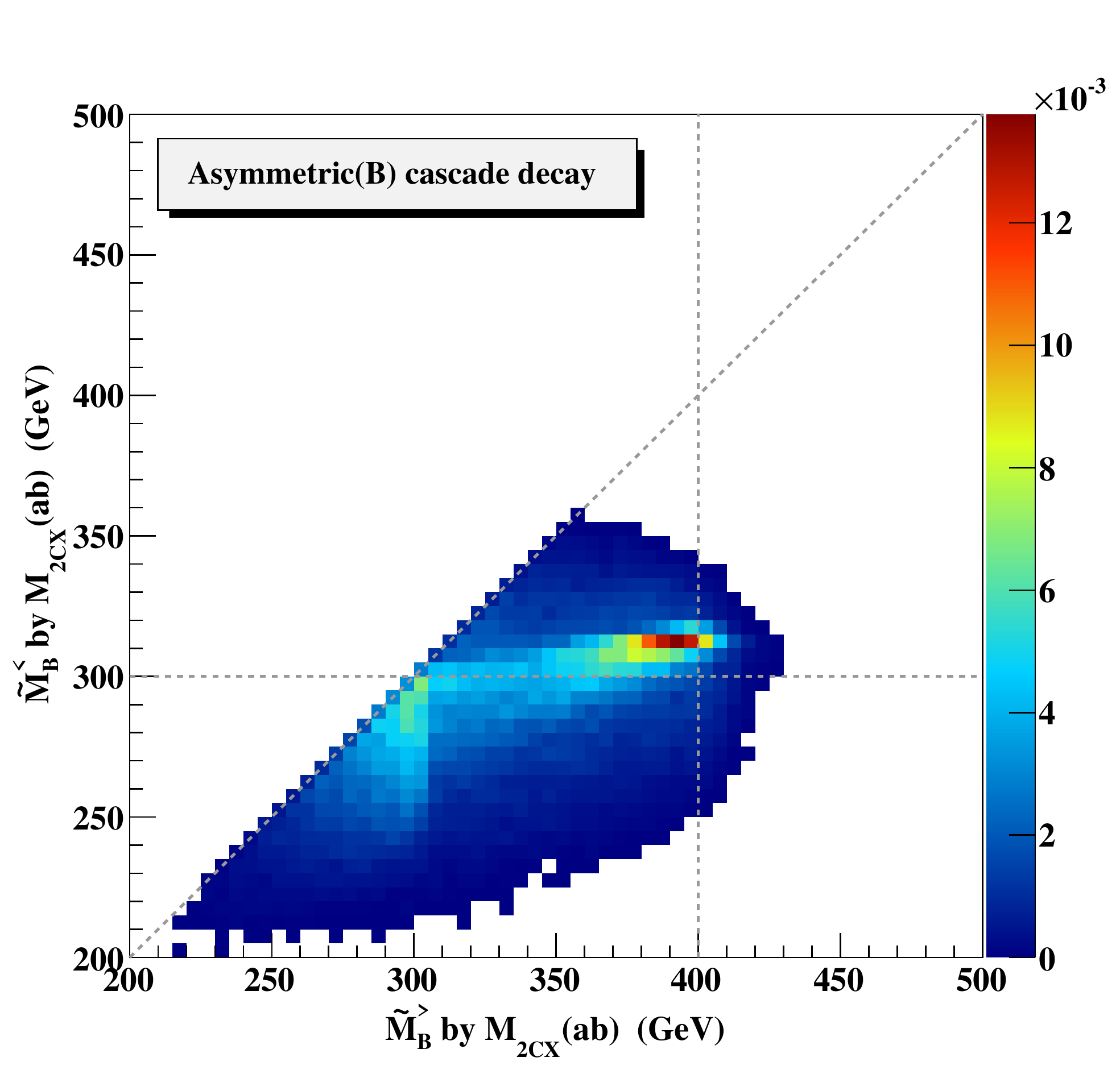}
\caption{\label{fig:3resonanceplots} Scatter plots of the reconstructed masses of the two intermediate resonances, 
$\tilde M_{B_1}$ and $\tilde M_{B_2}$, for the three scenarios from Fig.~\ref{fig:3endpoints},
with invisible momenta taken from the $M_{2CX}(ab)$ minimization.
The larger of the two reconstructed masses, $\tilde M_{B}^>$, is plotted on the $x$-axis, while
the smaller of the two reconstructed masses, $\tilde M_{B}^<$, is plotted on the $y$-axis.
The vertical and horizontal black dashed lines denote the true masses of the associated 
relative particles.}
\end{figure}

Since each event contains two decay chains, we will obtain two reconstructed values per event, 
$\tilde M_{B_1}$ and $\tilde M_{B_2}$, which we order as usual as 
\bea
\tilde M_{B}^> &=& \max\left\{ \tilde M_{B_1}, \tilde M_{B_2} \right\},  \\ [2mm]
\tilde M_{B}^< &=& \min\left\{ \tilde M_{B_1}, \tilde M_{B_2} \right\}.  
\eea
We then investigate the resonance structure of the corresponding scatter plot in the $(\tilde M_{B}^>,\tilde M_{B}^<)$ plane,
as shown in Fig.~\ref{fig:3resonanceplots}.

The left panel in Fig.~\ref{fig:3resonanceplots} represents the case of the event topology 
from Fig.~\ref{fig:DecayTopologies}(a), which has two identical intermediate resonances, $B_1$ and $B_2$.
Correspondingly, the scatter plot exhibits a distinct clustering of events near the diagonal line
($\tilde M_{B}^>=\tilde M_{B}^<$), indicating the presence of such identical resonances. Furthermore, 
we can also roughly read the mass scale as $M_{B_i}\sim 300$ GeV 
(compare with the true values marked with the black dashed lines). 
Now contrast this situation with the case of the event topology from 
Fig.~\ref{fig:DecayTopologies}(c), which is shown in the rightmost panel of Fig.~\ref{fig:3resonanceplots}.
Again, we find a narrow clustering of points, indicating the presence of intermediate $B_i$ 
resonances. Now, however, the cluster lies significantly far from the diagonal line, 
implying that the intermediate resonances are {\em different}. The location of the cluster 
is also consistent with the input mass spectrum ($m_{B_1}=300$ GeV, $m_{B_2}=400$ GeV, 
as indicated with the black dashed lines).

The third example, shown in the middle panel of Fig.~\ref{fig:3resonanceplots}, 
is the event topology from Fig.~\ref{fig:DecayTopologies}(b).
The two decay chains are the same, so we expect most of the events to end up near the 
diagonal line $\tilde M_{B}^>=\tilde M_{B}^<$. However, since there are no intermediate resonances, 
we do not expect a significant clustering in any particular location and instead would expect a broader 
distribution that in the previous two resonant cases.
These expectations are confirmed in Fig.~\ref{fig:3resonanceplots} ---
the middle panel exhibits a large population near the diagonal line whose 
structure differs from that in the left panel, allowing us to 
distinguish the topology of Fig.~\ref{fig:DecayTopologies}(b) from
the topology of Fig.~\ref{fig:DecayTopologies}(a). 
Again, we defer a more detailed shape analysis to future work.  

\section{Conclusions and outlook}
\label{sec:conclusions}

The main goal of this paper is to advocate a wider use of the 3+1-dimensional $M_2$-type variables,
which so far have been used only sporadically
~\cite{Ross:2007rm,Barr:2008ba,Barr:2011xt,Mahbubani:2012kx}.
In contrast, transverse mass variables like $M_T$, $M_{T2}$, $M_{CT}$, etc. 
have found widespread application in both precision measurements~\cite{Chatrchyan:2013boa,ATLAS:2012poa} and 
in searches for new physics~\cite{Chatrchyan:2012jx,TheATLAScollaboration:2013xha}.
There are two main advantages of the 3+1-dimensional formulation in terms of $M_2$:
\begin{enumerate}
\item It is very easy to impose various additional assumptions about the underlying event topology~\cite{Mahbubani:2012kx}.
In this paper we illustrated this feature with the addition of on-shell constraints for the relative particles,
which led us to two new variables, $M_{2XC}$ and $M_{2CC}$. The benefits from $M_{2XC}$ and $M_{2CC}$
are twofold --- first, the solution for the longitudinal invisible momenta is unique, as discussed in Sec.~\ref{sec:2CC},
and second, their distributions exhibit much sharper endpoints, as demonstrated in Sec.~\ref{sec:mass}.
\item The minimization procedure required to calculate the value of $M_2$ fixes {\em all} components 
of the invisible particle momenta, including the longitudinal components. This gives an event with fully 
determined kinematics, opening the door for a number of precision reconstruction studies. 
As an illustration, in Sec.~\ref{sec:peak}, we reconstructed the mass of the relative particle and showed that
the peak of the resulting distribution is nicely correlated with the true mass of the relative particle.
This provides a new technique for mass measurements in missing energy events,
which is complementary to the existing methods based on measuring kinematic endpoints.
\end{enumerate}

It is interesting to note that even for a process as simple as the one studied here (see Fig.~\ref{fig:process}),
we were able to define a relatively large number of $M_2$-type variables, summarized in Table~\ref{tab:M2variables}. 
While the casual reader might feel intimidated by this proliferation of kinematic mass variables, we emphasize that 
there is a great benefit in having such a large arsenal of kinematic variables at one's disposal.
The main reason why there are so many variables is that each involves different levels of assumptions.
Thus, by testing for consistency of the results obtained with two different variables, we 
are essentially checking the validity of the assumptions that are present in one of the variables but not the other.

Following this idea, we developed several tests for distinguishing among the alternative event topologies of
Fig.~\ref{fig:DecayTopologies}, which lead to the same final state:
\begin{itemize}
\item {\em Endpoint test.} In Sec.~\ref{sec:endpoint}, we proposed a test which compares the endpoints of 
the distributions of $M_2$ variables with and without relative constraints. If the constraints are satisfied in the 
event sample, the kinematic endpoints would match (even though the shapes of the distributions are generally different).
Conversely, if the mass constraints are not satisfied, the endpoints will be different, which is an indication
that our hypothesis regarding the event topology is wrong. We have checked that this test is applicable
even when one does not have precise knowledge of the daughter mass.
\item {\em Dalitz plot test.} In Sec.~\ref{sec:Dalitz}, we used the fact that the computation of the 
$M_2$ variables supplies values for the 4-momenta of the invisible particles and proposed to build
Dalitz-type plots of invariant mass combinations which include the invisible particles themselves (see Fig.~\ref{fig:3dalitzplots}).
We showed that the distinctive shape of the Dalitz scatter plots can be used to ascertain the
presence of intermediate resonances and to resolve the combinatorial ambiguity related to the
ordering of the visible final state particles along the decay chain.
\item {\em Resonance scatter plot test.} The invisible momenta supplied by $M_2$ found another
application in Sec.~\ref{sec:scatter}, where we were able to test for the symmetry of the events, i.e.,
whether the two decay chains are the same or not (see Fig.~\ref{fig:3resonanceplots}).
\end{itemize} 
These are just a few of the many potential applications of the $M_2$ variables --- for 
example, one could imagine spin measurements along the lines of~\cite{Cho:2008tj,Cheng:2010yy,Guadagnoli:2013xia},
using the invisible particle momenta supplied by the $M_2$ minimizations.
It is also possible to further extend the set of variables from Table~\ref{tab:M2variables}
to more complicated event topologies --- e.g.,~decay chains with more than one relative particle, 
decay chains with relatives of known mass, etc. One technical problem which will need to be addressed 
in the near future is the lack of a public code for the calculation of the on-shell constrained $M_2$ variables. 
The availability of such code would certainly encourage more experimentalists to make use 
of these variables whose benefits seem undeniable.


\acknowledgments
We would like to thank K.~Kong for stimulating discussions. This work is supported by DOE Grant No.~DE-FG02-97ER41990
and by the World Premier International Research Center Initiative (WPI Initiative), MEXT, Japan.
WC and DK are supported by LHCTI postdoctoral fellowships under grant NSF-PHY-0969510.

\begin{appendix}

\section{Proof that $M_{T2}=M_{2CX}$ with the method of Lagrange multipliers}
\label{app:proofMT2M2CX}

In this appendix, we show the equivalence between $M_{T2}$ and $M_{2CX}$ using the method of Lagrange multipliers. 
For concreteness, the formal proof is presented for the $(ab)$ subsystem, 
but the same argument can be applied to the other subsystems, $(a)$ and $(b)$, as well.

In order to calculate $M_{2CX}^2(ab)$, we must perform the minimization of the function
\bea
\max\left[M_{A_1}^2(\vec{q}_{1T},\;\Delta\eta_1;\;\tilde{m}),\;M_{A_2}^2(\vec{q}_{2T},\;\Delta\eta_2;\;\tilde{m})  \right],
\label{app:fn}
\eea
subject to the two constraints
\bea
M_{A_1}^2(\vec{q}_{1T},\;\Delta\eta_1;\;\tilde{m})&=&M_{A_2}^2(\vec{q}_{2T},\;\Delta\eta_2;\;\tilde{m}), 
\label{app:MAi}
\\ [2mm]
\vec{q}_{1T}+\vec{q}_{2T}&=& \mpt.
\label{app:mpt}
\eea
Here we have already assumed that the hypothesized masses of the daughter particles, $C_i$, are the same, 
so that there is a single input test mass, $\tilde m$. We have also expressed the parent masses, $M_{A_i}$, as functions of 
$\Delta\eta_i$ instead of $q_{iz}$, as in (\ref{eq:MAi}).

We can use the method of Lagrange multipliers to reformulate the
problem as the 
unconstrained minimization of a new target function 
\bea
f(\vec{q}_{1T},\vec{q}_{2T},\Delta\eta_1,\Delta\eta_2,\vec{\lambda}_T,\lambda_{\eta};\tilde{m})&& \nonumber \\
&&\hspace{-4cm}=\frac{1}{2}\left\{E_{v_1T}E_{q_1T}\cosh\Delta \eta_1-\vec{p}_{v_1T}\cdot\vec{q}_{1T}+E_{v_2T}E_{q_2T}\cosh\Delta \eta_2-\vec{p}_{v_2T}\cdot\vec{q}_{2T}\right\} \nonumber \\
&&\hspace{-4cm}+\lambda_{\eta}\left\{E_{v_1T}E_{q_1T}\cosh\Delta \eta_1-\vec{p}_{v_1T}\cdot\vec{q}_{1T}-E_{v_2T}E_{q_2T}\cosh\Delta \eta_2+\vec{p}_{v_2T}\cdot\vec{q}_{2T}\right\} \nonumber \\
&&\hspace{-4cm}+\vec{\lambda}_T\cdot(\vec{q}_{1T}+\vec{q}_{2T} - \mpt
), \label{eq:optimumeq}
\eea
which needs to be minimized over all of its arguments: 
$\vec{q}_{1T}$, $\vec{q}_{2T}$, $\Delta\eta_1$, $\Delta\eta_2$, $\vec{\lambda}_T$, and $\lambda_{\eta}$.
The constraint (\ref{app:MAi}) is implemented through the Lagrange multiplier $\lambda_\eta$,
while the constraint (\ref{app:mpt}) is incorporated through the Lagrange multiplier $\vec{\lambda}_T$.
In view of the constraint (\ref{app:MAi}) in the first term
we have replaced (\ref{app:fn}) with the average of $M_{A_1}^2$ and $M_{A_2}^2$.

The extremum conditions for $\vec{q}_{iT}$ and $\Delta\eta_i$ read:
\bea
\vec{\triangledown}_{q_{1T}}f&=&\left(\frac{1}{2}+\lambda_{\eta}\right)\left(\frac{E_{v_1T}}{E_{q_1T}}\cosh \Delta\eta_1\vec{q}_{1T}-\vec{p}_{v_1T}\right)+\vec{\lambda}_T =0, \label{eq:delq1}\\
\vec{\triangledown}_{q_{2T}}f&=&\left(\frac{1}{2}-\lambda_{\eta}\right)\left(\frac{E_{v_2T}}{E_{q_2T}}\cosh \Delta\eta_2\vec{q}_{2T}-\vec{p}_{v_2T}\right)+\vec{\lambda}_T =0, \label{eq:delq2}\\
\frac{\partial f}{\partial \Delta\eta_1}&=&\left(\frac{1}{2}+\lambda_{\eta}\right)E_{v_1T}E_{q_1T}\sinh \Delta\eta_1 =0, \label{eq:deleta1}\\
\frac{\partial f}{\partial \Delta\eta_2}&=&\left(\frac{1}{2}-\lambda_{\eta}\right)E_{v_2T}E_{q_2T}\sinh \Delta\eta_2 =0. \label{eq:deleta2}
\eea
There are two cases which can be considered separately:
i) $\lambda_{\eta}=1/2$ (or $\lambda_{\eta}=-1/2$), and ii) $\lambda_{\eta}\neq \pm1/2$. 

\paragraph{i) $\lambda_{\eta}=1/2$:} 
Since $\lambda_{\eta}=1/2$, (\ref{eq:deleta2}) is automatically solved, so 
$\Delta\eta_2$ remains arbitrary. Then (\ref{eq:delq2}) implies $\vec{\lambda}_T=0$,
and from (\ref{eq:deleta1}) it follows that $\Delta\eta_1=0$. Finally, (\ref{eq:delq1})
leads to $\vec{q}_{1T}=\frac{E_{q_1T}}{E_{v_1T}}\vec{p}_{v_1T}$. 
Substituting these results into (\ref{eq:optimumeq}), we get
\bea
f=E_{v_1T}E_{q_1T}-\vec{p}_{v_1T}\cdot\vec{q}_{1T}=M_{TA_1}^2(ab),
\eea
which is nothing but the transverse mass of $A_1$ as given in (\ref{eq:MTAi}). 
This implies that minimizing $f$ is equivalent to minimizing the transverse 
mass of $A_1$.

The same logic can be applied to the case $\lambda_{\eta}=-1/2$,
where one finds that the problem reduces to the minimization of the 
transverse mass of $A_2$. Thus we conclude that these two cases with $|\lambda|=\frac{1}{2}$ 
simply correspond to the {\it unbalanced} configuration of the $M_{T2}(ab)$ variable. 

\paragraph{ii) $\lambda_{\eta}\neq \pm1/2$:} 
Since $\lambda_{\eta}\neq \pm1/2$, the only way to satisfy Eqs.~(\ref{eq:deleta1}) 
and~(\ref{eq:deleta2}) is to have $\Delta\eta_i=0$. This reduces the function (\ref{eq:optimumeq}) to 
\bea
f&=&\frac{1}{2}(M_{TA_1}^2(ab)+M_{TA_2}^2(ab))+\lambda_{\eta}(M_{TA_1}^2(ab)-M_{TA_2}^2(ab))\nonumber \\
&+&\vec{\lambda}_T\cdot(\vec{q}_{1T}+\vec{q}_{2T}-\mpt),
\eea
which is nothing but the Lagrange function associated with the {\it balanced} solution of the corresponding $M_{T2}$ variable. 

\medskip

\noindent From i) and ii) we see that $M_{2CX}(ab)$ includes both the balanced and the unbalanced 
configurations of $M_{T2}(ab)$, thus it follows that $M_{2CX}(ab)=M_{T2}(ab)$. 


\end{appendix}

\end{document}